\pdfoutput=1
\documentclass{aa} % for printed version
\usepackage{natbib, my_def, graphicx} 
\bibpunct{(}{)}{;}{a}{}{,}
\bibstyle{aa-package/aa}     
\def \etal   {{~et~al.~}}
\begin{document}
\title{A sample of weak blazars at milli-arcsecond resolution}

%\author { ?? \inst{1,2} }
 
\author{ F. Mantovani \inst{1,2} \and
         M. Bondi     \inst{2} \and
	 K.-H. Mack   \inst{2} \and
	 W. Alef      \inst{1} \and
	 E. Ros       \inst{1,3,4} \and
      J.A. Zensus     \inst{1}  
        }
\offprints{Franco Mantovani\\
  \email{fmantovani@mpifr-bonn.mpg.de}}
\institute{Max-Planck-Institut f\"{u}r Radioastronomie, Auf dem H\"{u}gel 69, D-53121 Bonn, Germany
\and Istituto di Radioastronomia -- INAF, Via Gobetti 101, I--40129, Bologna, Italy
\and Observatorio Astron\`omico, Universitat de Val\`encia, Parc Cient\'ific, C. Catedr\'atico Jos\'e Beltr\'an 2, 
E-46980 Paterna, Val\`encia, Spain
\and Departament d'Astronomia i Astrof\'isica, Universitat de Val\`encia, C. Dr. Moliner 50, E-46100 Burjassot,Val\`encia, Spain    
}

\date{Received \today; accepted ???}

\abstract
% Context heading
{}
% Aims heading
{    We started a follow-up investigation of the ``Deep X-ray Radio Blazar Survey'' objects with declination
$>-10$\,deg to a better understand 
of the blazar phenomenon. We undertook a survey with the European Very Long Baseline 
Interferometry Network at 5\,GHz 
to make the first images of a complete sample of weak blazars, aiming at a follow-up 
comparison between high- and 
low-power samples of blazars.  
}
% Methods heading 
{  We observed 87 sources with the EVN at 5\,GHz during the period October 2009 to May 2013. 
The observations were correlated at the Max-Planck-Institut 
f\"ur Radioastronomie and at the Joint Institute for VLBI in Europe. 
The correlator output was analysed using both the ${\cal AIPS}$ and DIFMAP software packages.
}
% Results heading
{All of the sources observed were detected. Point-like sources are found in 39 cases on a 
milli-arcsecond scale, and 48 show core-jet structure. The total flux density distribution 
at 5\,GHz has a median value $\langle S \rangle$=44$^{+23}_{-10}$\,mJy.  A total flux density 
$\leq$150\,mJy is observed in 68 out of 87 sources. Their brightness temperature T$_b$ ranges 
between $10^7$\,K and $10^{12}$\,K. According to the spectral indices previously obtained with 
multi-frequency observations, 58 sources show a flat spectral index, and 29 sources show a 
steep spectrum or a spectrum peaking at a frequency around 1$-$2\,GHz. Adding to the DXRBS 
objects we observed those already observed with ATCA in the Southern sky, we found that 14 
blazars and a Steep Spectrum Radio Quasars, are associated to $\gamma$-ray emitters. 
}
% Conclusions heading 
{ We found that 56 sources can be considered blazars.  We also detected 2 flat spectrum 
narrow line radio galaxies. About 50\% of the blazars associated to a $\gamma$-ray object are 
BL\,Lacs, confirming that they are more likely detected among blazars $\gamma$-emitters. 
We confirm the correlation 
found between the source core flux density and the $\gamma$-ray photon fluxes down to fainter flux 
densities. We also found that weak blazars are also weaker $\gamma$-ray emitters compared to bright 
blazars. Twenty-two sources are SSRQs or Compact Steep-spectrum Sources, and 7 are GigaHz Peaked 
Sources. The 
available X-ray ROSAT observations allow us to suggest that CSS and GPS quasars are not obscured 
by large column of cold gas surrounding the nuclei. We did not find any significant difference in X-ray 
luminosity between CSS and GPS quasars.
}

%\authorrunning{?? et al.}
\authorrunning{F. Mantovani et al.}
\titlerunning{Weak blazars at milli-arcsecond resolution}
\maketitle
\section{Introduction} 

Less than 10\% of all the active galactic nuclei (AGN) is composed of flat spectrum radio quasars
(FSRQs) and BL Lacertae (BL\,Lac) objects. Together, these two classes of objects are
named blazars. Their broad band emission, mainly generated by synchrotron and inverse-Compton radiation,
extends from radio to $\gamma$ rays. Blazars are characterised by high luminosity, 
rapid variability, and high polarisation. The majority of
blazars are core-dominated objects that show apparent superluminal speeds. 
They are also well known $\gamma$-ray sources. The {\it Fermi} Gamma-ray Space Telescope has 
detected 1064 objects located at $|b|>10$ deg associated to blazars based on data accumulated over the 
first two years of the mission: second {\it Fermi}-LAT (2FGL) catalogue \citep{Nolan12}. 
The hint that relativistic Doppler boosting
is connected to $\gamma$-ray detection in AGN was suggested by many authors: e.g.
\citet{Kellermann04}, \citet{Kovalev05}, \citet{Taylor07}, \citet{Lister11}, and \citet{Ackermann11}.

Bright blazars are the target of many programmes for both single-dish and interferometric observations 
in the radio band. One of the largest samples of bright AGNs is the MOJAVE programme. It regularly 
observes with the Very Long Baseline Array sources with flux density $>$1.5\,Jy  
(see \citealt{Lister11} and references therein).

Several of the available small size blazar samples grounded for statistical studies were selected at 
flux densities $\sim$1\,Jy in the radio band and a few times 10$^{-13}$\,ergs\,cm$^{-2}$\,s$^{-1}$ in 
X-ray band. 

To increase the knowledge on blazars, a deeper, sizable sample has been assembled by \citet{Perlman98} 
and by \citet{Landt01}, the ``Deep X-ray Radio Blazar Survey" (DXRBS), which is later discussed in 
\citet{Padovani07}.  The DXRBS sample is currently the faintest and largest blazar sample with nearly 
complete optical identifications 
including both flat spectrum radio quasars (FSRQs) and BL\,Lac sources. The objects were selected 
cross-correlating all serendipitous X-ray sources in the ROSAT database WGACAT95 \citep{White95} with
several radio catalogues, namely GB6 \citep{Gregory96}, North20CM \citep{White92}, and the 6\,cm
Parks-MIT-NRAO catalogue PMN \citep{Griffith93}, plus a snapshot survey with the Australian Telescope
Compact Array (ATCA). The adopted selection criteria were 
a) X-ray flux in the range $\sim10^{-14}-\sim10^{-12}$\,ergs\,cm$^{-2}$\,sec$^{-1}$ 
depending on the region of the sky surveyed by ROSAT; 
b) a two-point spectral index $\alpha\leq0.7$; 
c) flux densities $>100$\,mJy at 20\,cm and $>50$\,mJy at 6\,cm.

For a deeper understanding of the blazar phenomenon, we started a follow-up investigation of the 
DXRBS objects,
aiming at a comparison between high- and low-power samples of blazars. We first
made simultaneous flux density measurements with the Effelsberg 100-m telescope at 2.64\,GHz, 4.85\,GHz,
8.35\,GHz, and/or 10.45\,GHz of all 
DXRBS sources with declination $>-$20 deg to properly compute their spectral index. Those measurements 
also 
allowed us to check for flux density variability comparing them with previous measurements found in 
the literature \citep{Mantovani11}.

Since it is reasonable to expect that the objects in the DXRBS are potential $\gamma$-ray sources 
detected 
by the {\it Fermi} Gamma-ray Space Telescope ({\it Fermi} GST), information on their radio structure at milli-arcsecond resolution 
is also essential for a comparison with bright blazar samples. For that reason we then undertook a survey with the European 
Very Long Baseline Interferometry Network (EVN) 
\footnote {The European VLBI Network is a joint facility of European, Chinese, South African and 
other radio astronomy institutes funded by their national research councils.} 
at 5\,GHz to obtain images of DXRBS objects, which is the subject of the present paper.  

In Section \ref{sec:observation} we summarise the observations and data
processing. Section \ref{sec:results} summarises the results achieved. Discussion and conclusions are
presented in Sections \ref{sec:discussion}  and \ref{sec:summary},
respectively.
\section{Observations and data reduction} 
\label{sec:observation}
 With the EVN we observed all the DXRBS sources with declination $>-10$\,deg. The list contains 
87 sources selected by applying the unique criteria of a cut in declination. The pointing positions 
were obtained 
from the FIRST catalogue for 63 objects, and for the remaining sources the coordinates were from 
the NVSS \citep{Mantovani11}.  
This sample of sources can therefore be considered as a ``complete'' sample that is statistically 
significant.

The observations made use of all the EVN telescopes available at the scheduled time. The observing setup 
was: frequency band 5\,GHz, dual circular polarization, recording bit rate 512 Mbps, and 2-bit sampling. 
Each source was tracked for four to five scans, each six minutes long at different hour angles for a 
better 
coverage of the {\it u,v} plane. About 30 minutes of observing time per source allow images to be
produced down to an rms noise of $\sim$0.02\,mJy/beam for good network performance. 
Information on the observing 
sessions from EM077a to EM097b are reported in Table~\ref{tab:sessions}.  
\subsection{Data reduction} 

The EVN observations were correlated at the DiFX software correlator \citep{Deller11} of the 
Max-Planck-Institut f\"ur Radioastronomie (Bonn, Germany), and those in real time (e-EVN) 
\footnote{e-VLBI research infrastructure in Europe is supported by the European Union's Seventh 
Framework Programme (FP7/2007-2013) under grant agreement number RI-261525 NEXPReS.} were correlated 
at the Joint Institute for VLBI in Europe (Dwingeloo, The Netherlands). 
The raw data, output from the correlator, were integrated for one second, corresponding to a field of 
view of about 11\arcsec. 
Table~\ref{tab:sessions} shows that the sample of sources was observed with patchy telescope arrays. 
Moreover, we had telescopes that were affected by bad weather during the observing sessions or by 
abnormal functioning of the data acquisition terminals. Data provided in those cases were not 
particularly good so were often discarded. As a result, the rms noise values in the images are 
rather disperse, as reported in Table~\ref{tab:parameters}.  

The correlator output was calibrated in amplitude and phase using ${\cal AIPS}$
\footnote {${\cal AIPS}$ is the NRAO's {\it Astronomical Image Processing System}} 
and imaged using DIFMAP 
\footnote {DIFMAP is part of the {\it Caltech VLBI software package}} \citep{Shepherd95}.
Total power measurements taken at the same times as the observations and the gain curve of the 
telescopes were applied in the amplitude calibration process.
The visibilities of each individual data set were displayed baseline-by-baseline and carefully 
edited, which gave a reading of bad data. To check for other problems that affect the data, 
closure 
phases, which are defined as the sum of visibility phases around a close triangle of three antennas, 
were also inspected. In all cases, the source showed up at the centre of the plotted field when 
generating the first dirty map. 
The self-calibration procedure, which uses closure amplitude to determine telescope amplitude 
corrections, was applied. The images were produced using the difference mapping technique, 
in which the observer converge on a model by progressively cleaning the residual map, and 
self-calibrating the phases using the latest model. The model was subtracted from the data in the 
{\it u-v} plane at any iteration, thus avoiding aliasing of the side lobe within the field of view. The 
cleaning was done on a small window around the brightest component found in the residual map. 
\subsection{Comments on individual observing sessions} 

The self-calibration procedure, which uses closure amplitudes to determine
telescope amplitude corrections, gave calibration for EM077a factors within
20\% of unity for all the telescopes except Sh, Ur, and Wb. The often insufficient coverage 
of the {\it u,v}
plane in particular for the longest baselines to Sh and Ur, may explain this behaviour. The reasons for 
the poor calibration of data provided by Wb are unclear.

EM077b and EM077c were rather unsuccessful. Several telescopes did not provide useful data. Consequently,
the images obtained from these observations were not considered. The sources in
both the observing lists were re-observed at a later stage, with
the exception of J1332.7$+$4722 and J1213.0$+$3248, for which
images were made using these data sets.

In EM077d four telescopes did not provide fringes. The calibration factors were within 20\%
of unity for the remaining successful telescopes.

In EM077e all telescopes worked properly and the calibration factors for the amplitude
corrections were within 20\% with the exception of Cm, Mc, and Jb, which required larger factors. The
correction factors found were consistent for both fringe finders observed during that session.

In EM077f all stations provided data. However, their quality was often
lower than expected. The closure amplitudes method showed that large calibration factors were
required for most of the stations. 

EM097a was quite successful. All stations provided good data. The amplitude corrections
were within 20\% for all telescopes but Jb and Hh.

Finally, EM097b was less successful. In particular the data acquired for the unique fringe finder,
namely 4C454.3, were rather poor. As a result, not only was the calibration in amplitude not 
satisfactory, but 
also many amplitude and phase solutions failed in fringe-fitting the data.

\section{Results} 
\label{sec:results} 
\subsection{Milli-arcsecond structures and spectral indices} 

All of the 87 sources observed have been detected.  Among them, point-like 
sources are found in 39 cases, and 48 show core-jet structure. Examples are presented 
in Fig.~\ref{fig:structure}.
The classification of the sources based on their milli-arcsecond (mas) structure 
and on the spectral indices
\footnote{The spectral index is defined as S$_{\nu}\propto\nu^{-\alpha}$.} 
previously reported by \citet{Mantovani11} is shown in Table~\ref{tab:struct-index}
where the spectral indices have been defined as
``Ultra Steep'' in case of $\alpha > 0.7$,
``Steep'' when $\alpha=0.7-0.5$,
``Steep-flat'' when $\alpha$ is steep at lower frequencies and flat
tens at higher frequencies,
``Flat-steep'' when $\alpha$ is flat at lower frequencies and steep 
at higher frequencies,
``Flat'' when $\alpha < 0.5$,  
Giga-Hertz Peaked (``GPS'') when the shape is convex with a peak at 
about the intermediate observing frequencies, and 
``Inverted'' when the spectral index shape is mainly straight and the flux density clearly 
increases with frequency. The 56 objects with flat or inverted
spectral index ($\alpha<0.5$) can be considered  bona fide blazars. Two are flat-spectrum 
narrow line radio galaxies (NLRGs). 
Sources that show ultra-steep and steep spectral index ($\alpha>0.5$), are most probably steep spectrum
radio quasars (SSRQs) or compact steep spectrum sources (CSSs).   

\vspace{-0.3cm}
\begin{figure}%[t]
\addtocounter{figure}{+0}
\centering
\resizebox{\hsize}{!}{\includegraphics{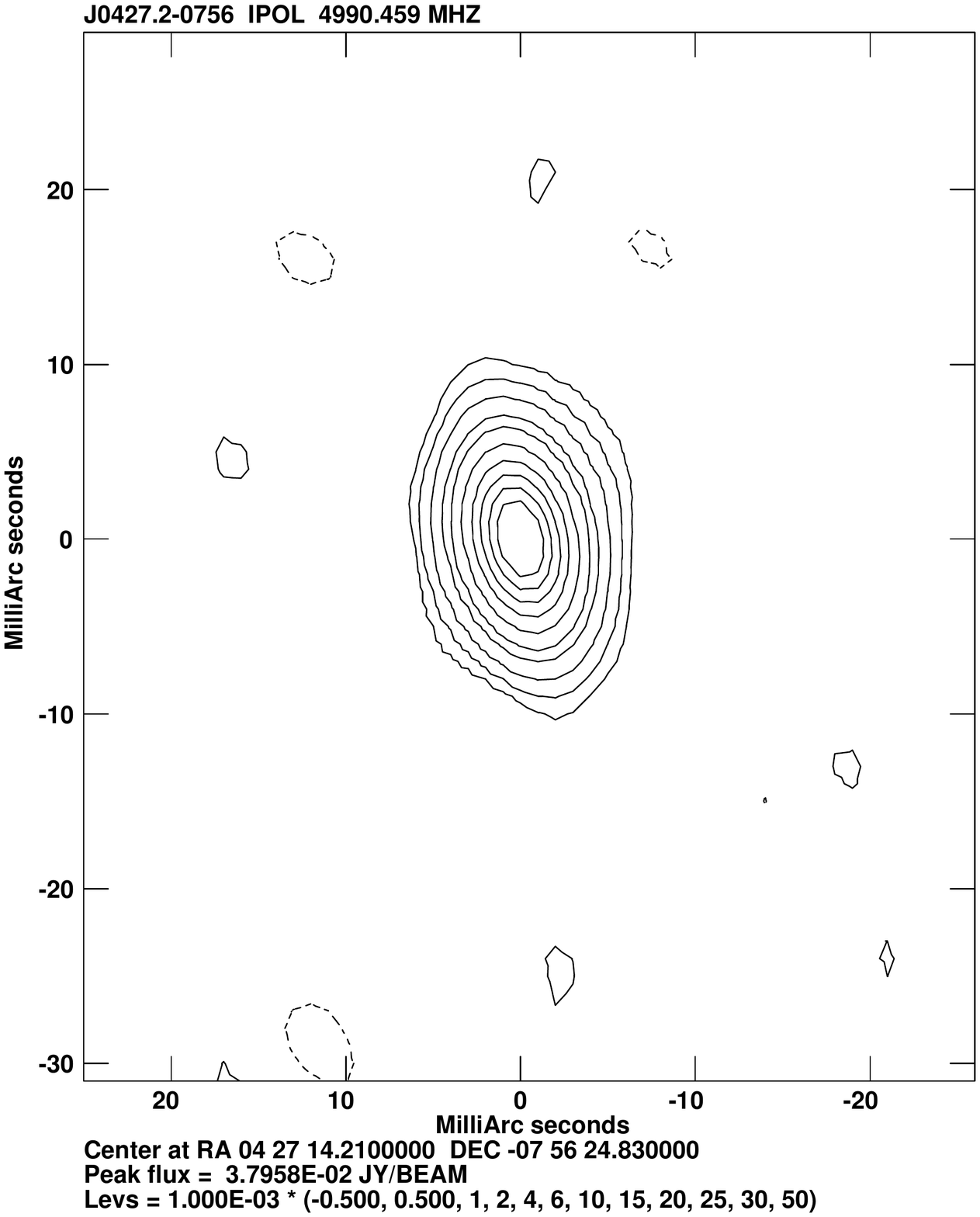}}
\resizebox{\hsize}{!}{\includegraphics{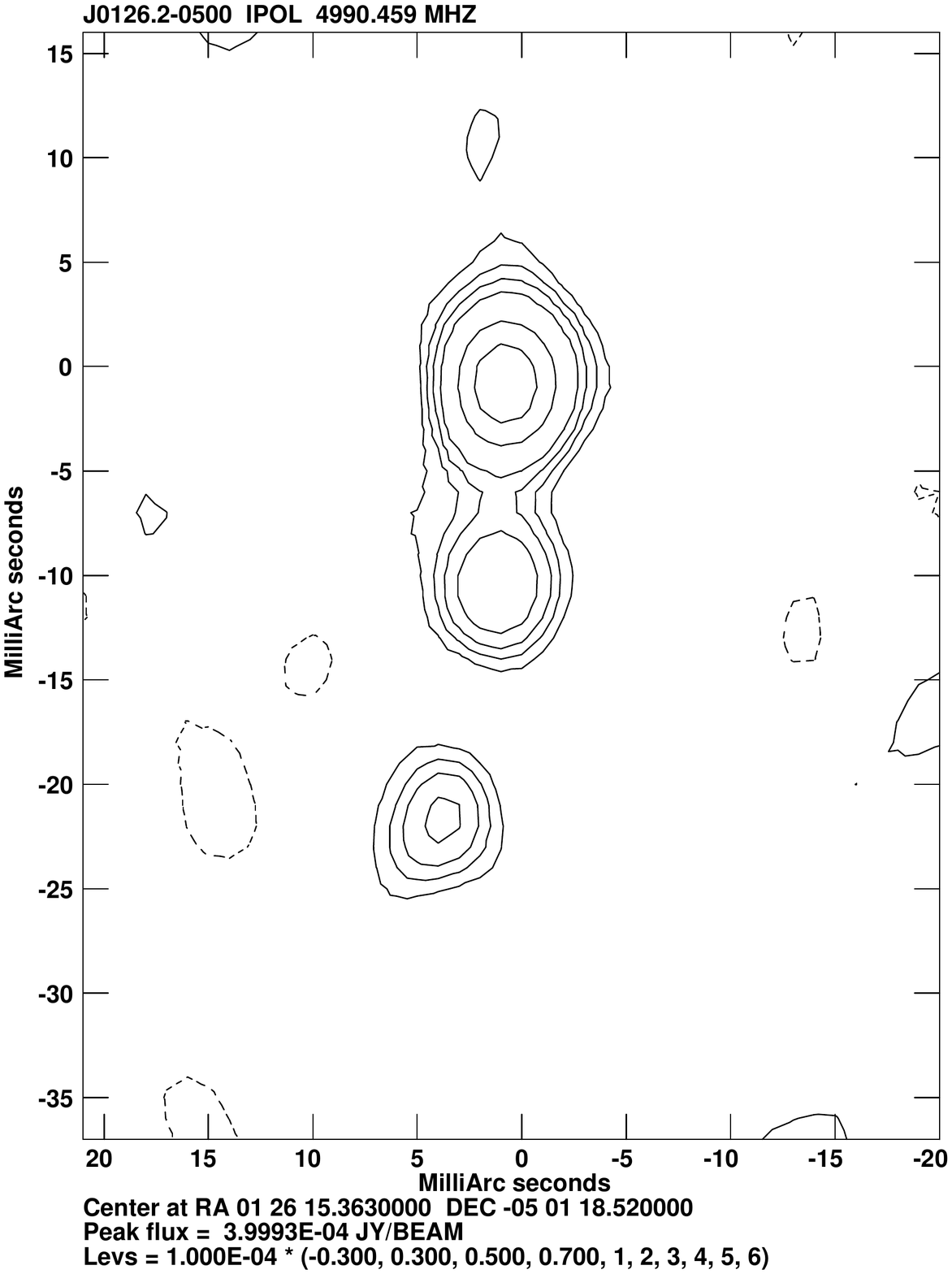}}
\caption{Total-intensity images of the point-like source
J0427.2$-$0756, the core-jet source J0126.2$-$0500.
%, and the triple source J1543.6$+$1847.
The convolution beams are 7.6$\times$4.3\,mas at 14.8\,deg, 6$\times$6\,mas , respectively.
%, and 2.5$\times$2.0\,mas at 5.4\,deg
\label{fig:structure}}
\end{figure}
\subsection{Derived parameters}

\begin{figure}
\resizebox{\hsize}{!}{\includegraphics{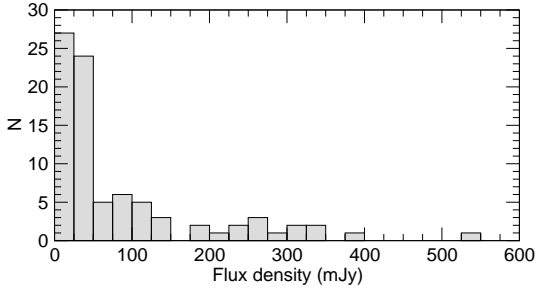}}
%\end{figure}
\caption{Histogram of the frequency distribution of the flux density measurements by the EVN at 5\,GHz. 
 Two points at 742.5\,mJy and 2378.0\,mJy are off the plot.
\label{fig:Stot_freq}}
\end{figure}

Source parameters derived from the present EVN observations are reported in 
Table~\ref{tab:parameters} for each source.
The total flux density distribution has a mean value of 120$\pm273$\,mJy and a median value 
$\langle S \rangle$ =44$^{+23}_{-10}$\,mJy. 
Sixty-eight out of 87 sources show a total flux density $\leq$150\,mJy (see Fig.~\ref{fig:Stot_freq}).
In Table~\ref{tab:parameters} we also report the core flux density and the core size as derived
using the ${\cal AIPS}$ task JMFIT. The ratio between the EVN flux density at 5\,GHz from the present 
observations to the Effelsberg flux density previously measured by \citet{Mantovani11}  
shows a median value $\langle R \rangle =0.36\pm0.08$, suggesting the presence of more extended 
structures around the compact
detected objects. The vast majority of sources shows $R<1$, while only 11 sources 
have $R>1$. The Effelsberg and the EVN flux density measurements were 
taken in different periods. Therefore, flux density variability and uncertainties in the EVN flux density 
measurements should be taken into account.
\subsection{Brightness temperature}

\begin{figure}
\resizebox{\hsize}{!}{\includegraphics{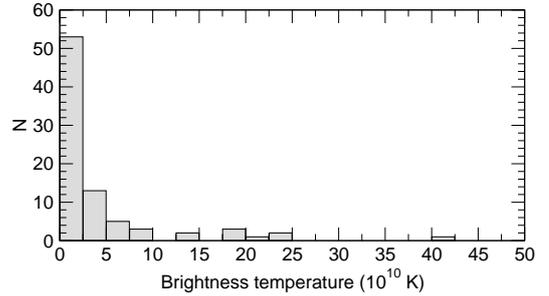}}
%\end{figure}
\caption{Histogram of the frequency distribution of the brightness temperature T$_b$. 
 Two points at 82.5$\times10^{10}$\,K and 110.00$\times10^{10}$\,K are off the plot.
\label{fig:Tb_freq}}
\end{figure}

The measurements of both flux density and size of the core at mas scale, together with the available 
red shift (\citealt{Perlman98}, \citealt{Landt01}) allowed us to calculate lower limits of the 
brightness 
temperature T$_b$. T$_b$ ranges between $10^7$\,K and $10^{12}$\,K. The mean value is
6$\pm16\times10^{10}$\,K and the median $\langle$T$_b\rangle$ = 0.9$^{+0.1}_{-0.6}$\,$
\times10^{10}$\,K. 
There are 19 sources that are unresolved when deconvolved by the beam. The mean and median values of 
their T$_b$ are 7$\pm7\times10^{10}$\,K and $\langle$T$_b\rangle$ = 3$^{+5}_{-1}$\,$\times10^{10}$\,K,
respectively. 
In Fig.~\ref{fig:Tb_freq}),
all the values of T$_b$ are plotted together. The large majority show a value of T$_b$ 
quite low, while 13 sources show T$_b>10^{11}$\,K.
These values might increase when observing sources at higher resolution. 
Only one object, J0937.1$+$5008, associated to a $\gamma$-ray object detected by {\it Fermi} 
(see below) has T$_b$ close to $10^{12}$\,K. This source also exhibits variability by a factor of 
about 2.3 in flux density at 5\,GHz when comparing the Effelsberg flux density to the EVN flux density, 
which were taken a few months apart.
%                                                                                                                    %                                                                        
%\vspace{-0.3cm}
%
%\begin{figure*}[t]
%\addtocounter{figure}{+0}
%\centering
%\end{figure*}
%
%
%
%\vspace{-0.3cm}
%
%\begin{figure*}[t]
%\addtocounter{figure}{+0}
%\centering
%\end{figure*}
%     
%            
\section{Discussion} 
\label{sec:discussion} 
It is widely accepted that the $\gamma$-ray emission is produced close to 
the origin of the jet in highly beamed relativistic jets associated with AGN
(\citealt{Dermer94}; \citealt{Fuhrmann14}). High angular-resolution images are therefore
critical for our understanding of the mechanism that generates such high-energy 
radiation, therefore a systematic comparison of the $\gamma$-ray and 
radio properties for a complete sample of blazars is mandatory.
Models have been suggested that predict that the  $\gamma$-ray sources
have more highly relativistic jets and are aligned close to the line of
sight (e.g. Lister \& Marscher 1997; Jorstad at al. 2001; Kellermann et 
al. 2004).  
Evidence that the observed $\gamma$-ray flares are associated with
the ejection of new superluminal components in AGN and that a $\gamma$-ray event 
occurs within the jet features was reported by Jorstad et al. (2001). To this
aim, the parsec-scale jet kinematic properties of AGN, and the correlation of the 
$\gamma$-ray flux with quasi-simultaneous parsec-scale radio flux density in the MOJAVE 
sample associated with bright $\gamma$-ray 
sources detected by the {\it Fermi} Gamma-ray Space Telescope, have been 
discussed by Lister et al. (2009) and Kovalev et al. (2009). As a consequence, 
about 100 sources selected amongst those
detected by the Large Area Telescope on-board the {\it Fermi} Gamma-ray Space 
Telescope with S$\mathrm{_{15GHz}}>100$\,mJy were added to the list of sources in 
the VLBA MOJAVE programme monitoring.

We made the first parsec-scale resolution observations 
of a complete sample of blazars selected from the faintest DXRBS by applying
the only criteria of a cut in declination at $>-$10 deg. The sample is formed by 87  
objects, $\sim13\%$ of which are BL\,Lacs. All sources were detected, confirming they
actually are compact objects. \citet{Mantovani11} made simultaneous multi-frequency
observations to better define the radio spectral index type for the DXRBS objects. We found
that 56 objects showing flat or inverted spectral index can be classified as blazars, 22 objects show 
steep spectral index, and 7 more objects are Giga-peaked spectral index. 
The 56 blazars show a median flux density $\langle S_{tot} \rangle$=75$^{+15}_{-27}$\,mJy. Of them 42 
show a flux
density S$<$150\,mJy at 5\,GHz. They potentially represent the appropriate sample for a comparison 
between faint and bright blazar samples. 
The sample also contains two flat-spectrum NLRGs, J1120.4$+$5855 and J1229.5$+$2711.

\subsection{Association to a $\gamma$-ray object}

The objects in the DXRBS are distributed over the whole sky.
To find any possible association between them and the 2FGL objects, we cross-checked their 
positions. 
For the 87 DXRBS objects observed in the northern sky, we improved their positional accuracy 
by deriving the positions from the FIRST \citep{Becker94} and from the NVSS \citep{Condon98} 
(see Mantovani et al. 2011). 
For 91 DXRBS objects in the southern sky with declination $<-10$ degrees, the NVSS or ATCA positions are 
taken from \citet{Landt01} and \citet{Perlman98}. 
Nine of the sources observed with the EVN show a good positional overlapping, which holds for an 
association 
between these radio sources and $\gamma$ objects. Eight of the $\gamma$-ray emitters are FSRQs or 
BLLacs, 
and one is a SSRQ. Since all the DXRBS objects observed with the EVN were detected, we assume 
that the 
DXRBS sources observed with ATCA at 2 arcsecond resolution are  also compact objects detectable with mas 
interferometric observations. Six out of the 91 southern sources also show a good positional 
overlapping with 
2FGL objects. The associations are presented in Table~\ref{tab:association} with other useful 
information. 
All of the associations have already been  suggested in the 2FGL. However, we add further 
characteristics for 
these sources suitable to possibly confirm radio-gamma identification.

Considering an association of a blazar with a $\gamma$-ray source as an identification 
also requires evidence of a correlation between the radio and $\gamma$ light curves, the former often 
missing in case of weak radio sources. 
Three of the blazars in Table~\ref{tab:association} are variable in both radio and $\gamma$ rays. Two 
sources only vary in the radio band and two in the $\gamma$ band. For two more objects, 
there is no evidence of any variability so far. 
Amongst sources observed with the EVN, all but one of the associated radio-gamma objects are among 
the 56 blazars we found, suggesting that the 14\% of weak blazars are also $\gamma$-ray emitters.
No information is available about variability in the radio band, 
while only one is reported as variable in $\gamma$-rays.

If we now take all the suggested radio-$\gamma$ associations into account, about 50\% of the objects 
listed in Table~\ref{tab:association} are classified as BL\,Lac. The DXRBS contains about 13\% of 
BL\,Lac objects in total, showing that BL\,Lacs are more likely to be detected as a $\gamma$ 
emitter than 
FSRQs. As pointed out by \citet{Ackermann11} investigating a much larger sample, this is mainly due 
to the different limiting flux in the $\gamma$ regime for FSRQs and BL\,Lacs, which show soft and hard 
spectral indices, respectively.

It is worth noting that the source J0204.8+1514 is the only
SSRQ amongst those associated to a $\gamma$-ray object. This source is also classified as a 
possible CSS in Table~\ref{tab:css-gps}. 
The angular size reported by \citet{Cooper07} for J0204.8+1514 observed with the VLA A-array at 
1.4\,GHz after beam 
deconvolution is 1.58 arcsec as major axis, 1.42 arcsec as minor axis, at a position angle 
of $-$40.5\,deg. 
This corresponds to a linear size LS =4.7\,kpc, which classifies it as CSS. However, \citet{Cooper07} 
also find a faint extended 
emission above the 3$\sigma$ noise level of their image. \citet{Mantovani11} and \citet{Nolan12} 
found the source variable 
in the radio- and $\gamma$-ray bands, respectively, which is rather peculiar in a CSS. J0204.8+1514 
was included in the 
MOJAVE programme selected because it was classified flat spectrum source, which does not seem to be 
the case. VLBA observations at 15\,GHz 
show a mas core-jet structure for J0204.8+1514.   

\subsection{Comparison between high- and low-power samples of blazars}

An interesting by-product of this investigation is the preliminary comparison 
between high- and low-power samples of blazars. One of the most extended samples of
bright flat spectrum blazars is the MOJAVE sample (\citet{Lister05}, \citet{Lister09}) 
regularly observed with the Very Long Baseline Array
\footnote{The VLBA is a facility of the National Science Foundation operated by
The National Radio Astronomy Observatory under cooperative agreement with the Associated 
Universities, Inc.}.
It includes all the objects with VLBA flux density exceeding 
1.5\,Jy at 15\,GHz, declination $>-30$\,deg, plus all known AGNs above declination 
$>-20$\,deg detected by {\it Fermi} and with compact flux density of at least 100\,mJy
at 15\,GHz. An investigation of the $\gamma$-ray and 15\,GHz radio properties of bright
blazars reflecting a wide range of spectral energy distribution parameters 
was possible for blazars included in the MOJAVE programme. The synchrotron self-Compton 
model is favored for the $\gamma$-ray emission in BL\,Lac objects over external seed-photon models.
FSRQs are shown to have a different behaviour. 

The association between
relativistic beaming and the derived properties suggests that high-synchrotron peaked
BL\,Lac have lower Doppler factors than lower-synchrotron peaked BL\,Lac objects and FSRQs
\citep{Lister11}. 
The present sample of weak blazars, 41 of which show S$_{core}<$100\,mJy, for the time being
lacks the monitoring programme needed to search for 
structural changes and to evaluate Doppler factors. However, the existing mas resolution observations
allowed us to derive useful parameters that can be used to complement previous findings for 
bright blazars.
      
A highly significant correlation  with a formal probability of a chance correlation of 
$5\times10^{-6}$  
was found by \citet{Pushkarev10} between the source core flux density at 15\,GHz
obtained by the MOJAVE VLBA programme and the integrated 0.1$-$100\,GeV $\gamma$-ray photon fluxes taken 
from the {\it Fermi} 1FGL catalogue. They analysed a large sample of 183 LAT-detected AGNs. 
Because the correlation is weaker if 
the jet flux density is used instead, they suggest that the core is the site of the $\gamma$-ray 
emission. 
Furthermore, they suggest 
that the region where most of $\gamma$-ray photons are produced is located within the compact 
opaque parsec-scale 
core, at a typical distance between the $\gamma$-ray production region and the 15 GHz radio 
core of ~7\,pc, in good agreement with \citet{Fuhrmann14}.

The presence of this correlation in this highly variable population further suggests that Doppler 
beaming is likely the cause and that Doppler factors of the individual jets are not changing 
substantially over time \citep{Pushkarev10}. Because of the selection criteria followed for sources 
observed in the MOJAVE VLBA programme, there is a limit in the plot
about 100\,mJy in the core flux density. Most of the sources associated with {\it Fermi} 2FGL objects
we observed with the EVN reaches a much lower level in flux densities at 5\,GHz. 
The plot of Fig.\,2 by \citet{Pushkarev10}
can be updated with these new data pairs taking two limiting factors into account: 
a) Interferometric observations for our sample at 15\,GHz are missing. However, all but one source 
listed in Table~\ref{tab:association} have a flat spectral index. We can assume that their flux 
density at 15\,GHz is similar to or even lower than the one we measured with the EVN at 5\,GHz 
owing to the 
better resolution achieved with the VLBA at that frequency;
b) The flux densities by \citet{Pushkarev10} are those measured after the LAT event with a delay of 
about 2.5 months. Again, the information for structural changes with time for the DXRBS sources are 
still not available. 

\begin{figure}
\resizebox{\hsize}{!}{\includegraphics{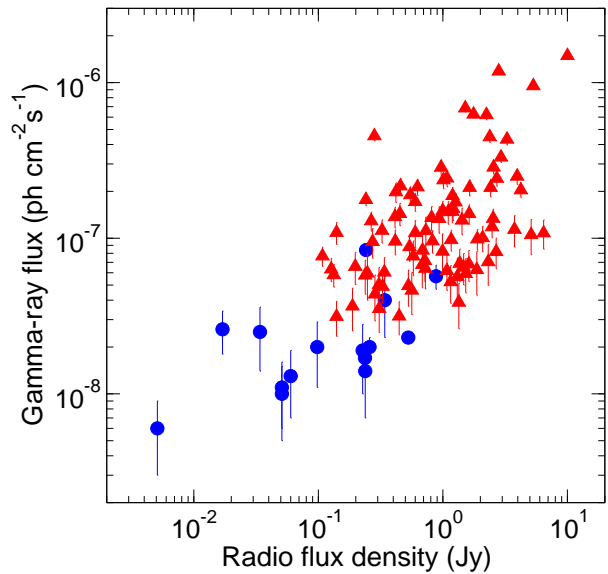}}
%\end{figure}
\caption{Integrated 0.1-100GeV {\it Fermi} LAT photon flux vs. 15\,GHz VLBA core flux density
as in Figure 2 of \cite{Pushkarev10} (triangles), complemented with data pairs from the present 
work (circles).
\label{fig:S-gamma}}
\end{figure}

Keeping that in mind, we updated the \citet{Pushkarev10} plot by adding data pairs that have the radio 
flux density taken from
our EVN measurements and the ATCA measurements made by \citet{Perlman98} and \citet{Landt01} and 
the integrated $\gamma$-ray 
photon flux of the associated {\it Fermi} objects, evaluated from the fit of the full band 
(100\,MeV$-$100\,GeV), 
taken from the 2FGL catalogue \citep{Nolan12}. The updated plot is shown in Fig.~\ref{fig:S-gamma}. 
Taking the uncertainties in the data pairs we added into account, the correlation is clearly confirmed.
The ATCA flux densities should be considered as upper limits. Shifting those points to left in the plot
should improve the correlation.
%                                                                        
%\vspace{-0.3cm}
%
%\begin{figure*}[t]
%\addtocounter{figure}{+0}
%\centering
%\includegraphics[width=8cm]{flux_flux.eps}
%\end{figure*}
%
To make sure that the method adopted for updating the plot in Fig.~\ref{fig:S-gamma} is convincing, 
we also compared our data pairs with those of Fig.\,1 in \citet{Linford12}, in which they plot the 
$\gamma$-ray photon flux
versus the total VLBA radio flux density at 5\,GHz for 232 sources. For the 15 associated objects of 
the present work,
we derived the total flux densities 
at 5\,GHz from the EVN images and from the ATCA measurements, plotting them versus the corresponding
integrated $\gamma$-ray photon flux from the 2FGL catalogue. The new data pairs fit nicely into the 
\citet{Linford12} plot,
which also show that a strong correlation between radio flux density and $\gamma$-ray flux of 
blazars is in place.

\citet{Lister10} plotted the 11-month median {\it Fermi} energy flux against VLBA 15\,GHz flux 
density for the first
Fermi-MOJAVE sample (1FM). The bottom left-hand quadrant of Fig.\,1 of that paper shows that more 
than half of the 
$\gamma$-ray-limited 1FM
sample consists of AGN that are not in the flux-limited MOJAVE sample, implying that they 
have lower-than-average 
radio-to-$\gamma$-ray flux ratio. All of them lie below the MOJAVE limit of 1.5\,Jy. Lister suggests 
that a possible
explanation is that the weaker $\gamma$-ray emission from blazars is boosted by a higher Doppler 
factor than their
radio emission, as for the proposed external Compton model by \citet{Dermer95}. 
Adding the data pairs from the present work to Lister's plot, they fill the bottom left-hand quadrant, 
a clear indication that weak blazars are also weak $\gamma$-ray emitters. Therefore invoking the 
external Compton model to explain the lack of faint $\gamma$-ray emitters may not be needed.
The {\it Fermi} observations (1FGL 2008-08-04 -- 2009-07-04; 2FGL 2008-08-04 - 2010-08-01) and 
the EVN observations (2009-10-22 -- 2013-05-27) can almost be considered as concurrent. However, 
before considering it 
unnecessary to apply the external Compton model, it is better to keep in mind the uncertainties 
incidental to data pairs from the present work.

\subsection{Steep-spectrum and GigaHz-Peaked sources}

Twenty-nine sources in our sample show a steep spectrum or a spectrum peaking at a frequency 
around 1$-$2\,GHz 
(GigaHz Peaked Sources, GPSs), according to \citet{Mantovani11}. 
We extracted their arcsecond scale sizes from the FIRST and NVSS and computed their projected linear 
size LS. 
This sub-sample of objects is listed in Table~\ref{tab:css-gps}, together with other parameters 
collected from 
literature.  Column 2 in Table~\ref{tab:css-gps} reports which class of objects they belong to, signally 
SSRQs, CSSs, GPSs, and BL\,Lacs. 

Sources with LS $\leq 20$\,kpc and power $P\mathrm{_{1.4GHz}}>10^{25}$\,W\,Hz$^{-1}$ are classified as 
CSSs or GPSs, which 
are young radio sources with ages $<10^7$\,yr \citep{Fanti90}. Their projected linear size LS was 
computed using
H$_{\circ}$=100\,km\,sec$^{-1}$\,Mpc$^{-1}$ and q$_{\circ}$=1,
following the cosmology adopted in that paper. The source J0435.1$-$0811 was classified as GPS 
with a question mark, showing a peaked spectral index and an LS quite larger than 20\,kpc.

Since X-ray \textrm{ROSAT} observations are available for the DXRBS objects, this sample of 
CSS and GPS quasars
can be used for a systematic study of their X-ray properties. The X-ray luminosity is found in the range 
$10^{43}-10^{46}$ ergs sec$^{-1}$ with a median value of 
$\langle L\mathrm{_X} \rangle = 9^{+42}_{-7}\times10^{44}$\,ergs\,sec$^{-1}$, a value that is a bit lower 
than the luminosity of $10^{45}-10^{46}$\,ergs\,sec$^{-1}$ 
reported by \citet{O'Dea98} for CSS and GPS quasars by X-ray measurements taken by different observers. 
Moreover, we do not find any significant difference in X-ray luminosity between CSSs and GPS quasars.
The derived column densities of hydrogen nH, resemble the Galactic nH column density, 
ignoring the presence of X-ray absorbing material in excess, which suggests that CSS and GPS quasars 
are not obscured by large column of cold gas surrounding the nuclei.
The percentage of polarised emission at 8.35 or 10.45 GHz is $<$1\% in almost all cases 
(see Table~\ref{tab:css-gps}), 
in good agreement with the finding of \citet{Mantovani09}, who made single-dish observations with the 
100-m Effelsberg telescope of a complete sample of CSSs.

Sources listed in Table~\ref{tab:css-gps}, showing a projected LS $>20$\,kpc are classified as SSRQs. 
These are 
objects in which the lobe emission dominates the radio core emission. Their jets are
viewed at larger angles than blazars, resulting in weaker beaming
effects. They show properties between those of both FSRQs and
radio quiet quasars. Clearly, the faint lobe emission is resolved out by the EVN observations 
presented in this paper.

Table~\ref{tab:css-gps} also lists three objects classified as BL\,Lacs. This is quite unexpected 
since BL\,Lacs 
show flat spectral index in the radio band. It is worth mentioning that the optical classification 
of these 
three objects as BL\,Lacs is questionable (see Perlman et al. 1998; Landt et al. 2001)
\section{Summary and conclusions}
\label{sec:summary}

We have presented EVN observations at 5\,GHz of a complete sample of 87 weak objects from
the DXRBS selected with a declination $>-10$\,deg as the unique criterion.
All of the sources were detected: 39 are point-like, 48 show a core-jet structure.

Their brightness temperature T$_b$ ranges between $10^7$\,K and $10^{12}$\,K. The large majority of them
show a very low value of T$_b$. Thirteen sources show T$_b>10^{11}$\,K. 
There are 19 sources that are unresolved when deconvolved by the beam. The mean and median values of 
their T$_b$ are 7$\pm7\times10^{10}$\,K and $\langle$T$_b\rangle$ = 3$^{+5}_{-1}$\,$\times10^{10}$\,K, 
respectively.
Only J0937.1$+$5008, associated to a $\gamma$-ray object, has T$_b$ close to $10^{12}$\,K.

Amongst the 87 sources observed, 56 can be considered as blazars, 22 show a steep spectral index, 
and 7 more objects are 
classified as Giga-Peaked Sources. The sample also contains two NLRGs, namely J1120.4$+$5855 and 
J1229.5$+$2711.
The 56 blazars potentially represent an appropriate sample for a direct comparison between 
faint- and bright- blazar samples. They show a median flux density 
$\langle S_{tot} \rangle$=75$^{+15}_{-27}$\,mJy. 
Of them 42 show a flux density S$<$150\,mJy at 5\,GHz.  

We found that 14 DXRBS flat spectrum objects and 1 steep spectrum object are associated to 
$\gamma$-ray emitters from
the 2FGL. Among the radio-$\gamma$ associations, about 50\% of the objects are BL\,Lacs. 
The DXRBS containes about 
13\% of BL\,Lac in total. This confirms that BL\,Lacs are more likely to be detected among blazars 
$\gamma$-emitters.

The correlation found by \citet{Pushkarev10} between the source core flux density at 15\,GHz
obtained by the MOJAVE VLBA programme and the 0.1$-$100\,GeV $\gamma$-ray photon fluxes taken 
from the {\it Fermi} 1FGL
extends to lower fluxes when adding the new data pairs of weak blazars we collected to the plot.
After updating the plot by \citet{Lister10}, data pairs from the present work fill the bottom 
left quadrant of the plot, indicating that weak blazars are also weaker $\gamma$-ray emitters 
compared to bright blazars, suggesting 
that they may not be needed to invoke the external Compton model. Before considering this as a 
firm statement, we have to keep the constraints on the added data pairs in mind.

About half of the steep or convex spectrum sources are point-like, and half show a core-jet
in the FIRST or NVSS images. With linear size $\leq 20$\,kpc and power 
$P\mathrm{_{1.4GHz}}>10^{25}$\,W\,Hz$^{-1}$, 
they can be classified as CSSs or GPSs, i.e. young radio sources 
with ages $<10^7$\,yr.  This sub-sample of objects with X-ray ROSAT observations 
can be used for a systematic study of the X-ray properties in CSS and GPS quasars. 
The X-ray luminosity range is 
$10^{43}-10^{46}$\,ergs\,sec$^{-1}$ with a median value of 
$\langle L_{X} \rangle = 9^{+42}_{-7}\times10^{44}$\,ergs\,sec$^{-1}$, a bit lower than a luminosity of 
$10^{45}-10^{46}$\,ergs\,sec$^{-1}$ as reported by \citet{O'Dea98}.
Any difference in X-ray luminosity between CSSs and GPS quasars is not significant. The nH column 
densities resemble the Galactic nH column density, suggesting that CSS and GPS quasars are not 
obscured by a large column of 
cold gas surrounding the nuclei. All of these objects are rather weakly polarised.

The DXRBS sample has the great advantage over the MOJAVE sample because it facilitates a 
direct comparison between $\gamma$-ray detected and non-detected sources in the same 
radio flux-limited sample. The present investigation can be considered as a useful 
basis for further monitoring programmes aiming to observe structure changes, to detect
flux density variability, and to derive other parameters for weak blazars.

\begin{acknowledgements}
The research leading to these results has received funding from the European Commission 
Seventh Framework Programme (FP/2007-2013) under grant agreement No 283393 (RadioNet3).
E.R. acknowledges partial support by the Spanish MINECO project AYA2012-38941-C02-01 and the
Generalitat Valenciana project PROMETEOII/2012/057.
This research has made use of data from the MOJAVE database that is maintained by the
MOJAVE team \citep{Lister09}. We are gratful to Alaksander Pushkarev for providing
his data and to Lars Fuhrmann for all the suggestions while we were drafting the paper. 
This work made use of the Swinburne University of Technology software correlator, developed 
as part of the Australian Major National Research Facilities Programme and operated under licence.
It has also used the NASA/IPAC Extragalactic Database (NED) which is operated by
the Jet Propulsion Laboratory, California Institute of Technology,
under contract with the National Aeronautics and Space Administration,
and NASA's Astrophysics Data System.  
We are very grateful to an anonymous referee for very helpful comments
and suggestions and for a careful reading of the manuscript of this
paper.

\end{acknowledgements} 
\newpage
\tabcolsep0.1cm
\begin{table*}[h]
\tiny
\centering
\caption{ EVN observing sessions}
\label{tab:sessions}
\begin{tabular}{lllll}
\hline
\hline
Code     &  Observing date & Telescopes                                              & Fringe finders             & Notes \\
\hline
EM077a   & 22 Oct 2009     & (Da),Ef,Jb,(Kn),Mc,Nt,On,Sh,Tr,Ur,Wb,Ys                 & J2005$+$77, 4C39.25, DA193 &           \\
EM077b   & 30 May 2010     & (Cm), Ef, Jb, (Kn), Mc, On, Sh, (Tr), (Ur), (Wb)        & OQ208                      &          \\
EM077c   & 31 May 2010     & Cm, Ef, Jb, (Kn), Mc, On, Sh, Tr, Ur, Wb                & 3C273                      &          \\
EM077d   & 23 Nov 2010     & Ef, Jb, Mc, On, Tr, Cm, (Kn), (Da), (Ar), Wb, Ys, (Sh)  & J1419+3821                 & e-EVN     \\
EM077e   & 15 Dec 2010     & Ef, Jb, Mc, On, Tr, Wb, Ys, Cm, Hh                      & 4C39.25, DA193             & e-EVN     \\
EM077f   & 31 May 2011     & Bb, Ef, Hh, Jb, Mc, On, Sh, Sv, Tr, Ur, Wb, Ys, Zc      & 3C273, 3C345               &           \\
EM097a   & 24 Feb 2013     & Ef,Hh,Jb,Mc,Nt,On,(Sh),Sv,Tr,(Ur),Wb,Ys,Zc              & J0319$+$4130, J0449$+$1121 &           \\
EM097b   & 27 May 2013     & Ef,Hh,Jb,Mc,Nt,On,(Sh),Sv,Tr,Ur,Wb,Ys,Zc                & J0449$+$1121, 3C454.3      &            \\    
\hline
\end{tabular}
\normalfont
\smallskip\noindent
\flushleft{\normalsize {
Column 1: EVN code; 
2: observing date;
3: observing array. Telescopes: 
Ar Arecibo, Bd Badary, Cm Cambridge, Da Darnhall, Eb Effelsberg, Hh  Hartebeesthoek,
Jb Jodrell Bank, Kn Knockin, Mc Medicina, Nt Noto, On Onsala,
Sh Shanghai, Sv Svetloe, Tr Torun, Ur Urumqi, Wb Westerbork, Ys Yebes,
Zc Zelenchukskaya. The telescope in brackets did not provide useful data;
4: Fringe finders;
5: Notes. e-EVN: VLBI observations correlated in real time at the Joint Institute for VLBI in 
Europe (JIVE).
}}
\end{table*}
\tabcolsep0.1cm
\begin{table*}[h]
%\tiny
\centering
\caption{Source structure and spectral index}
\label{tab:struct-index}
\begin{tabular}{lllllllll}
\hline
Structure & Ultra steep    & Steep                 & Steep-flat  & Flat-steep & Flat         & GPS  & Inverted & Not available \\
          & $\alpha>0.7$   &$\alpha=0.7-0.5$       &             &            &              &      &          &                 \\
\hline
Point     & 6 (1)          & 4 (1)                 &  9 (1)      &    1       & 13 (2)       & 2    &  5       &  1            \\
Core-jet  & 8              & 3                     &  5 (1+1NLRG)&    3 (1)   & 18 (3+1NLRG) & 5 (1)&  1       & $-$           \\
Triple    & 1              &  $-$                  &  2          &   $-$      & $-$          & $-$  & $-$      & $-$           \\
\hline                                                                     
\end{tabular}                                                            
\normalfont                                                                
\smallskip\noindent                                                      
\flushleft{\normalsize {                                                 
Notes: in brackets the number of BL Lac objects; the spectral index information for J0513.8$+$0156 is missing in \citet{Mantovani11}.                                                                     
}}                                                                        
\end{table*}     
\newpage
\tabcolsep0.1cm
%\begin{table*}[h]
\begin{table*}
%\tiny
%\centering
\caption{Derived parameters}
\label{tab:parameters}
\begin{tabular}{llrcrrrlrrrrrr}
\hline
\\
Name           & Code &S$_{tot}$  &Im rms&$\theta\mathrm{_{max}}$&$\theta\mathrm{_{min}}$& PA &Str&S$\mathrm{_{core}}$&$\theta\mathrm{_{max}}$&$\theta\mathrm{_{min}}$& PA   & T$_b$ &  R \\
               &      &[mJy]     &[mJy/b]& [mas]       &  [mas]       &[deg]&     &[mJy]     & [mas]       &  [mas]      &[deg] & [10$^{10}$\,K] \\
\hline                                                                                   
J0015.5+3052   &EM097b&     3.8  &   0.02  & 6.0  &  2.3  &$-$44.5  &   cj   &     1.2  & 5.6  &   4.3  &   22  &    0.0006 &  0.04  \\ 
J0029.0+0509   &EM097b&    98.2  &   0.70  & 5.5  &  4.4  &$-$19.0  &   cj   &    87.1  & 4.7  &   2.3  &   70  &    0.1040 &  0.35  \\ 
J0125.0+0146   &EM097b&    48.1  &   0.20  & 5.5  &  3.5  &$-$11.7  &   cj   &    47.2  & 4.5  &   3.3  &   19  &    0.0406 &  0.35  \\ 
J0126.2$-$0500 &EM097b&     0.5  &   0.02  & 5.5  &  3.7  &    2.2  &   cj   &     0.4  & 2.9  &   2.3  &  120  &    0.0004 &  0.02  \\ 
J0204.8+1514   &EM097b&   527.7  &   2.70  & 5.5  &  3.1  &$-$17.3  &   p    &   527.7  & 5.7  &   4.5  &  100  &    0.4700 &  0.25  \\ 
J0227.5$-$0847 &EM097b&    40.5  &   0.40  & 6.5  &  3.8  &$-$ 3.8  &   p    &    40.5  & 3.9  &   2.4  &  107  &    0.2281 &  0.37  \\ 
J0245.2+1047   &EM097b&     4.8  &   0.05  & 6.5  &  3.3  &$-$11.5  &   p    &     4.8  & 5.4  &   4.1  &  147  &    0.0011 &  0.03  \\ 
J0304.9+0002   &EM077a&     6.8  &   0.10  & 6.4  &  1.4  &    7.6  &   cj   &     6.6  & 1.2  &   0.0  &   27  &    0.1170 &  0.13  \\ 
J0421.5+1433   &EM097a&     1.3  &   0.04  & 4.1  &  3.7  &    6.4  &   p    &     1.3  & 2.1  &   1.0  &   75  &    0.0156 &  0.01  \\ 
J0427.2$-$0756 &EM097b&    43.2  &   0.20  & 7.6  &  4.3  &   14.8  &   p    &    43.2  & 2.4  &   1.5  &  161  &    0.4655 &  0.85  \\ 
J0435.1$-$0811 &EM097a&    37.9  &   0.20  & 6.6  &  3.8  &$-$28.1  &   cj   &    35.7  & 2.0  &   0.0  &  140  &    0.4122 &  0.41  \\ 
J0447.9$-$0322 &EM097a&     8.2  &   0.10  & 5.8  &  3.9  &$-$25.7  &   cj   &     6.5  & 6.1  &   2.1  &   11  &    0.0044 &  0.42  \\ 
J0502.5+1338   &EM097b&   277.8  &   0.70  & 5.2  &  3.7  &   13.0  &   cj   &   266.7  & 2.4  &   0.3  &   66  &    2.7226 &  0.65  \\ 
J0510.0+1800   &EM097a&   341.5  &   1.40  & 4.5  &  4.1  &$-$26.3  &   p    &   341.0  & 1.5  &   0.7  &  126  &    7.5111 &  0.48  \\ 
J0513.8+0156   &EM097a&     2.3  &   0.06  & 4.8  &  3.7  & $-$1.4  &   p    &     1.8  & 3.1  &   1.2  &  130  &    0.0086 &  $-$   \\ 
J0518.2+0624   &EM097a&    26.3  &   0.60  & 5.4  &  4.5  &$-$27.4  &   p    &    24.8  & 7.7  &   3.7  &  105  &    0.0269 &  0.11  \\ 
J0535.1$-$0239 &EM097a&    49.4  &   0.20  & 5.9  &  3.9  &$-$33.6  &   p    &    49.2  & 2.1  &   0.7  &  163  &    1.1114 &  0.57  \\ 
J0646.8+6807   &EM097a&     7.7  &   0.10  & 6.0  &  3.5  &   70.6  &   p    &     7.7  & 1.8  &   1.4  &   50  &    0.0962 &  0.11  \\ 
J0651.9+6955   &EM097a&    39.1  &   0.20  & 4.4  &  4.1  &   68.2  &   cj   &    30.3  & 2.0  &   0.7  &  104  &    0.2510 &  0.26  \\ 
J0724.3$-$0715 &EM097a&   347.6  &   1.40  & 6.2  &  4.3  &$-$10.6  &   p    &   347.6  & 1.5  &   1.3  &  116  &    3.6987 &  1.26  \\ 
J0744.8+2920   &EM077a&   102.4  &   0.44  & 2.8  &  1.0  &   14.2  &   cj   &    86.6  & 0.6  &   0.4  &  121  &    3.8332 &  0.45  \\ 
J0816.0$-$0736 &EM097a&    11.9  &   0.07  & 6.2  &  3.6  & $-$8.4  &   cj   &    11.9  &13.4  &   5.4  &  172  &    0.0008 &  0.29  \\ 
J0829.5+0858   &EM077a&    45.5  &   0.38  & 4.9  &  1.4  &    1.7  &   cj   &    39.8  & 0.6  &   0.5  &   18  &    1.3043 &  0.27  \\ 
J0847.2+1133   &EM077a&     5.0  &   0.06  & 5.0  &  5.0  &         &   p    &     5.1  & 1.2  &   0.0  &   29  &    0.1554 &  0.28  \\ 
J0853.0+2004   &EM077a&    23.7  &   0.13  & 4.5  &  1.2  & $-$1.0  &   p    &    24.9  & 0.7  &   0.0  &   74  &   13.7920 &  0.35  \\ 
J0908.2+5031   &EM077a&    20.7  &   0.14  & 2.1  &  1.0  &   48.0  &   cj   &    17.1  & 0.7  &   0.2  &   82  &    1.0853 &  0.27  \\ 
J0927.7$-$0900 &EM097a&    83.3  &   0.40  & 6.9  &  3.5  &$-$20.4  &   cj   &    59.9  & 3.5  &   1.1  &  142  &    0.0956 &  0.91  \\ 
J0931.9+5533   &EM097a&     1.8  &   0.03  & 4.1  &  3.0  &$-$44.4  &   cj   &     1.5  & 3.1  &   1.6  &   78  &    0.0019 &  0.03  \\ 
J0937.1+5008   &EM077a&   257.6  &   1.60  & 2.2  &  1.0  &   46.8  &   p    &   258.5  & 0.5  &   0.0  &   39  &  107.6653 &  2.33  \\ 
J0940.2+2603   &EM077a&   257.1  &   0.44  & 4.7  &  1.5  &    3.6  &   cj   &   232.5  & 1.2  &   0.4  &  146  &    3.5554 &  0.57  \\ 
J1006.1+3236   &EM077a&    32.8  &   0.12  & 4.9  &  1.6  &    5.1  &   cj?  &    32.3  & 2.3  &   1.1  &  133  &    0.1264 &  0.26  \\ 
J1006.5+0509   &EM077e&   319.5  &   4.20  & 7.0  &  1.8  &   80.1  &   p    &   329.5  & 3.2  &   0.5  &   64  &    7.4538 &  2.53  \\ 
J1010.8$-$0201 &EM077e&    83.0  &   1.70  & 6.8  &  1.5  &   75.8  &   cj   &    50.9  & 1.0  &   0.0  &   73  &    3.1525 &  0.15  \\ 
J1011.5$-$0423 &EM077e&   227.9  &   2.20  & 7.4  &  2.3  &   75.5  &   p    &   289.9  &10.2  &   1.4  &   49  &    0.8581 &  1.89  \\ 
J1025.9+1253   &EM077e&  2378.0  &   20.0  & 8.7  &  2.5  &   82.0  &   p    &  2409.2  & 4.1  &   0.6  &   85  &   26.6014 &  3.64  \\ 
J1026.4+6746   &EM097a&    15.5  &   0.20  & 4.9  &  3.1  &$-$54.5  &   p    &    15.5  & 2.6  &   2.2  &   73  &    0.0965 &  0.12  \\ 
J1028.5$-$0236 &EM077e&   318.6  &   3.40  & 8.3  &  2.6  &   75.5  &   p    &   380.4  & 6.6  &   0.0  &   33  &    5.3470 &  2.41  \\ 
J1028.6$-$0336 &EM077e&    43.7  &   0.80  & 6.5  &  1.7  &   78.5  &   p    &    33.1  & 0.4  &   0.0  &   61  &   22.1103 &  0.72  \\ 
J1101.8+6241   &EM077a&   131.6  &   0.19  & 2.9  &  1.3  &   65.8  &   p    &   129.8  & 2.0  &   0.4  &   17  &    1.3221 &  0.38  \\ 
J1116.1+0828   &EM077e&   742.5  &   9.00  & 7.7  &  2.4  &   79.6  &   p    &  1130.7  &12.0  &   0.0  &   49  &    6.2575 &  1.87  \\ 
J1120.4+5855   &EM077a&     5.6  &   0.07  & 2.9  &  1.4  &   67.9  &   cj   &     4.8  & 0.3  &   0.0  &   64  &    0.7205 &  0.11  \\ 
J1150.4+0156   &EM077e&    90.1  &   1.60  & 7.2  &  3.1  &   83.4  &   p    &    91.3  & 1.9  &   0.0  &   43  &    6.3346 &  0.69  \\ 
J1204.2$-$0710 &EM077f&    74.9  &   1.80  & 1.7  &  1.4  &   35.0  &   cj   &    34.0  & 0.0  &   0.0  &       &    8.2950 &  0.50  \\ 
J1206.2+2823   &EM077e&    59.5  &   1.50  & 7.3  &  1.6  &$-$86.0  &   p    &    46.6  & 0.0  &   0.0  &       &    3.3910 &  3.02  \\ 
J1213.0+3248   &EM077c&    28.5  &   0.43  & 5.0  &  1.2  &   15.0  &   p    &    28.5  & 0.0  &   0.0  &       &    8.1509 &  0.43  \\ 
J1213.2+1443   &EM077f&    13.5  &   1.50  & 1.9  &  1.7  &   48.7  &   cj   &     7.0  & 0.0  &   0.0  &       &    1.8244 &  0.26  \\ 
J1217.1+2925   &EM077e&    44.5  &   1.40  & 5.7  &  1.9  &   88.7  &   p    &    37.7  & 0.0  &   0.0  &       &    3.3671 &  1.00  \\ 
J1222.6+2934   &EM077e&   195.2  &   2.30  & 7.2  &  2.3  &   86.7  &   p    &   194.4  & 4.1  &   1.2  &   66  &    0.9042 &  1.76  \\ 
J1223.9+0650   &EM077f&   116.7  &   3.50  & 1.7  &  1.5  &   45.4  &   cj   &    54.5  & 0.0  &   0.0  &       &   22.9244 &  0.46  \\ 
J1224.5+2613   &EM077e&    46.5  &   1.20  & 5.9  &  1.8  &   88.1  &   p    &    37.6  & 0.0  &   0.0  &       &    2.9267 &  0.20  \\ 
J1225.5+0715   &EM077e&    50.5  &   1.40  & 7.6  &  2.3  &   83.7  &   p    &    47.0  & 0.0  &   0.0  &       &    2.7931 &  0.89  \\  
J1229.5+2711   &EM077f&    36.0  &   0.86  & 2.0  &  1.8  &$-$22.6  &   cj   &    18.7  & 0.0  &   0.0  &       &    3.7925 &  0.29  \\ 
J1231.7+2848   &EM077f&    28.1  &   0.65  & 2.0  &  1.6  & $-$5.6  &   cj   &    17.0  & 0.0  &   0.0  &       &    3.9047 &  0.23  \\ 
J1311.3$-$0521 &EM077f&     9.1  &   0.29  & 2.0  &  1.4  &    3.9  &   cj   &     7.9  & 1.3  &   0.0  &  139  &    0.2467 &  0.21  \\ 
J1315.1+2841   &EM077f&    19.4  &   0.57  & 2.0  &  1.7  &    2.5  &   cj   &     0.6  & 1.3  &   0.0  &   47  &    0.3430 &  0.19  \\ 
J1320.4+0140   &EM077f&    77.8  &   2.30  & 1.9  &  1.5  &   17.6  &   cj   &    50.6  & 0.0  &   0.0  &       &   19.4437 &  0.16  \\     
J1329.0+5009   &EM077d&    78.2  &   1.90  &19.1  &  5.0  &   40.1  &   cj   &    71.7  & 6.0  &   1.7  &   36  &    2.1600 &  0.44  \\ 
J1332.7+4722   &EM077b&   112.4  &   3.30  &20.7  &  4.8  &   38.0  &   p    &    98.2  & 5.6  &   0.0  &   32  &    0.9953 &  0.59  \\ 
J1359.6+4010   &EM077d&   267.0  &   3.90  &16.4  &  4.4  &   61.2  &   cj   &   157.6  & 4.6  &   2.7  &   78  &    0.0875 &  1.00  \\
\hline
\end{tabular}
\end{table*}
\newpage
\begin{table*}
\tabcolsep0.1cm
%\tiny
%\centering
%\caption{Derived parameters}
%\label{tab:parameters}
\begin{tabular}{llrcrrrlrrrrrr}
\hline
\\
Name           & Code &S$_{tot}$  &Im rms&$\theta\mathrm{_{max}}$&$\theta\mathrm{_{min}}$& PA &Str&S$\mathrm{_{core}}$&$\theta\mathrm{_{max}}$&$\theta\mathrm{_{min}}$& PA   & T$_b$ &  R \\
               &      &[mJy]     &[mJy/b]& [mas]       &  [mas]       &[deg]&     &[mJy]     & [mas]       &  [mas]      &[deg] & [10$^{10}$\,K] \\
\hline
\\  
J1400.7+0425   &EM077f&   103.0   &  2.20  &  2.1  &  1.7  &   45.1  &   cj &    37.8   &  0.0  &   0.0  &       &   18.4182 &   0.67 \\ 
J1404.2+3413   &EM077d&    12.1   &  0.70  & 14.8  &  3.9  &   54.1  &   cj &    12.0   &  0.0  &   0.0  &       &    1.1973 &   0.22 \\ 
J1406.9+3433   &EM077d&   216.9   &  6.20  & 16.2  &  5.2  &   51.6  &   cj &   206.3   &  8.2  &   3.1  &   76  &    0.1414 &   0.78 \\ 
J1416.4+1242   &EM077f&    33.4   &  0.88  &  2.4  &  1.9  &   35.7  &   cj &    18.5   &  0.0  &   0.0  &       &    2.6539 &   0.41 \\ 
J1417.5+2645   &EM077d&   104.4   &  2.40  & 16.2  &  5.4  &   60.9  &   cj &    39.6   &  5.3  &   2.8  &   77  &    0.0320 &   1.19 \\ 
J1419.1+0603   &EM077f&   148.9   &  2.50  &  2.0  &  1.6  &   29.6  &   cj &    33.9   &  0.0  &   0.0  &       &   17.5921 &   0.68 \\ 
J1420.6+0650   &EM077f&    21.6   &  0.60  &  2.3  &  1.7  &   15.8  &   cj &    11.8   &  1.0  &   0.0  &  151  &    0.4204 &   0.11 \\ 
J1423.3+4830   &EM077d&    41.4   &  1.40  & 15.9  &  5.6  &   45.1  &   cj &    36.4   &  2.8  &   0.0  &   59  &    0.1785 &   0.28 \\ 
J1427.9+3247   &EM077d&    17.4   &  0.80  & 11.5  &  4.9  &   59.9  &   cj &    17.1   &  1.1  &   0.0  &   50  &    0.2332 &   0.29 \\ 
J1442.3+5236   &EM077d&    34.8   &  0.50  & 14.7  &  4.6  &   46.8  &   cj &    19.2   &  3.5  &   0.0  &   32  &    0.1636 &   0.37 \\ 
J1507.9+6214   &EM077a&    11.5   &  0.14  &  4.0  &  2.0  &   60.3  &   cj &     9.5   &  1.3  &   0.0  &   32  &    0.4437 &   0.05 \\ 
J1539.1$-$0658 &EM077f&    48.0   &  1.00  &  2.3  &  1.5  &   68.9  &   cj &    16.1   &  0.0  &   0.0  &       &    3.4300 &   0.57 \\ 
J1543.6+1847   &EM077f&    44.4   &  0.70  &  2.5  &  2.0  &    5.4  &   cj &    19.3   &  1.3  &   0.0  &       &    0.8715 &   0.27 \\ 
J1606.0+2031   &EM077f&    20.3   &  0.27  &  2.0  &  1.7  &    7.7  &   cj &     5.5   &  0.0  &   0.0  &       &    0.1096 &   0.25 \\ 
J1626.6+5809   &EM077a&   180.5   &  0.60  &  2.1  &  1.1  &   12.7  &   cj &   165.7   &  0.5  &   0.0  &  163  &   27.4517 &   0.50 \\ 
J1629.7+2117   &EM077f&    10.3   &  0.15  &  2.2  &  1.5  &$-$19.7  &   cj &     5.3   &  0.0  &   0.0  &       &    1.4425 &   0.11 \\ 
J1648.4+4104   &EM077a&   392.6   &  0.48  &  6.0  &  2.3  &   59.4  &   p  &   390.5   &  0.6  &   0.0  &   48  &   80.2037 &   0.79 \\ 
J1656.6+5321   &EM077a&    74.8   &  0.51  &  3.3  &  1.6  &$-$79.3  &   p  &    75.5   &  0.5  &   0.0  &  138  &   41.0253 &   0.56 \\ 
J1656.8+6012   &EM077a&   237.0   &  0.50  &  2.9  &  1.4  &$-$85.4  &   p  &   237.0   &  0.9  &   0.6  &  157  &   14.0301 &   0.49 \\ 
J1722.3+3103   &EM077d&    27.0   &  0.70  & 10.1  &  4.2  &   66.5  &   cj &    16.7   &  0.0  &   0.0  &    0  &    0.2517 &   0.52 \\ 
J1804.7+1755   &EM097b&     6.1   &  0.07  &  6.9  &  4.2  &   67.4  &   p  &     6.1   &  4.5  &   1.2  &   27  &    0.0265 &   0.06 \\ 
J1840.9+5452   &EM097b&    27.2   &  0.20  &  4.4  &  3.0  &   69.9  &   p  &    27.2   &  2.9  &   2.0  &   14  &    0.1099 &   0.14 \\ 
J2239.7$-$0631 &EM097b&    25.6   &  0.20  &  6.9  &  4.4  &$-$21.2  &   p  &    25.6   &  4.4  &   2.0  &   38  &    0.0601 &   0.28 \\ 
J2320.6+0032   &EM077a&    67.2   &  0.47  &  6.0  &  1.3  &    5.9  &   cj &    62.5   &  0.3  &   0.0  &  160  &   22.7253 &   0.70 \\ 
J2322.0+2114   &EM097b&    13.2   &  0.10  &  5.2  &  3.8  &$-$51.6  &   p  &    13.2   &  5.4  &   1.8  &   71  &    0.0379 &   0.17 \\ 
J2329.0+0834   &EM097b&    43.8   &  0.30  &  5.4  &  4.1  &$-$32.2  &   p  &    43.8   &  4.2  &   1.6  &   55  &    0.2074 &   0.15 \\ 
J2333.2$-$0131 &EM077a&   126.9   &  0.35  &  6.2  &  1.2  &    5.6  &   cj &   114.1   &  1.0  &   0.2  &   30  &    5.7642 &   0.88 \\ 
J2347.6+0852   &EM097b&     1.8   &  0.02  &  7.3  &  3.2  &$-$29.0  &   p  &     1.8   &  5.7  &   2.0  &   32  &    0.0033 &   0.03 \\ 
\hline                                                                                  
\end{tabular}
\normalfont
\smallskip\noindent
\flushleft{\normalsize {
Table $-$                                                                          
column 1: source name;                                                           
2: EVN code;                                              
3: total flux density in mJy;                                               
4: image rms error;                              
5: beam major axis;
6: beam minor axis;
7: position angle of the beam major axis;                                                
8: source structure: p point, cj core-jet, t triple;             
9: core flux density in mJy obtained by JMFIT;
10: core major axis after beam deconvolution;
11: core minor axis after beam deconvolution;                                        
12: major axis position angle;                     
13: T$_b$ 10$^{10}$\,K.                                                           
           T$_b$ = 4.9 $\times$ 10$^{10}$ S$_{core}$(1+z)/($\theta\mathrm{_{max}}$ $\times$ $\theta\mathrm{_{min}}$).         
           S$\mathrm{_{core}}$ Jy; $\theta$ mas; in case of unresolved source, 30\% of the        
           beam size is taken; when not available $z$=0.5 assumed;
14: R is the EVN to Effelsberg flux density ratio.
}}
\end{table*}
\newpage
\tabcolsep0.1cm
\begin{table*}[h]
\tiny
\centering
\caption{List of steep spectrum sources}
\label{tab:css-gps}
\begin{tabular}{llrrrcrcllrrrrr}
\hline
Name           & Class    &Maj.ax. &min.ax. & P.A. &    & LS       &EVN    &     ROSAT        &  X-lumin     & Col.dens.& m       &  m     & RM            \\
               &          &[arcsec]&[arcsec]&[deg]&    &[kpc]     &struct &[erg/cm$^2$/sec]&[erg/sec]    & [nH cm$^2$]& 8.35    & 10.45  &[rad/m$^2$]    \\
J0015.5+3052   &  CSS?    &$<$16.7&$<$16.3& $-$  &N   & $<$57.3  & cj    &  2.82E$-$13    &  5.09E45     &  5.74E+20 &         &  $<1$  & 14.4            \\
J0126.2$-$0500 &  SSRQ    &   48.3&$<$17.1& 43.3 &N   &   145.0  & cj    &  0.31E$-$12    &  1.90E44     &  4.08E+20 &         &  $<1$  &            \\
J0204.8+1514   &  CSS?    &$<$18.9&$<$16.8& $-$  &N   & $<$56.4  & p     &  2.60E$-$13    &  1.54E44     &  5.53E+20 &         &  $<1$  & 62.9            \\
J0227.5$-$0847 &   GPS    & 1.38  & 1.02  &136.8 &F   &     4.3  & p     &  0.22E$-$12    &  8.76E45     &  3.39E+20 &         &  $<1$  &            \\
J0245.2+1047   &  BL\,Lac &   87.8&   29.4&$-$36.0& N &    78.0  & p     &  7.80E$-$13    &  9.39E42     & 10.20E+20 &         &  $<1$  &            \\
J0304.9+0002   &  SSRQ    &   8.12&   4.14&162.1 &F   &    27.2  & cj    &  4.02E$-$13    &  5.27E44     &  6.51E+20 &         &  $<1$  &            \\
J0421.5+1433   &  BL\,Lac &   67.9& 18.5  & 66.6 &N   &   219.4  & p     &  3.51E$-$14    &  3.45E43     & 15.90E+20 & 10.8    &        &$-$68.6            \\
J0435.1$-$0811 &   GPS    & 21.2  &$<$17.2& 31.6 &N   &    76.0  & cj    &  1.26E$-$13    &  3.81E44     &  6.17E+20 &         &  $<1$  &            \\
J0447.9$-$0322 &  CSS?    &$<$18.2&$<$16.0& $-$  &N   & $<$65.1  & cj    &  1.38E$-$12    &  3.96E45     &  3.80E+20 &         &  $<1$  &            \\
J0518.2+0624   &  CSS?    &$<$18.8&$<$17.5& $-$  &N   & $<$68.1  & p     &  9.69E$-$14    &  3.94E44     & 12.00E+20 &  2.6    &        & 95.4            \\
J0931.9+5533   &   GPS    &  23.8 & 18.9  & 64.4 &F   &    57.4  & cj    &  0.43E$-$12    &  9.56E43     &  2.22E+20 &         &  $<1$  &            \\
J1006.1+3236   &   CSS    &  3.38 & 1.91  &132.1 &F   &    12.4  & cj?   &  0.43E$-$12    &  2.45E45     &  1.47E+20 &         &  $<1$  &            \\
J1101.8+6241   &   CSS    & 1.89  & 0.75  & 13.7 &F   &     6.6  & p     &  0.46E$-$12    &  9.00E44     &  0.90E+20 &         &  $<1$  & $-$8.1            \\
J1116.1+0828   &   GPS    & 0.68  & 0.49  & 31.3 &F   &     2.2  & p     &  0.22E$-$12    &  2.02E44     &  2.85E+20 &         &  $<1$  &            \\
J1213.2+1443   &  SSRQ    & 9.27  & 8.20  & 12.3 &F   &    32.8  & t     &  0.48E$-$12    &  1.14E45     &  2.82E+20 &   $<1$  &        &            \\
J1224.5+2613   &  SSRQ    & 14.08 &10.28  &118.9 &F   &    49.4  & p     &  1.04E$-$13    &  2.22E44     &  1.77E+20 &   $<1$  &        &            \\
J1225.5+0715   &  SSRQ    &  7.17 & 3.55  & 46.9 &F   &    26.0  & p     &  0.84E$-$13    &  6.05E44     &  1.72E+20 &         &  $<1$  &            \\
J1404.2+3413   &  SSRQ    & 22.3  &$<$18.7&  5.8 &N   &    81.0  & cj    &  0.78E$-$13    &  3.60E44     &  1.23E+20 &         &  $<1$  &            \\
J1406.9+3433   &   GPS    &  0.87 & 0.35  & 11.9 &F   &     2.6  & cj    &  0.11E$-$12    &  6.13E45     &  1.31E+20 &         &  $<1$  &            \\
J1420.6+0650   &   CSS    &  0.85 & 0.0   &179.4 &F   &     1.9  & cj    &  0.33E$-$12    &  5.58E43     &  2.18E+20 &         &        &            \\
J1427.9+3247   &   GPS    &  3.42 & 3.05  & 26.9 &F   &    11.5  & cj    &  0.38E$-$12    &  5.09E44     &  0.97E+20 &  4.5    &        &  0.5?            \\
J1442.3+5236   &   CSS    &  4.80 & 1.21  &119.1 &F   &    16.0  & cj    &  0.14E$-$12    &  3.29E45     &  1.46E+20 &         &  $<1$  &            \\
J1507.9+6214   &   CSS    &  2.30 & 1.21  & 2.7  &F   &     8.0  & cj    &  0.12E$-$12    &  1.73E45     &  1.55E+20 &         &  $<1$  &            \\
J1539.1$-$0658 &  BL\,Lac &  1.02 & 0.61  & 81.2 &F   &    3.3   & cj    &  0.70E$-$13    &  6.88E43     &  9.02E+20 &   $<1$  &        &            \\
J1629.7+2117   &  CSS?    &$<$17.1&$<$15.6& $-$  &N   & $<$61.6  & cj    &  0.14E$-$12    &  4.82E44     &  4.11E+20 &   $<1$  &        &            \\
J1722.3+3103   &  SSRQ    & 42.48 & 13.63 &104.0 &F   &   110.6  & cj    &  0.16E$-$12    &  4.88E43     &  3.15E+20 &   $<1$  &        &            \\
J1804.7+1755   &  SSRQ    & 40.30 &$<$19.6& 36.3 &N   &   123.8  & p     &  0.11E$-$12    &  7.74E43     &  8.27E+20 &   $<1$  &        &            \\
J2322.0+2114   &  SSRQ    & 75.2  &$<$17.5& 57.3 &N   &   265.3  & p     &  1.49E$-$13    &  3.42E44     &  4.47E+20 &   $<1$  &        &            \\
J2347.6+0852   &  SSRQ    & 42.7  & 21.5  &$-$34.9&N  &   108.6  & p     &  6.40E$-$13    &  1.76E44     &  5.75E+20 &   2.8   &        &            \\
\hline                                                                     
\end{tabular}                                                            
\normalfont                                                                
\smallskip\noindent                                                      
\flushleft{\normalsize {                                                 
Table $-$                                                                     
column 1: source name;                                                    
2: source class;
3$-$5: major axis, minor axis in arcseconds and position angle of the major axis in degrees; 
6: source angular size extracted from FIRST (F) or NVSS (N) catalogues; 
7: linear size computed using H$_{\circ}$=100 km/sec/Mpc, 
q$_{\circ}$=1)
8: EVN structure; 
9: ROSAT X-ray flux (Perlman et al. 1998;  Landt et al. 2001);                                       
10: X-ray luminosity computed using $www.astro.soton.ac.uk/\sim td/flux\_convert.html$ 
           with cosmological parameters
           {$H_0=70\,{\rm km}\, {\rm s}^{-1}\, {\rm Mpc}^{-1}, \Omega_{\rm m}=0.27, \Omega_{\rm vac}=0.73$}
11: nH column density (computed for cone radius of 0.5 deg from input position using heasarc.gsfc.nasa.gov/cgi-bin/Tools/;
          numbers are (LAB Average + DL Average nH) / 2);
12$-$13: polarisation percentage at 8.35\,GHz and 10.45\,GHz (Mantovani et al. 2011). The median value of m for CSS at 8.35\,GHz is 
          $<$1\% (Mantovani et al. 2009);                       
14: rotation measure (Mantovani et al. 2009).     
}}                                                                        
\end{table*}                                                                                        
\newpage
\tabcolsep0.1cm
\begin{table*}[h]
\tiny
\centering
\caption{List of DXRBS sources possibly associated to Fermi objects}
\label{tab:association}
\begin{tabular}{lllccclcc}
\hline
Name FIRST      &  Name 2FGL      & O.I   &2FGL R.A.   &2FGL Dec.   &Radio-$\gamma$ Sep.& Association               & $\gamma$-var.& Radio-var.  \\   
~~~~~~~~~NVSS   &                 &       &rms[arcmin] &rms[arcmin] & [arcmin]          &                           &              &             \\
\hline
 
J0204.8+1514    &  J0205.0+1514   &SSRQ   &   8.1      &   6.5      &      2.7          & 4C15.05                   & T            & yes         \\
J0510.0+1800    &  J0509.9+1802   &FSRQ   &   6.5      &   5.1      &      1.7          & PKS 0507$+$17             & T            & no          \\
J0847.2+1133    &  J0847.2+1134   &BL\,Lac&   5.3      &   4.6      &      0.8          & RX J0847.1$+$1133         &              & no          \\
J0937.1+5008    &  J0937.6+5009   &FSRQ   &   8.6      &   2.3      &      7.4          & GB6 J0937$+$5008          & T            & yes         \\
J1010.8$-$0201  &  J1010.8$-$0158 &FSRQ   &   8.3      &   7.7      &      4.1          & PKS 1008$-$01             &              & yes         \\   
J1204.2$-$0710  &  J1204.3$-$0711 &BL\,Lac&   8.0      &   6.9      &      1.8          & 1RXS J120417.0$-$070959   &              & no          \\ 
J1231.7+2848    &  J1231.7+2848   &BL\,Lac&   2.3      &   2.2      &      0.4          & B2 1229$+$29              & T            & no          \\
J1332.7+4722    &  J1332.7+4725   &FSRQ   &  13.6      &  13.1      &     10.4          & B3 1330$+$476             & T            & yes         \\
J1656.8+6012    &  J1656.5+6012   &FSRQ   &   8.4      &   7.7      &      4.5          & 87GB 165604.4$+$601702    &              & yes         \\
\hline
% \\             
Name ATCA   \\  %  &   Name 2FGL     & O.I &2FGL R.A.   &2FGL Dec.   &Radio-$\gamma$ Sep.& Association               & $\gamma$-Var. & Radio-Var. \\  
               % &                 &     &rms[arcmin] &rms[arcmin] &  [arcmin]         &                           &              &             \\
\hline
J0448.2$-$2110  &  J0448.6$-$2118 &FSRQ   &  14.8      &  10.3      &      9.9          & PKS 0446$-$212            &              &              \\
J0449.4$-$4349  &  J0449.4$-$4350 &BL\,Lac&   1.2      &   1.1      &      1.1          & PKS 0447$-$439                & T            &              \\
J1610.3$-$3958  &  J1610.6$-$4002 &FSRQ   &  18.5      &  11.5      &      6.1          & PMN J1610$-$3958          &              &             \\
J1936.8$-$4719  &  J1936.8$-$4721 &BL\,Lac&   4.1      &   8.5      &      2.0          & PMN J1936$-$4719          &              &              \\
J2258.3$-$5525  &  J2258.8$-$5524 &BL\,Lac&  10.2      &   9.0      &      9.1          & PMN 2258$-$5526           &              &              \\
J2330.6$-$3724  &  J2330.6$-$3723 &BL\,Lac&   7.1      &   6.5      &      0.5          & PKS 2327$-$376            &              &             \\
\hline                                                                     
\end{tabular}                                                            
\normalfont                                                                
\smallskip\noindent                                                      
\flushleft{\normalsize {                                                 
Table $-$                                                                     
column 1: FIRST, NVSS or ATCA source name;                                                    
2: 2FLG source name;
3: optical Identification;
4: right ascention rms from the 2FGL in arcmin;
5: declination rms from the 2FGL in arcmin;                                       
6: separation beetween FIRST, NVSS or ATCA object to 2FGL object in arcmin;                 
7: Name of the associated radio source as suggested in 2FGL;
8: variability as in 2FGL: T indicates $<1$\% chance of being a steady source;                       
9: radio variability as in \citet{Mantovani11}.     
}}                                                                       
\end{table*}                                                                                        
% 
% 	online material
\Online

\begin{appendix} %First online appendix
\section{Images}
In this Appendix we present the images of the 87 weak blazars observed with the EVN at 5\,GHz.
Map peak in mJy/beam, contours in percentage of the peak flux density in the map, and the
full width half maximum (FWHM) of the beam in milli-arcseconds and the position angles of
the beam major axis in degrees are reported for each image.
\begin{figure*}
%\centering
\includegraphics[width=7cm,clip]{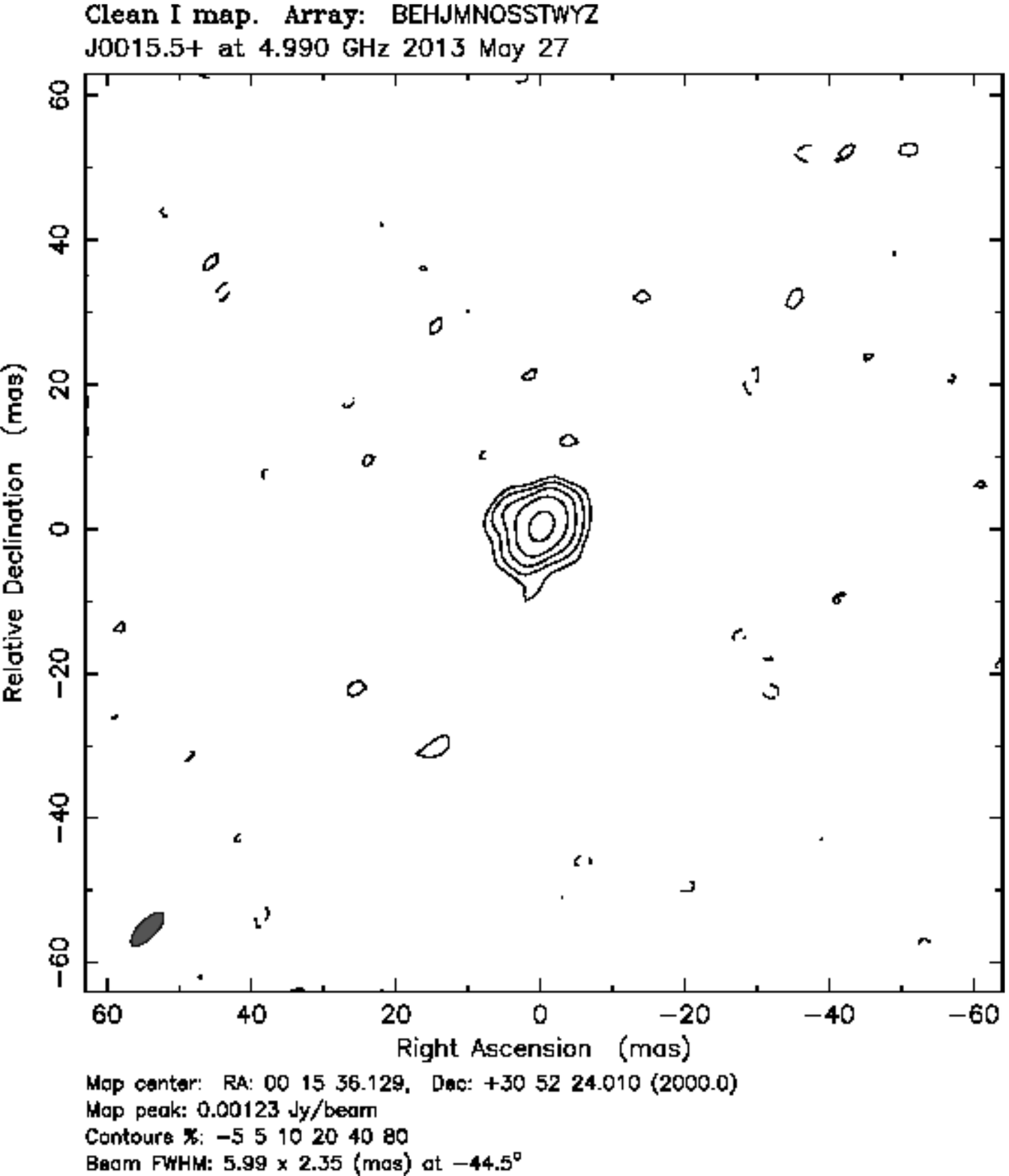}
\includegraphics[width=7cm,clip]{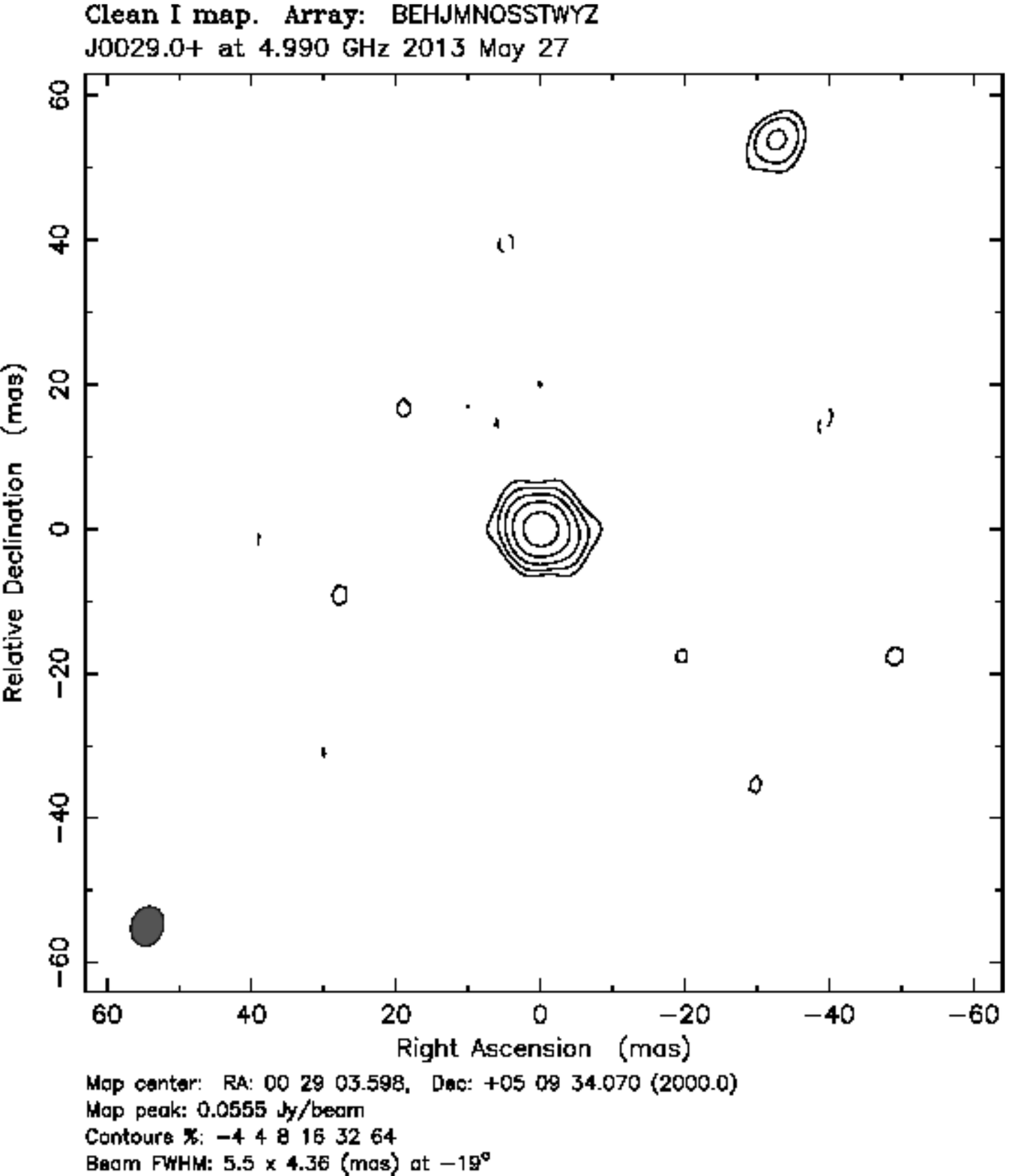} \\
\newline
\includegraphics[width=7cm,clip]{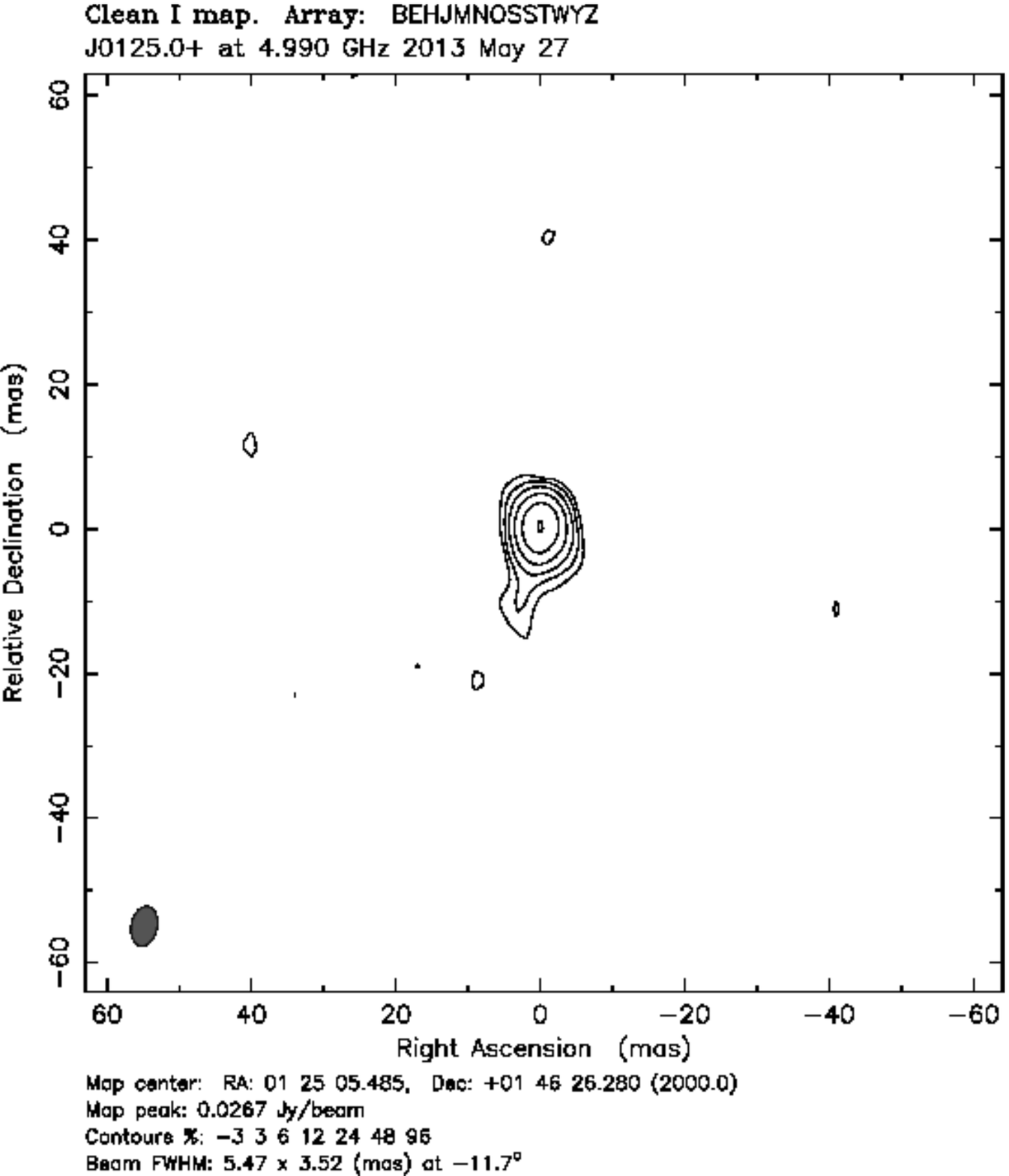} 
\includegraphics[width=7cm,clip]{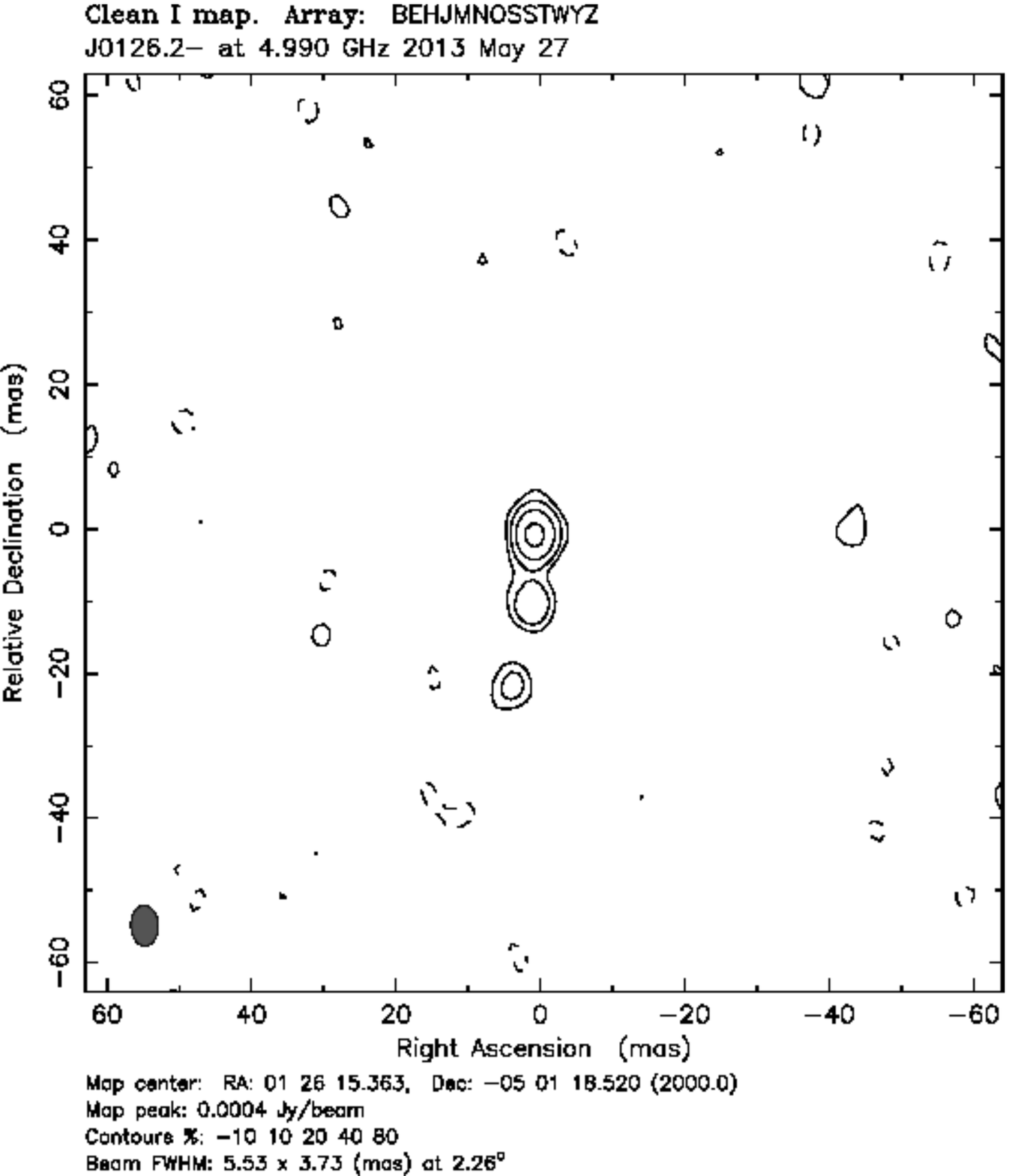} \\
\newline
\includegraphics[width=7cm,clip]{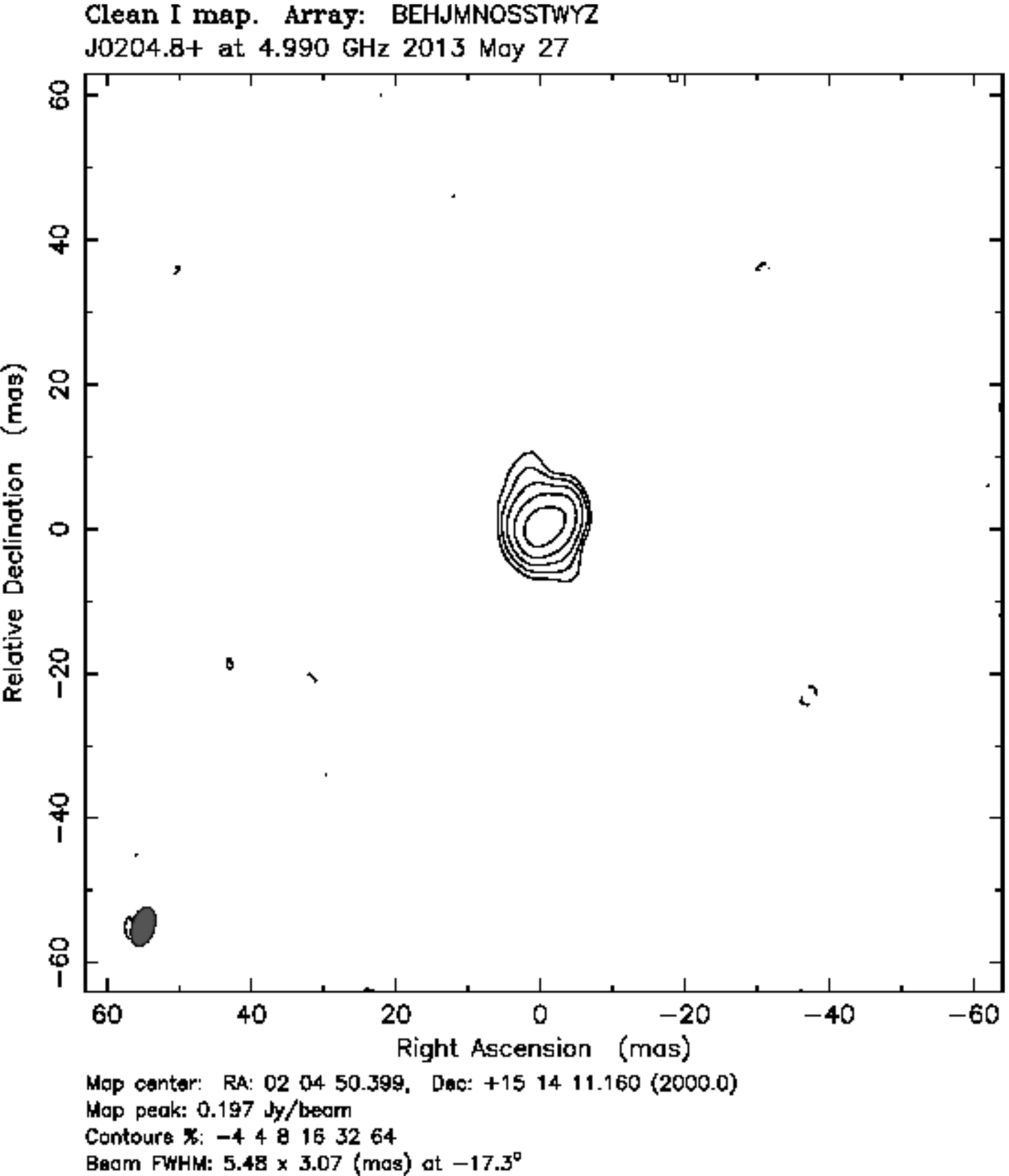}
\includegraphics[width=7cm,clip]{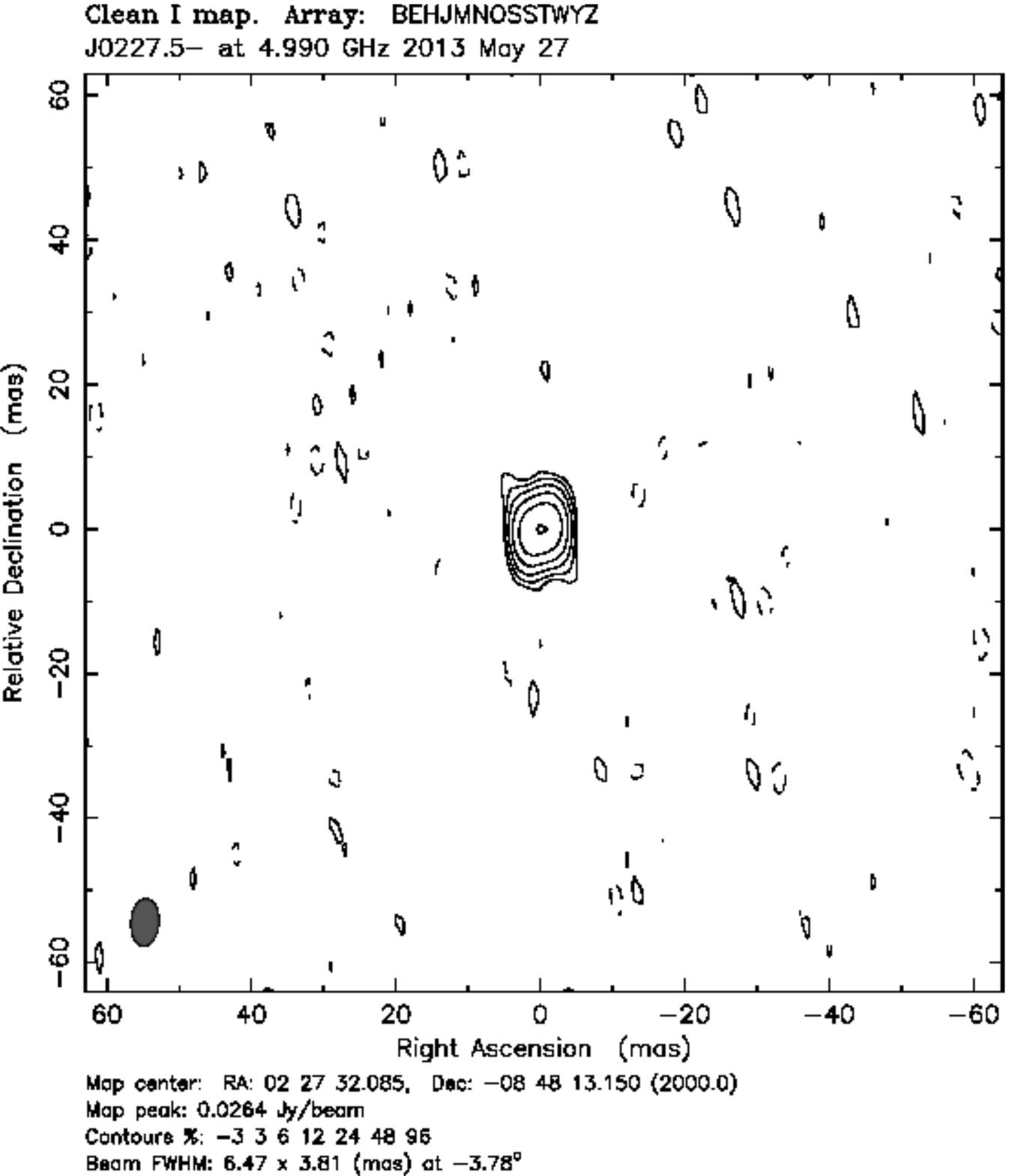}
\caption{EVN images at 5\,GHz.}\label{appfig}
\end{figure*}
\begin{figure*}
\includegraphics[width=7cm,clip]{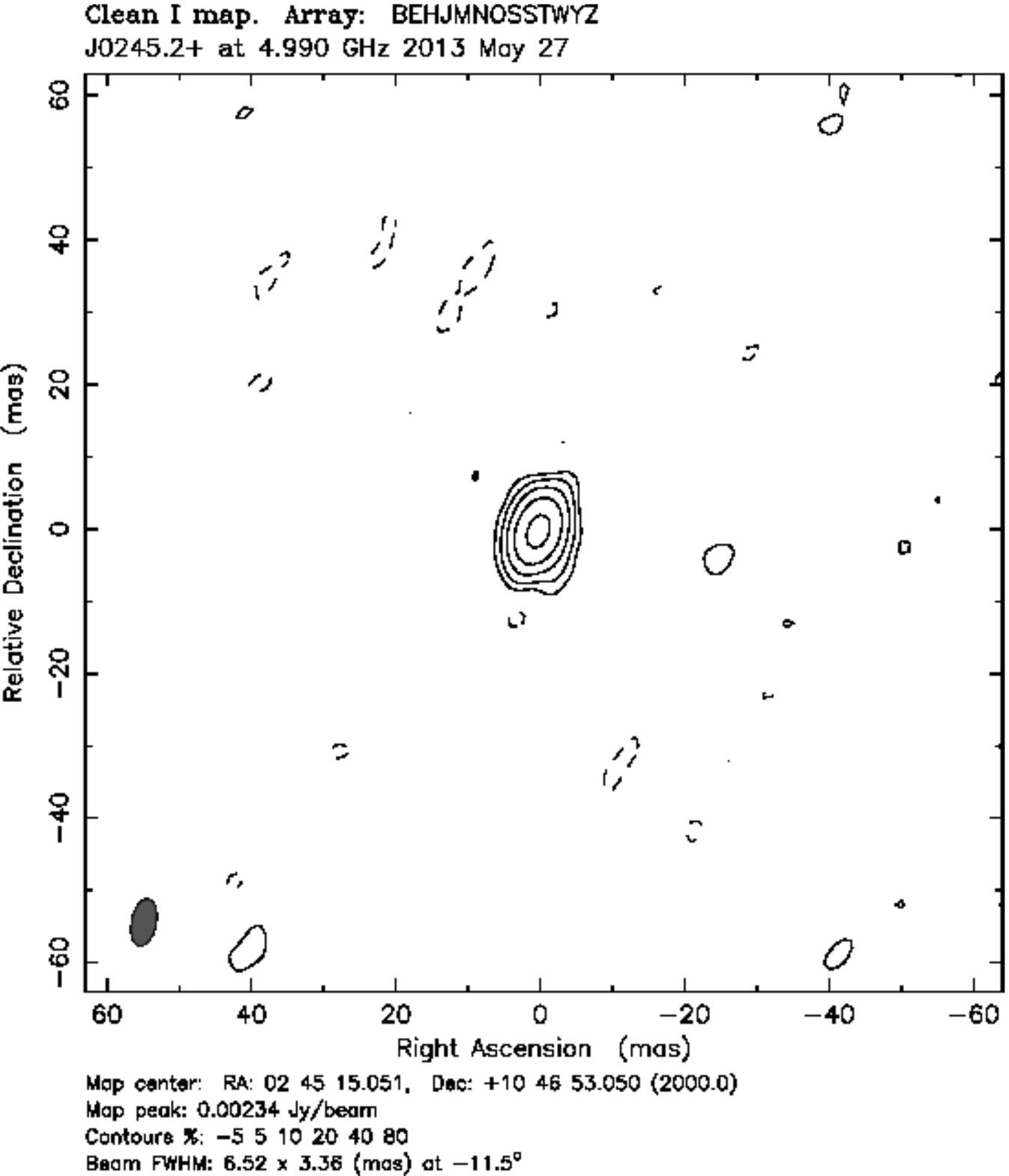} 
\includegraphics[width=7cm,clip]{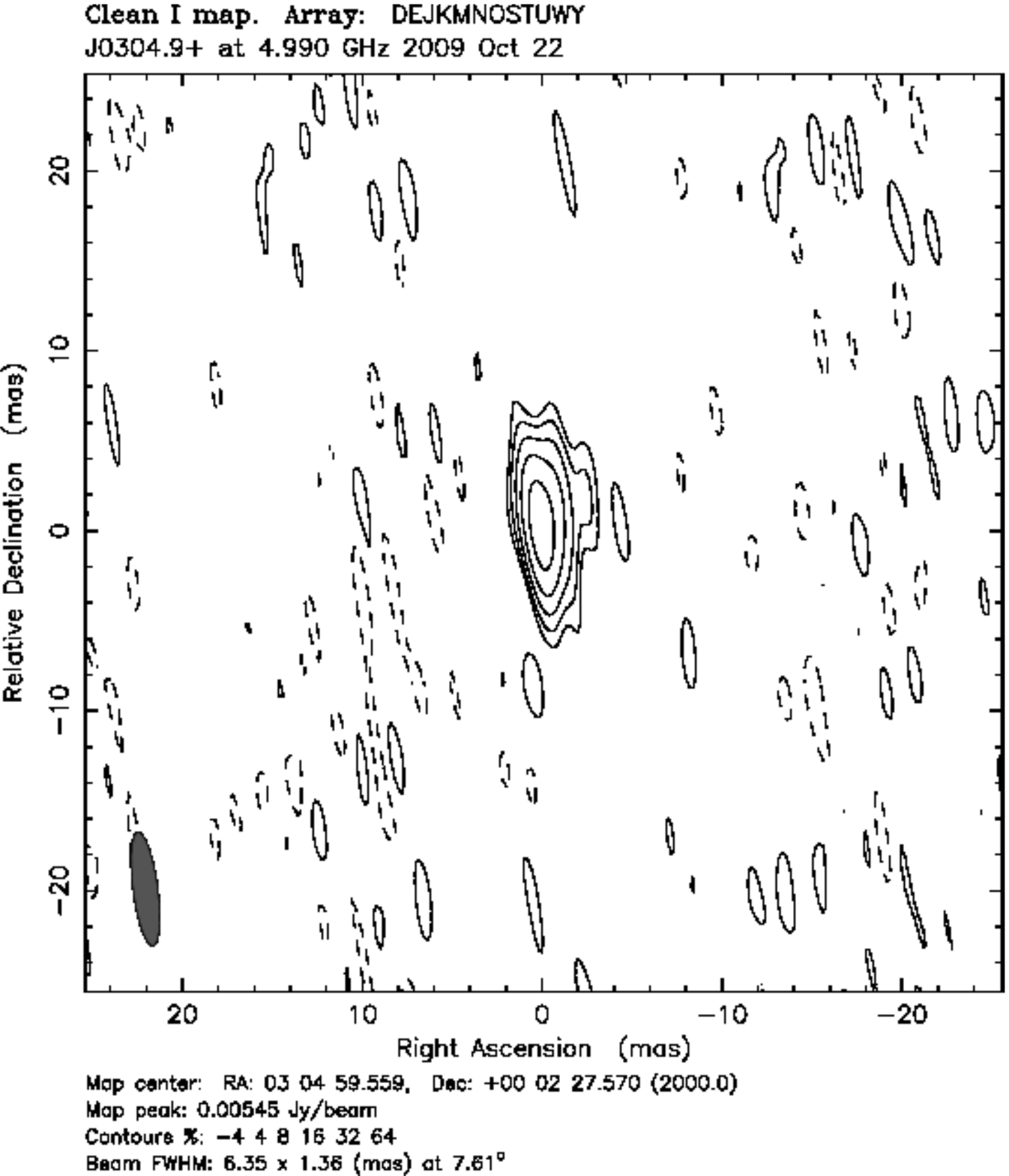} \\
\newline
\includegraphics[width=7cm,clip]{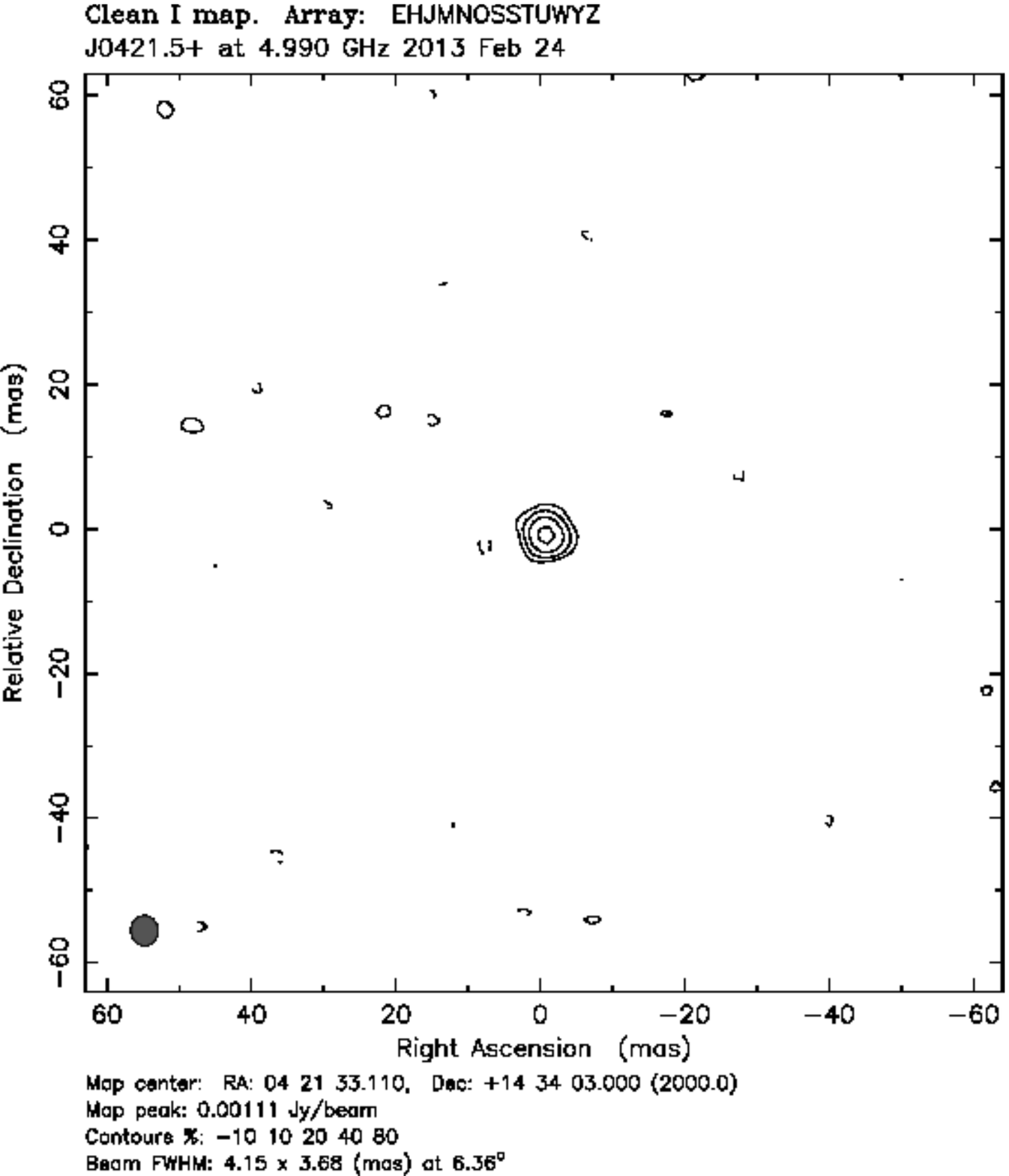}
\includegraphics[width=7cm,clip]{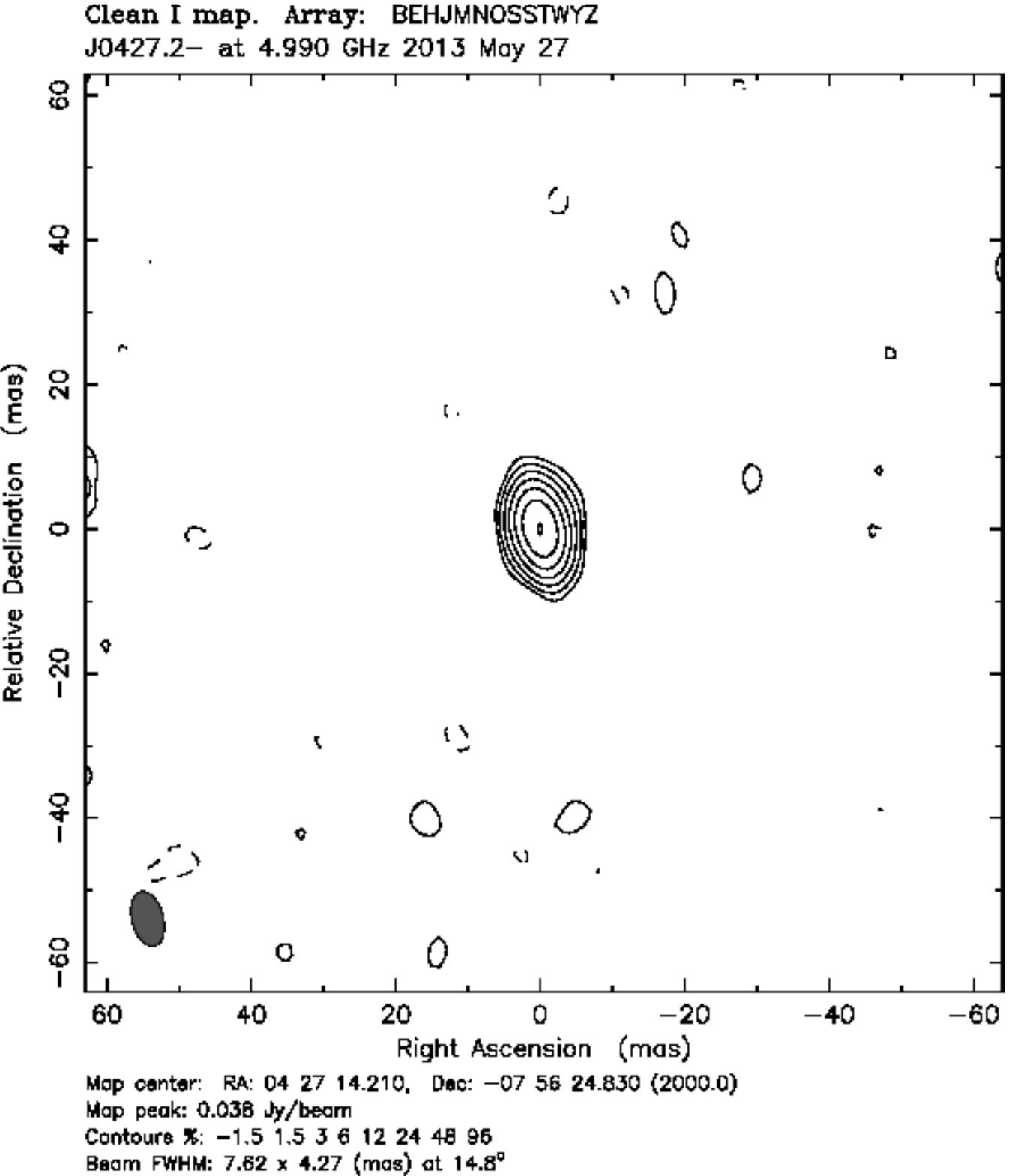} \\
\newline
\includegraphics[width=7cm,clip]{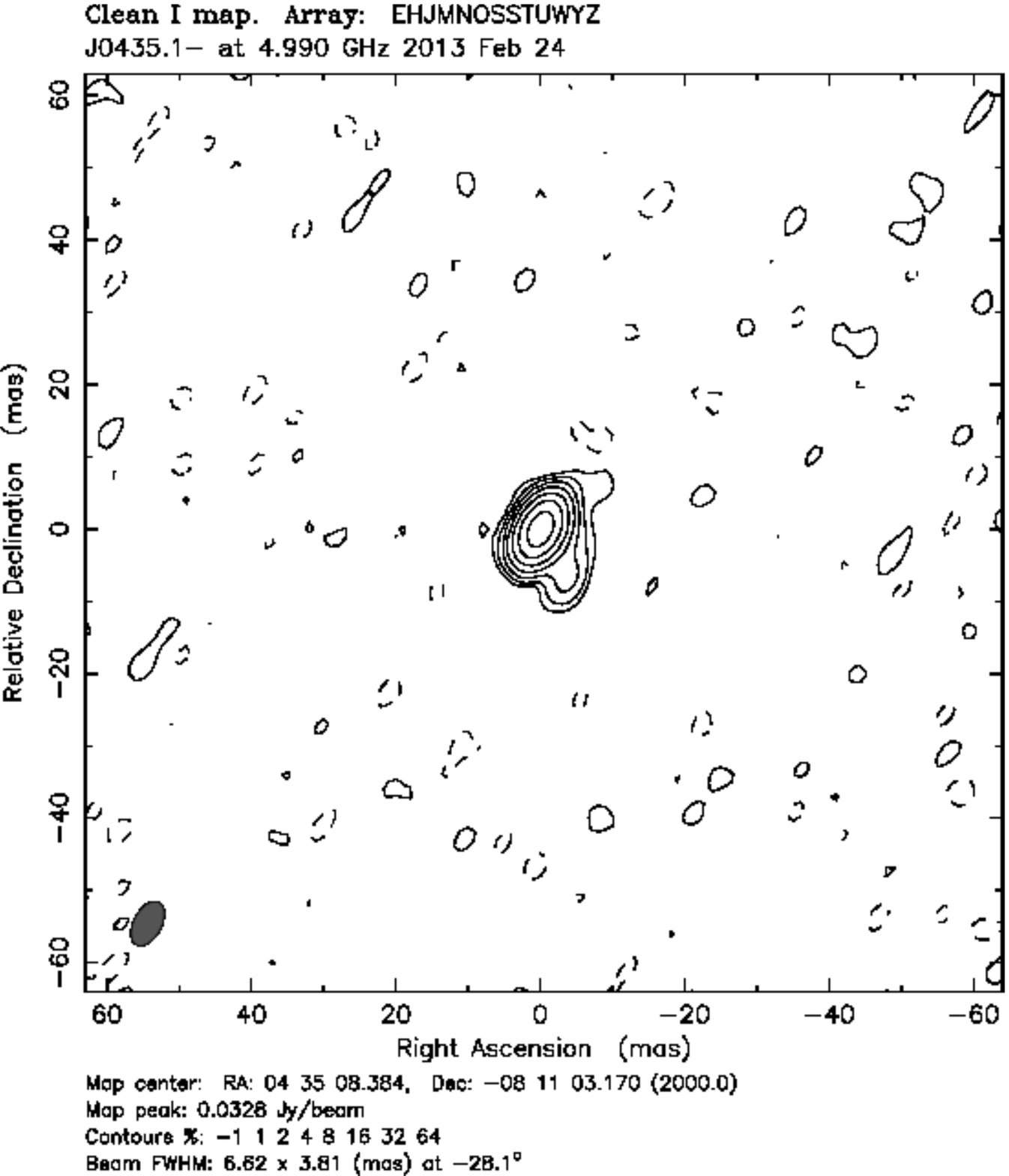}
\includegraphics[width=7cm,clip]{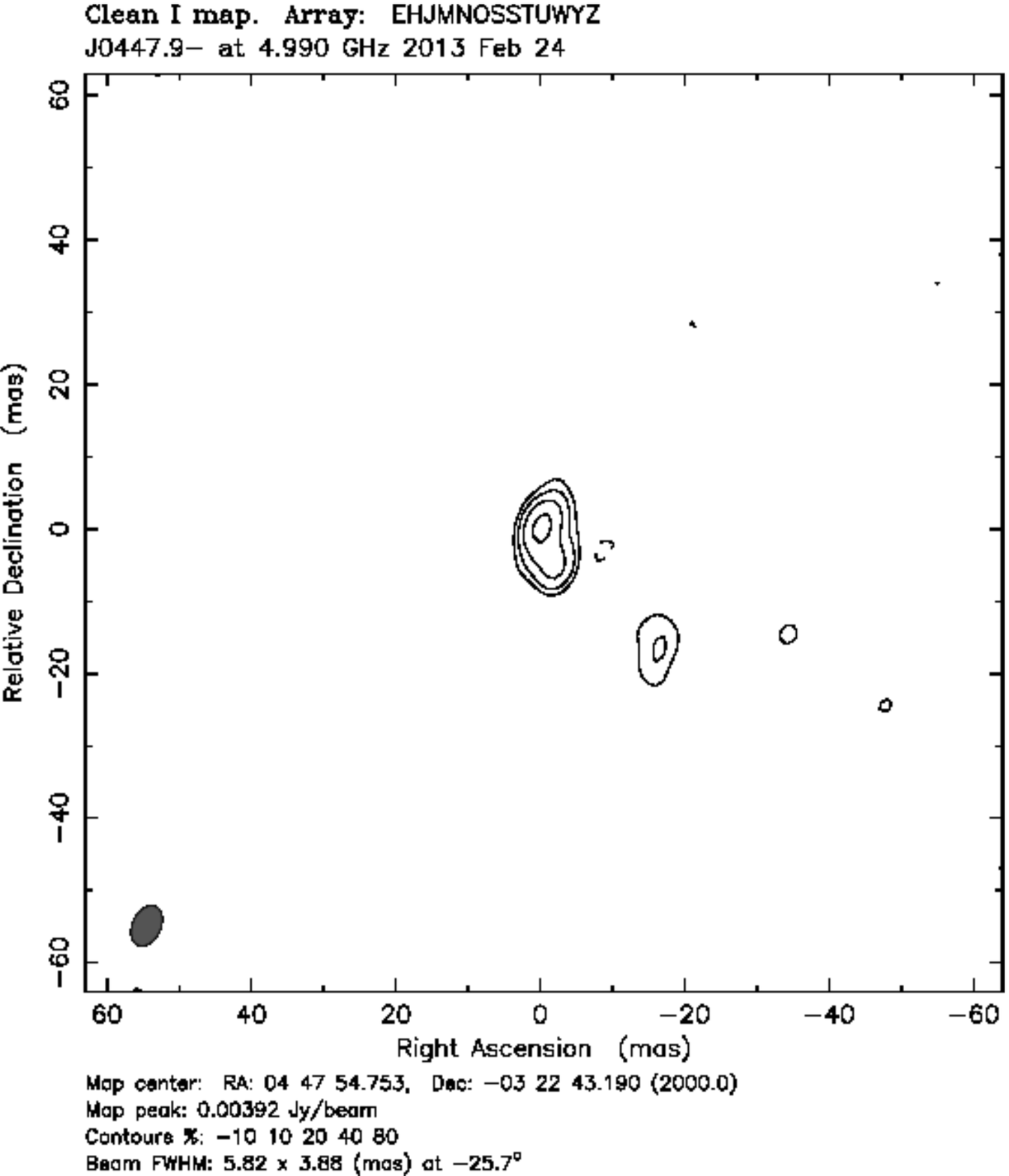}
\caption{EVN images at 5\,GHz. Continued}\label{appfig}
\end{figure*}
\begin{figure*}
\includegraphics[width=7cm,clip]{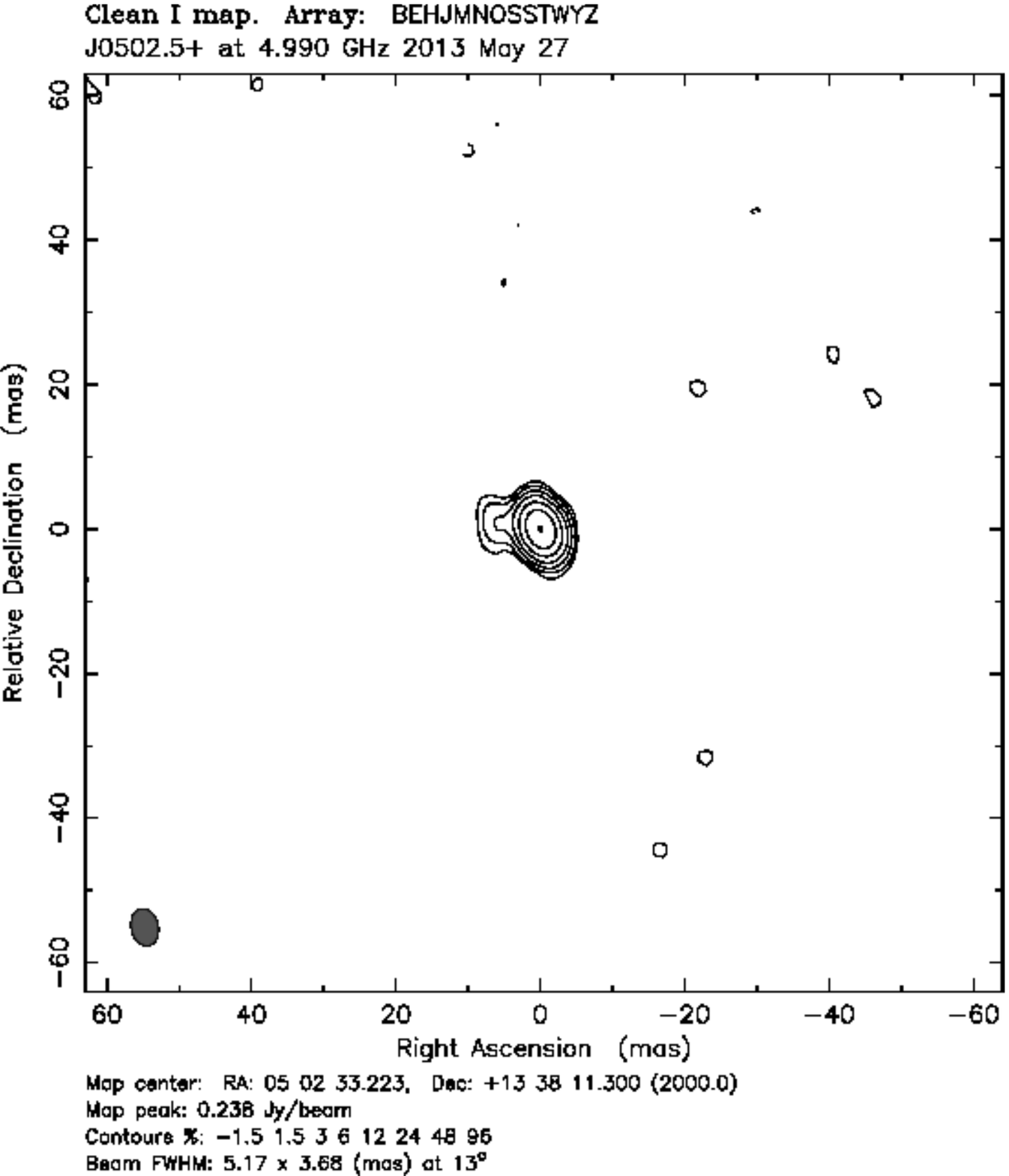}
\includegraphics[width=7cm,clip]{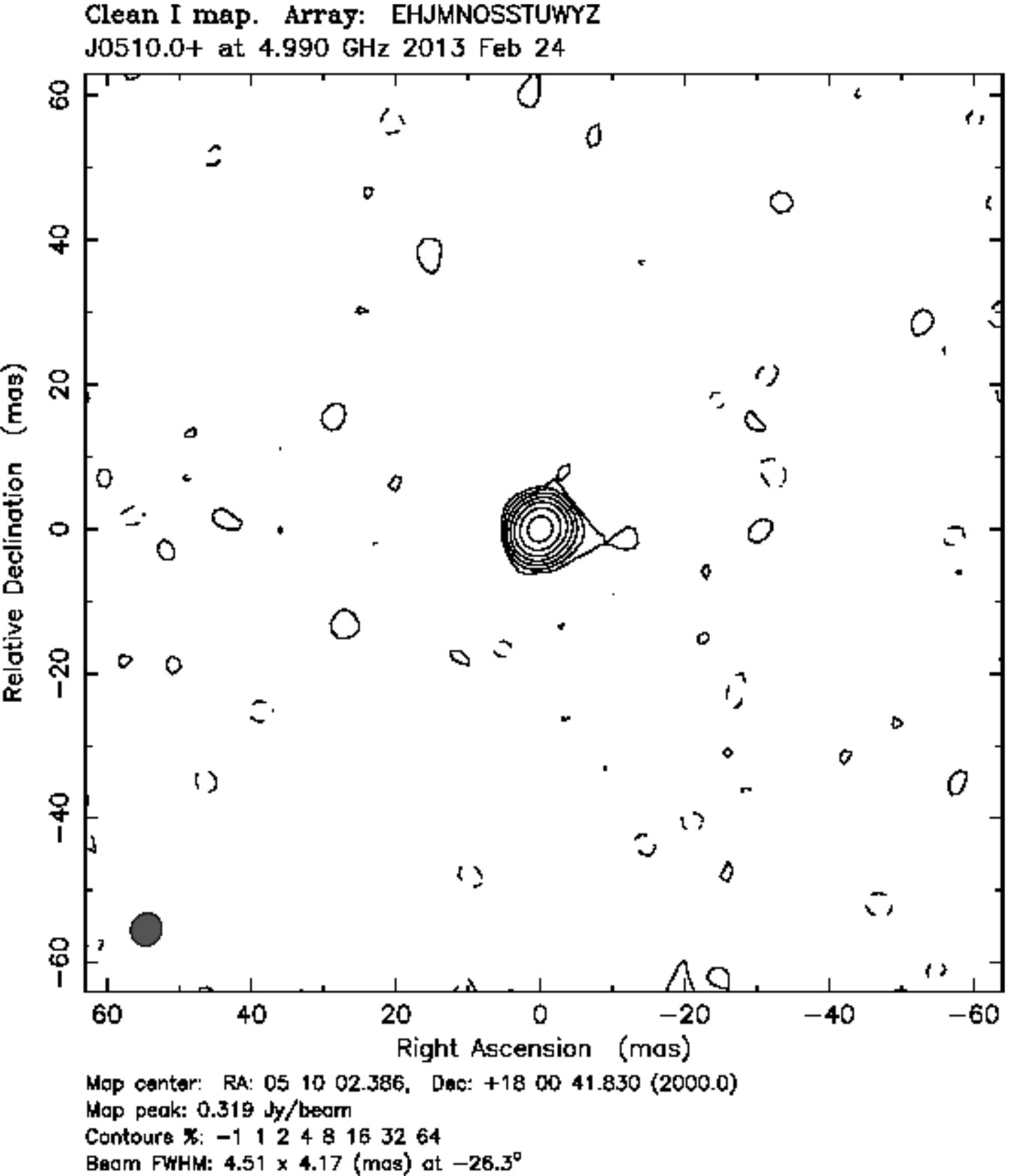} \\
\newline
\includegraphics[width=7cm,clip]{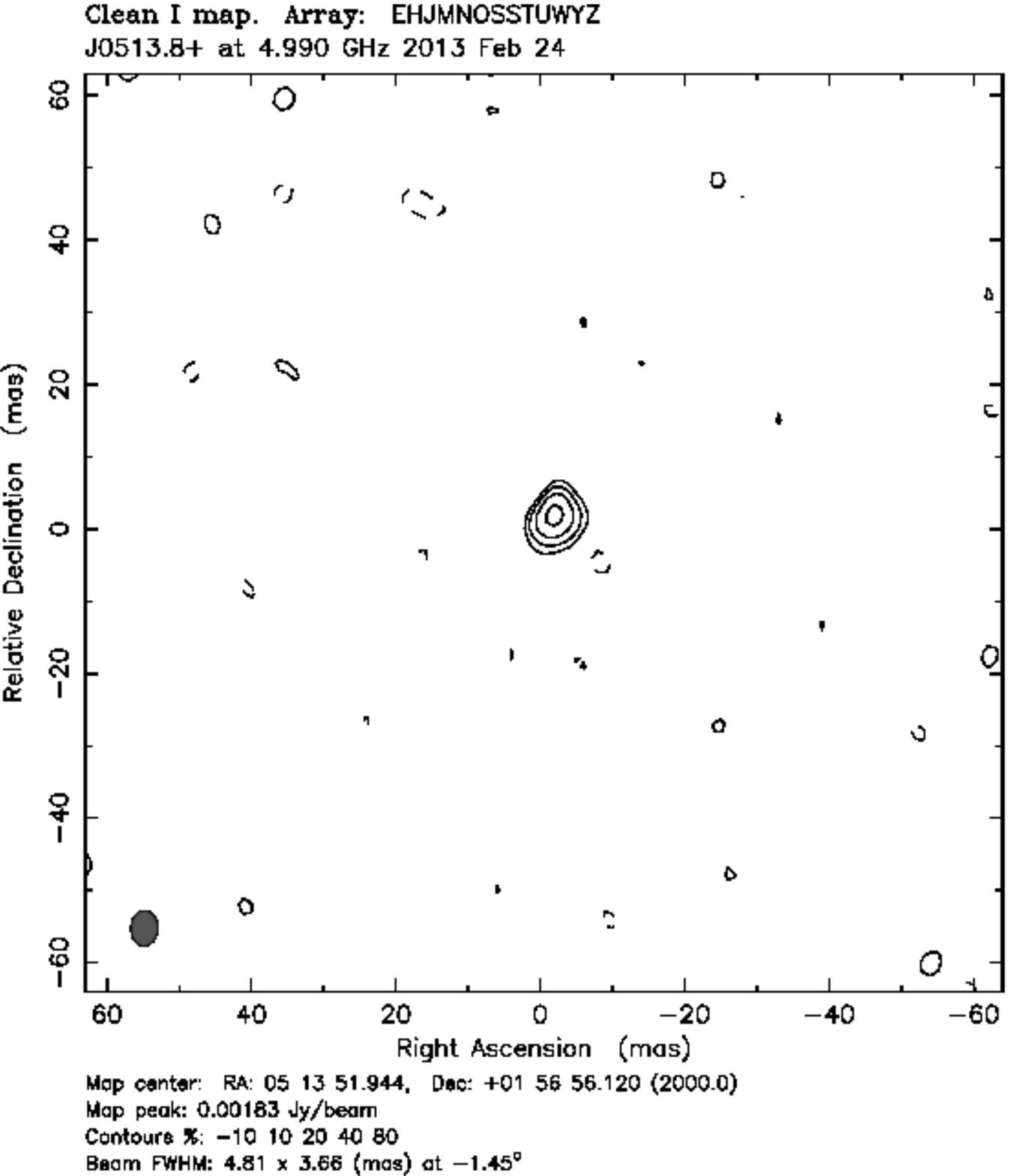}
\includegraphics[width=7cm,clip]{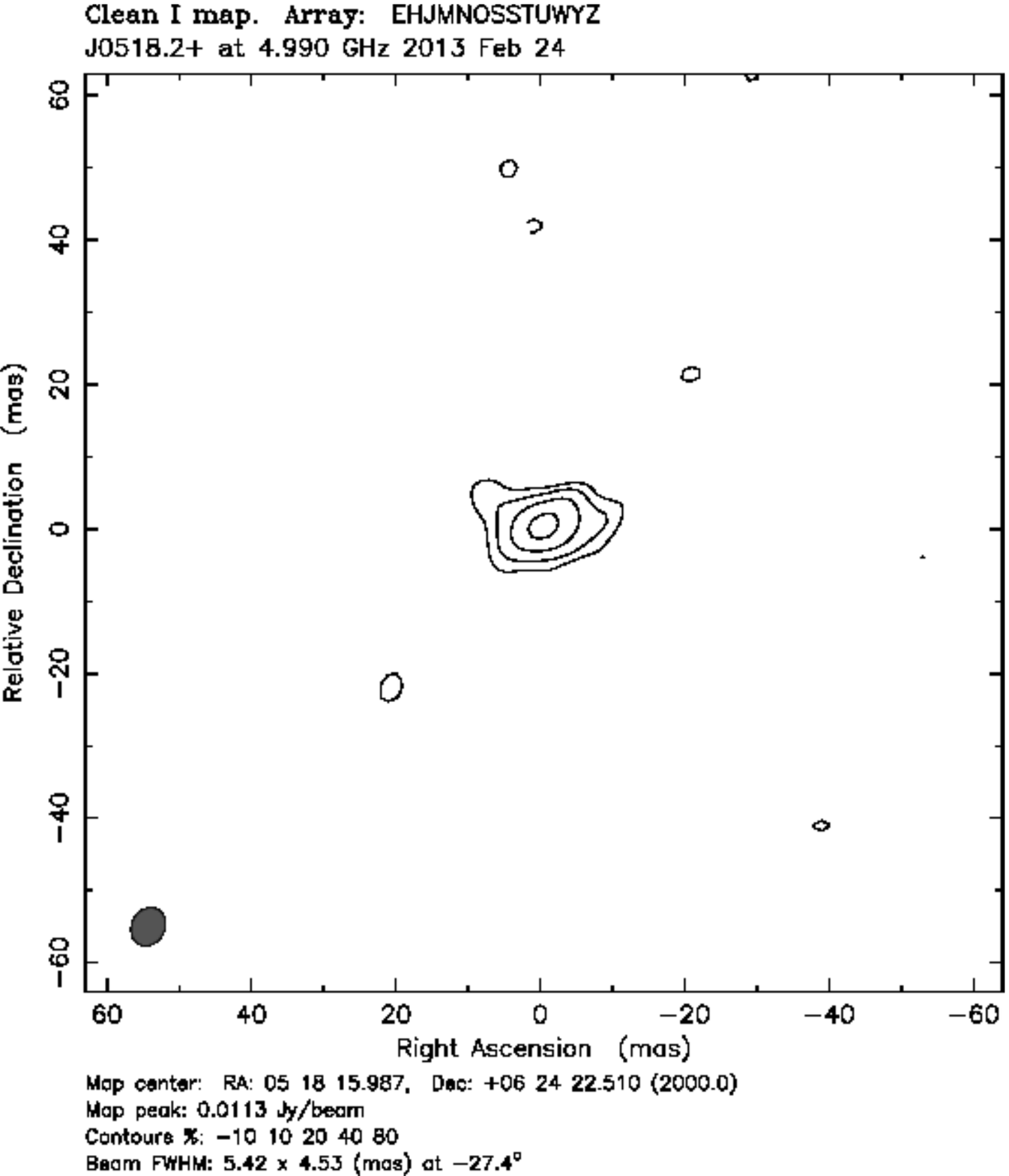} \\
\newline
\includegraphics[width=7cm,clip]{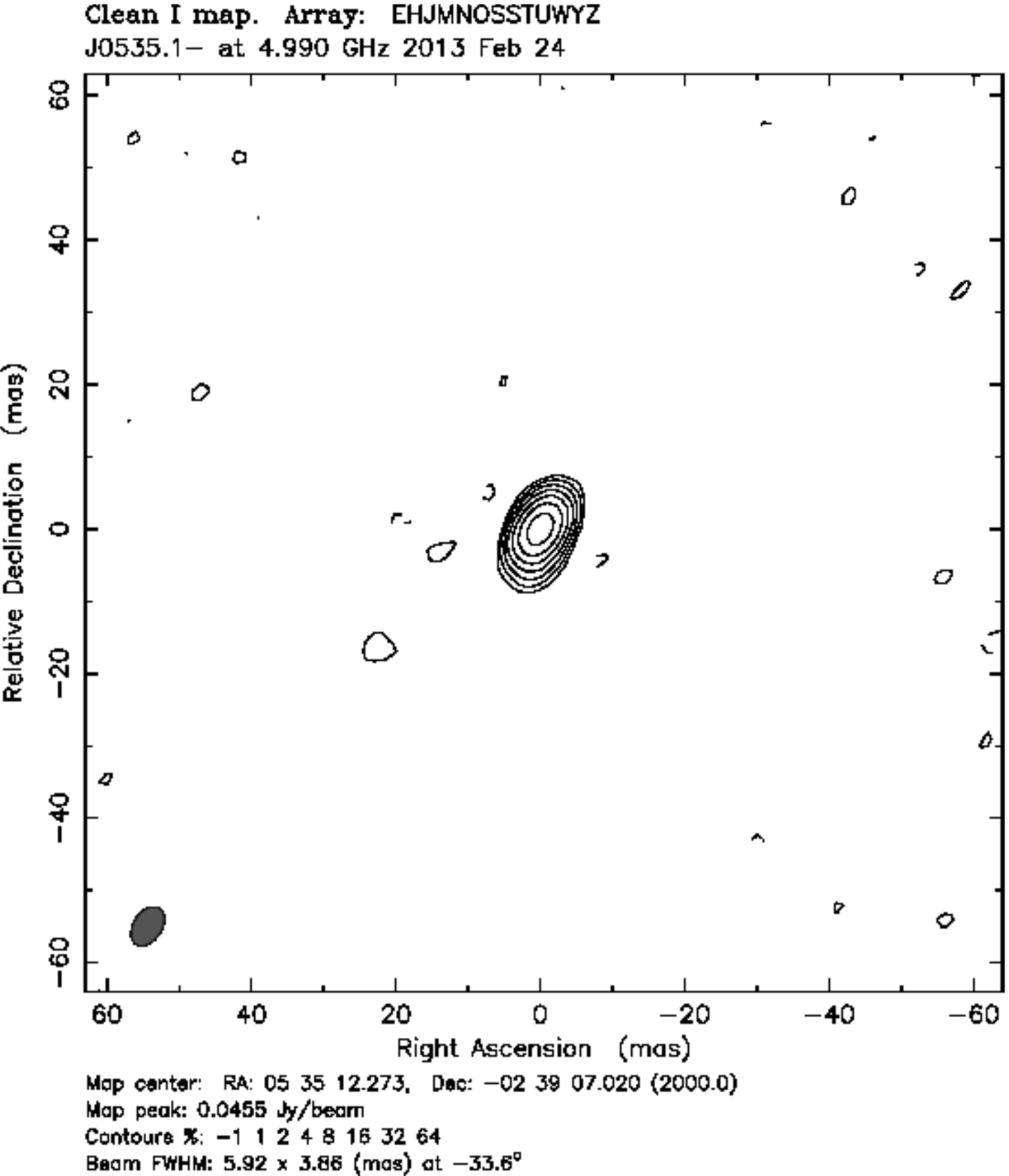}
\includegraphics[width=7cm,clip]{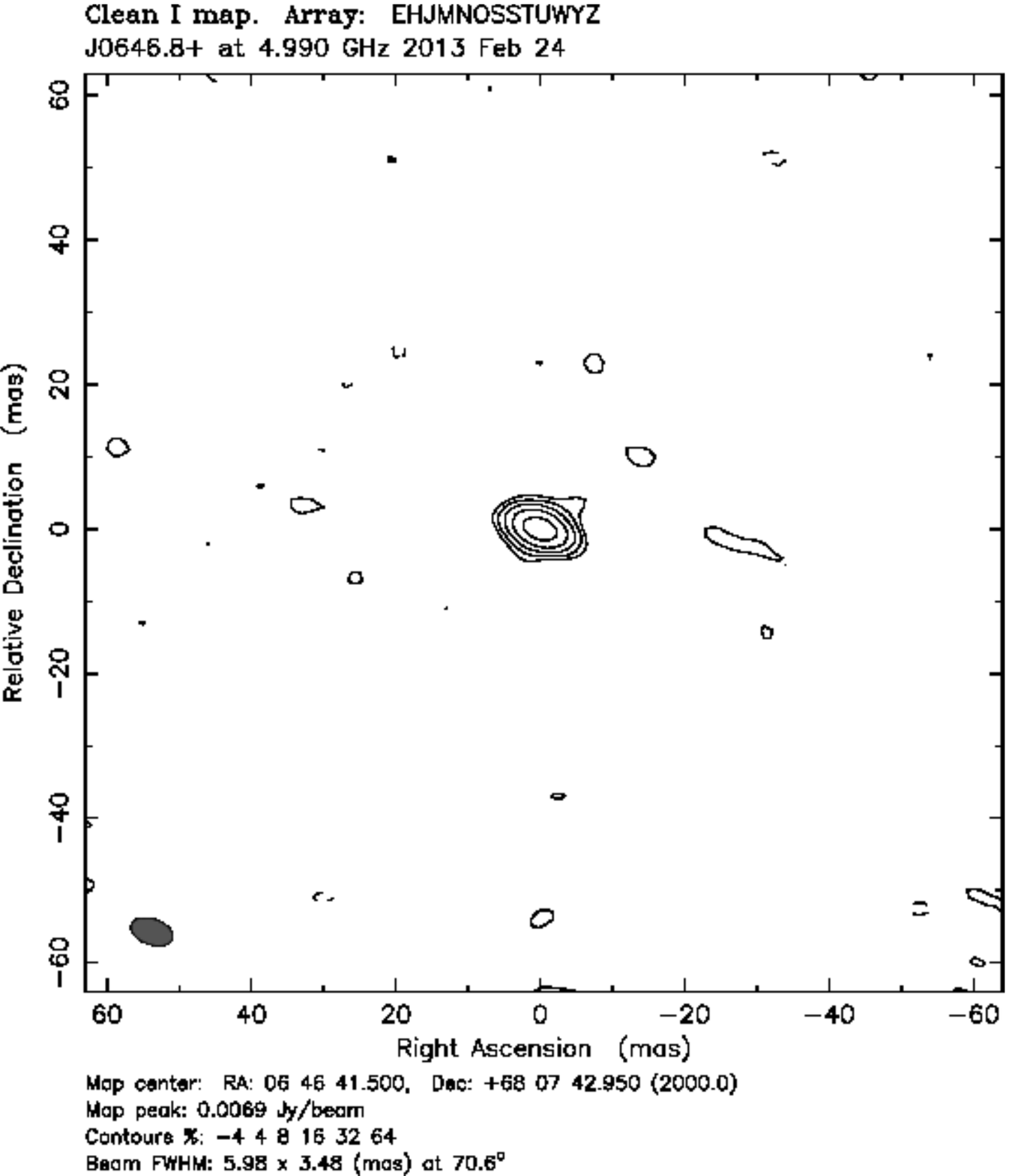}
\caption{EVN images at 5\,GHz. Continued}\label{appfig}
\end{figure*}
\begin{figure*}
\includegraphics[width=7cm,clip]{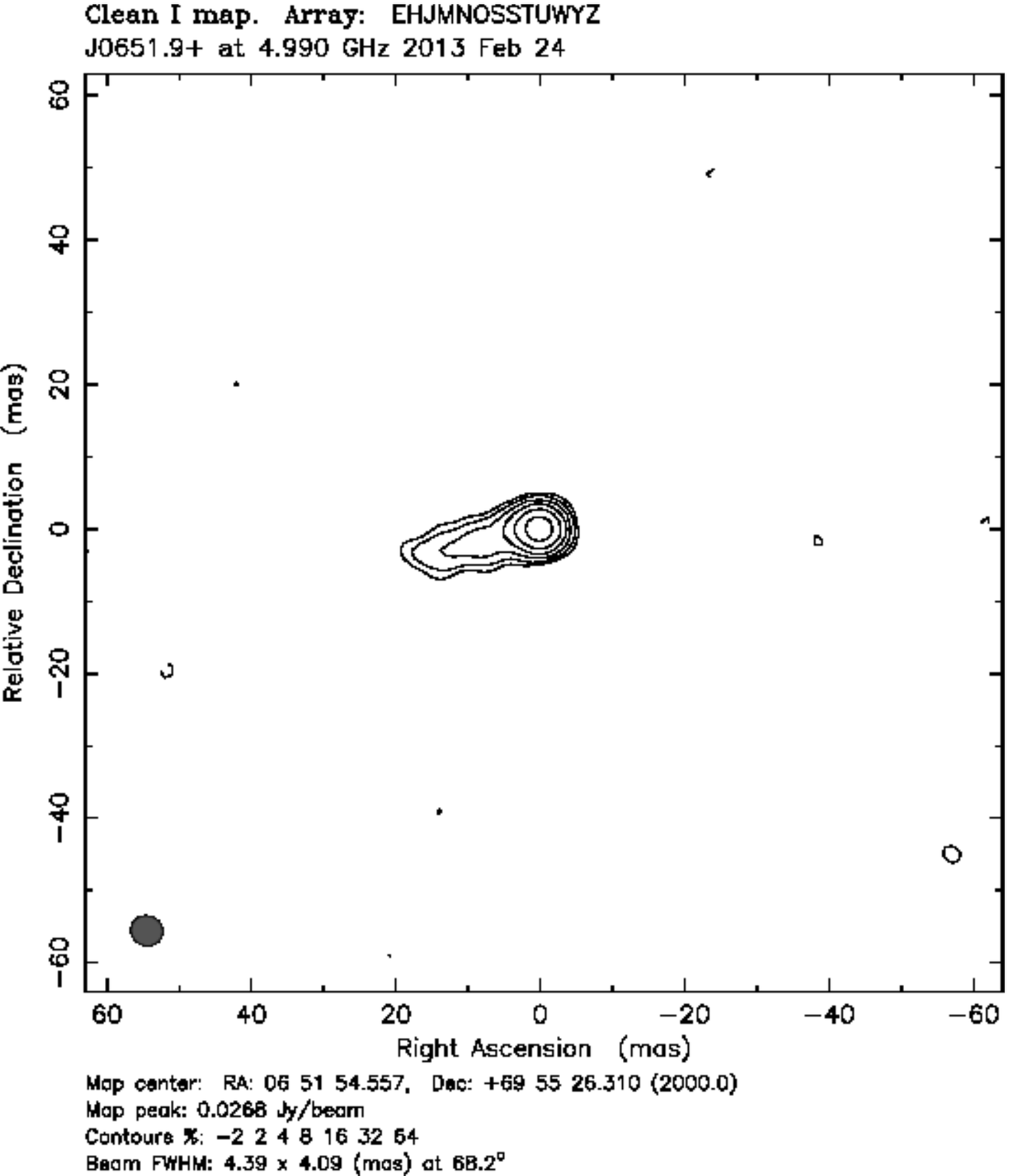}
\includegraphics[width=7cm,clip]{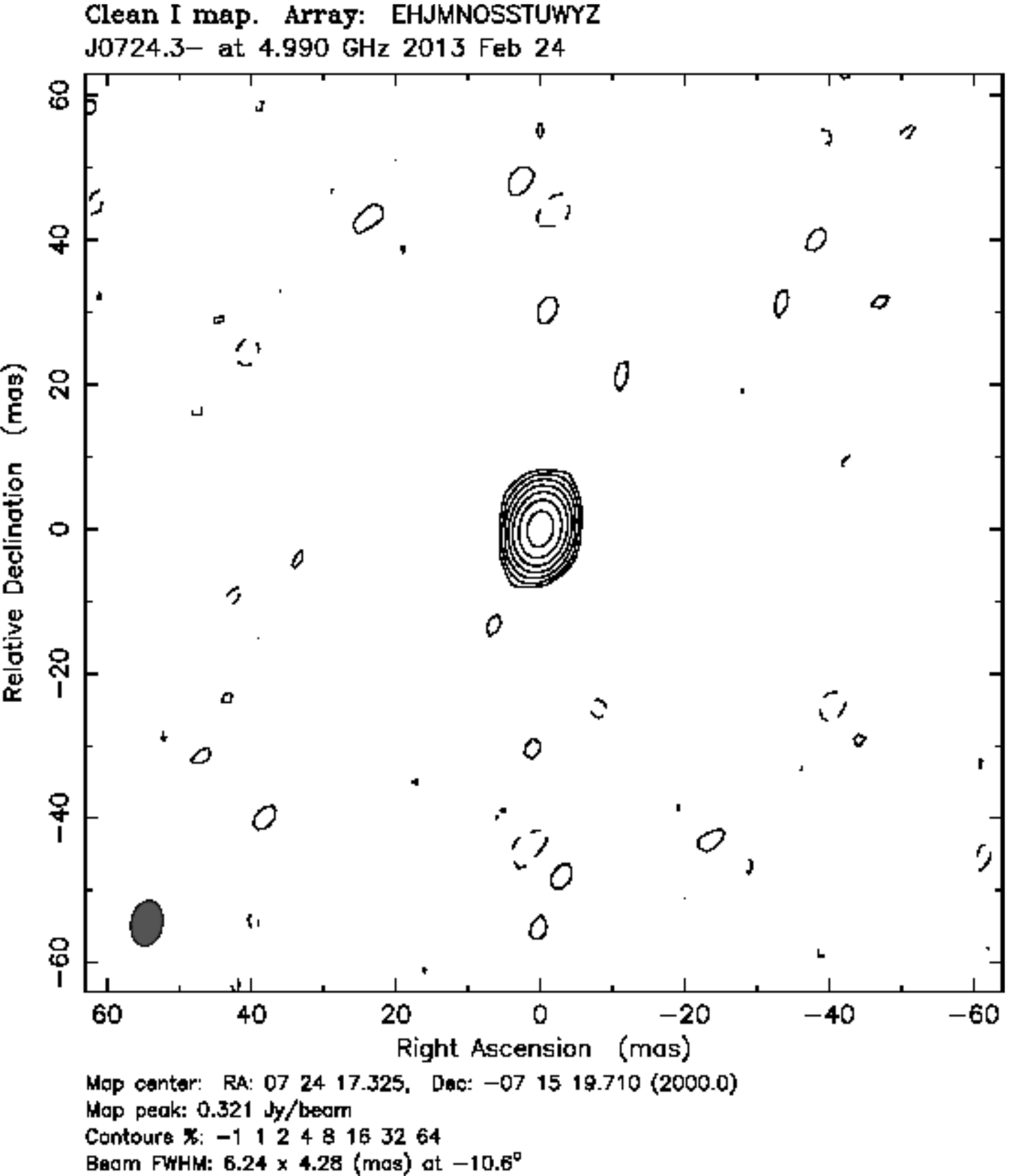} \\
\newline
\includegraphics[width=7cm,clip]{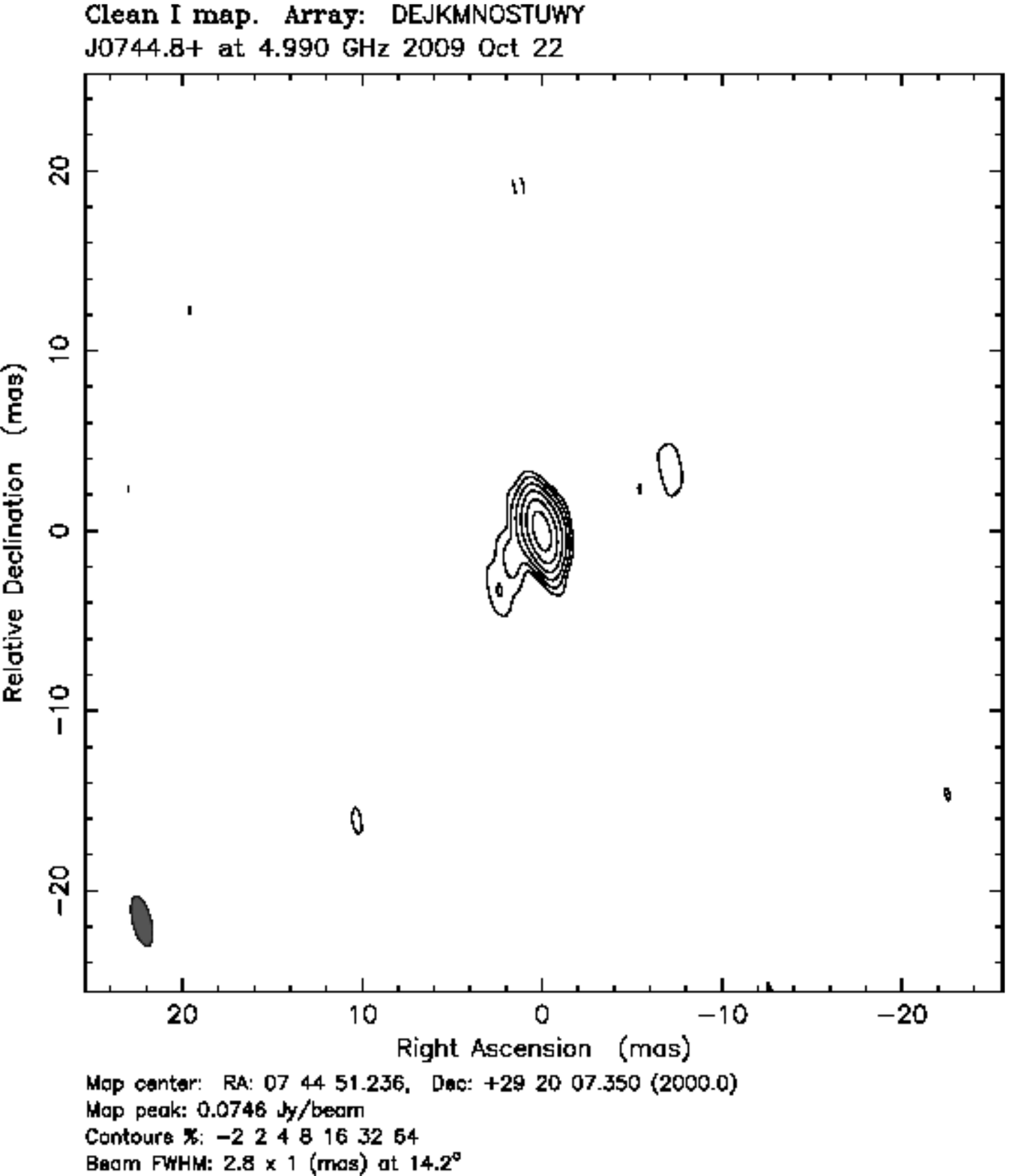}
\includegraphics[width=7cm,clip]{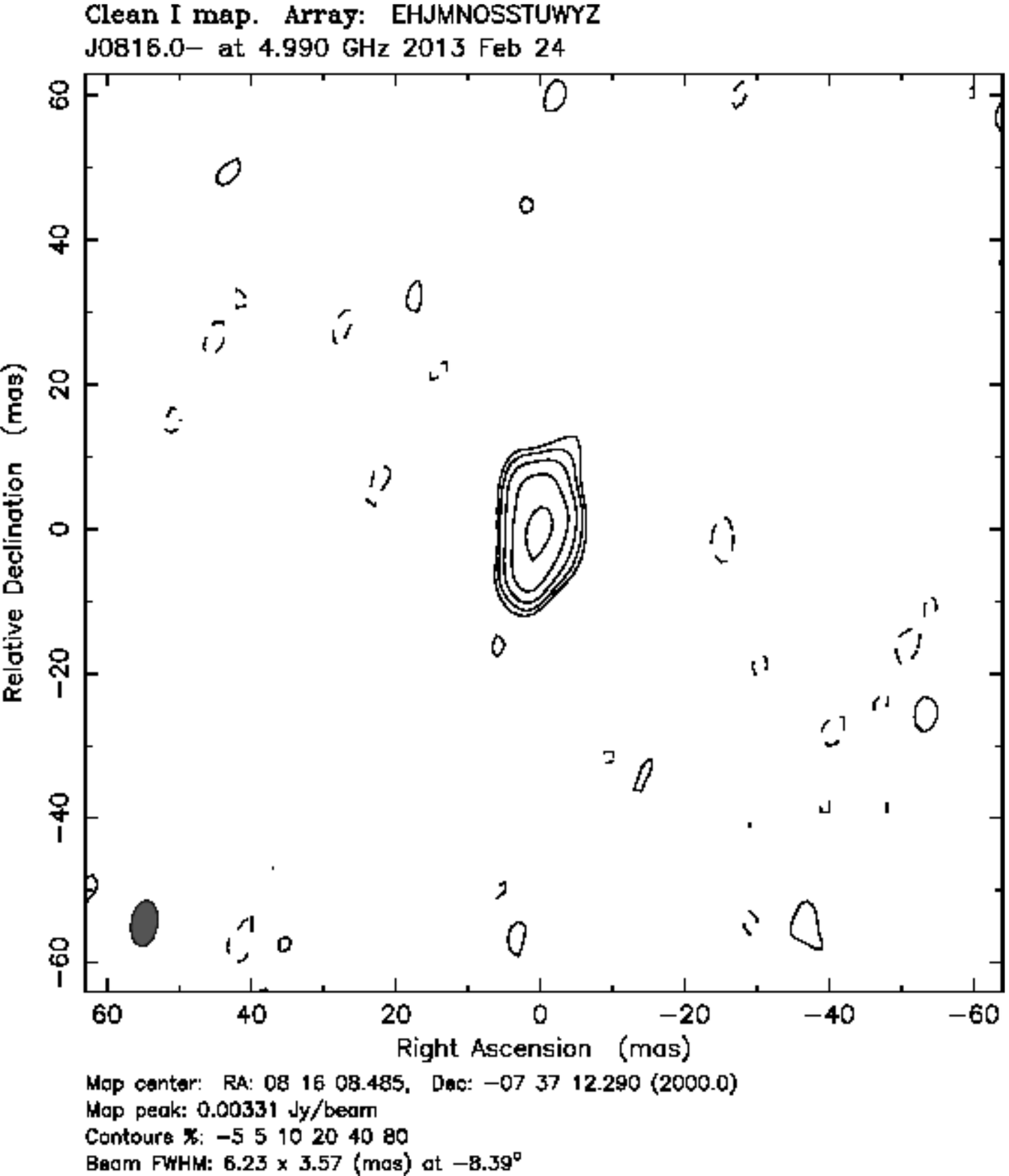} \\
\newline
\includegraphics[width=7cm,clip]{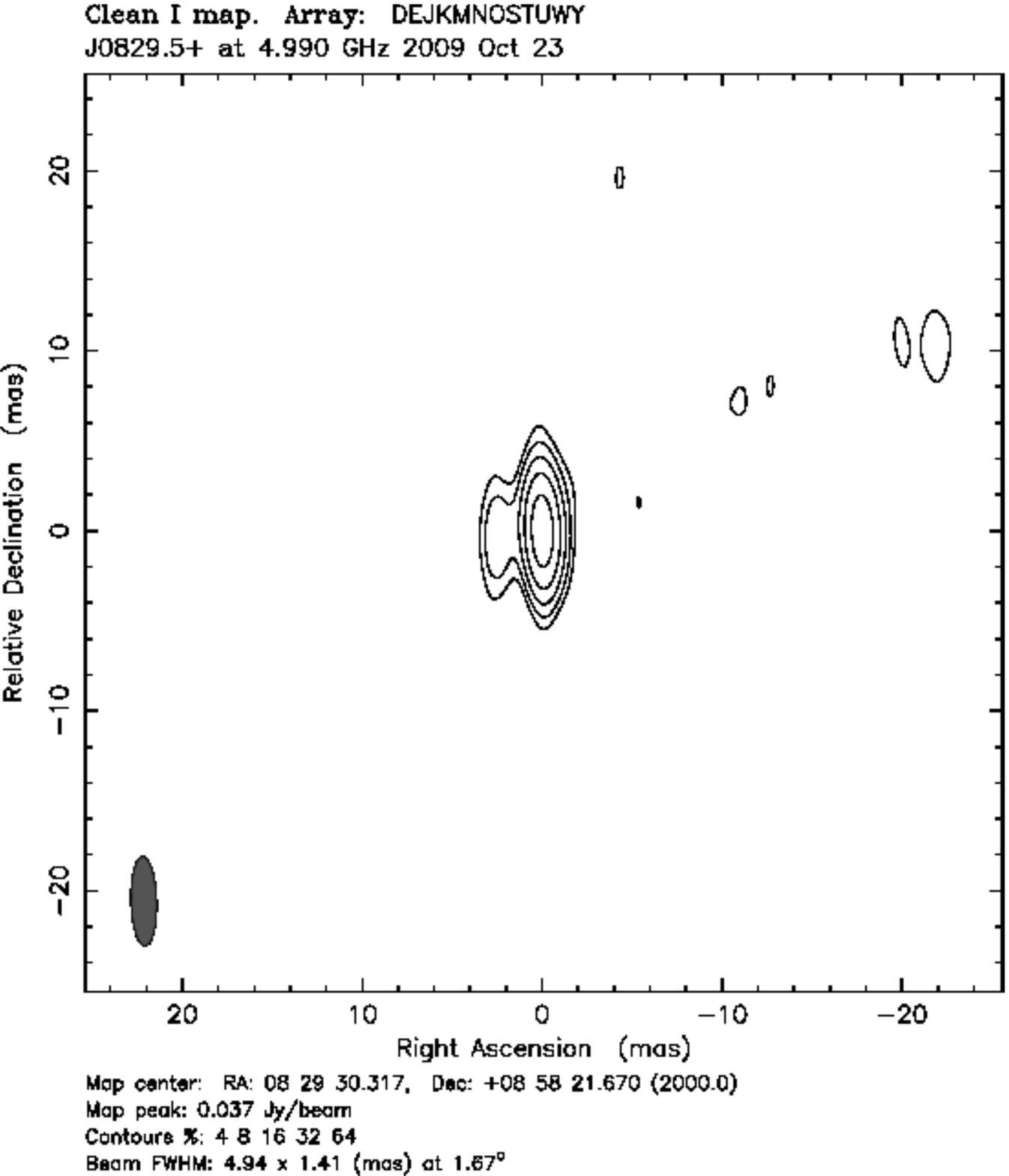}
\includegraphics[width=7cm,clip]{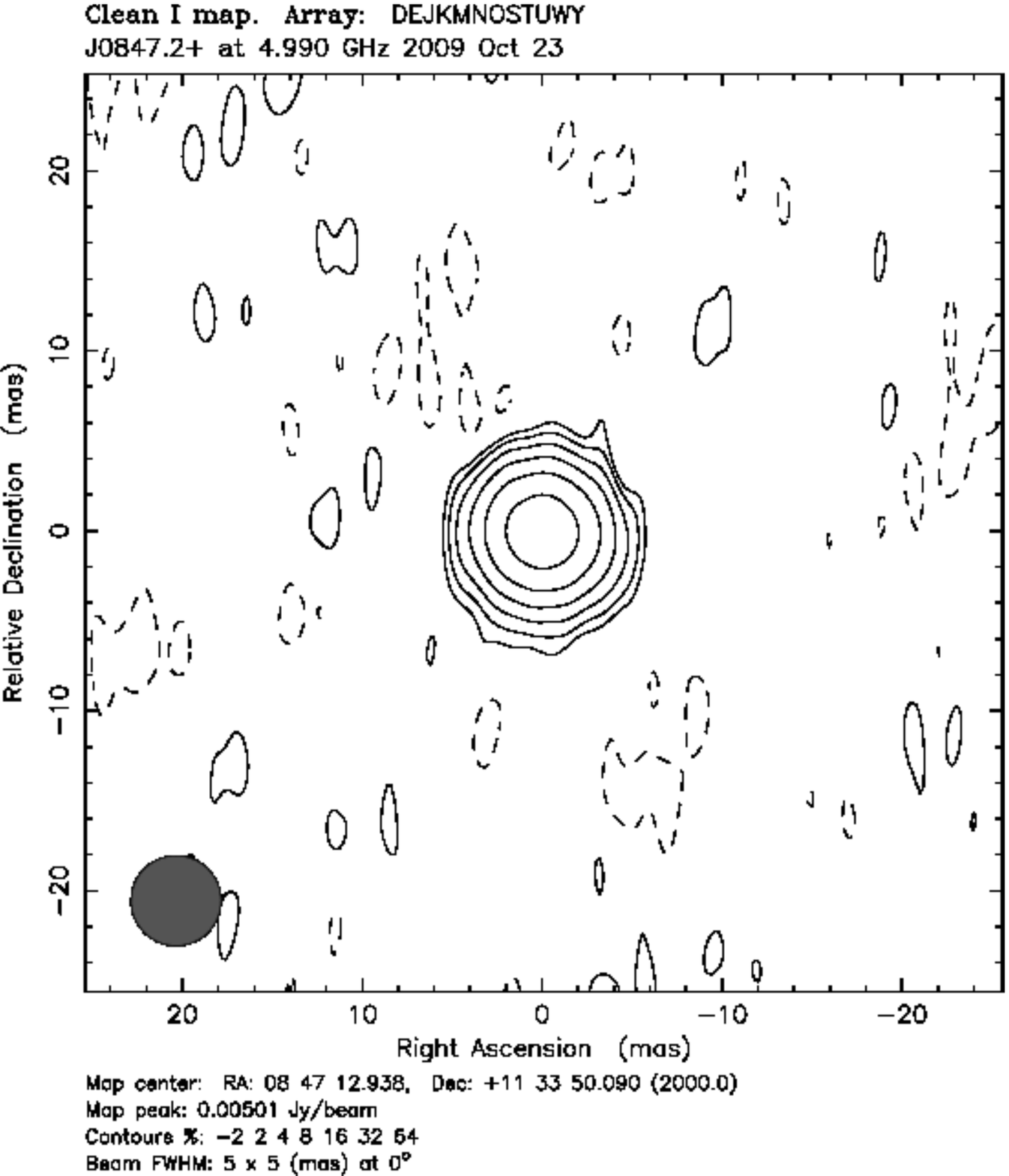}
\caption{EVN images at 5\,GHz. Continue}\label{appfig}
\end{figure*}
\begin{figure*}
\includegraphics[width=7cm,clip]{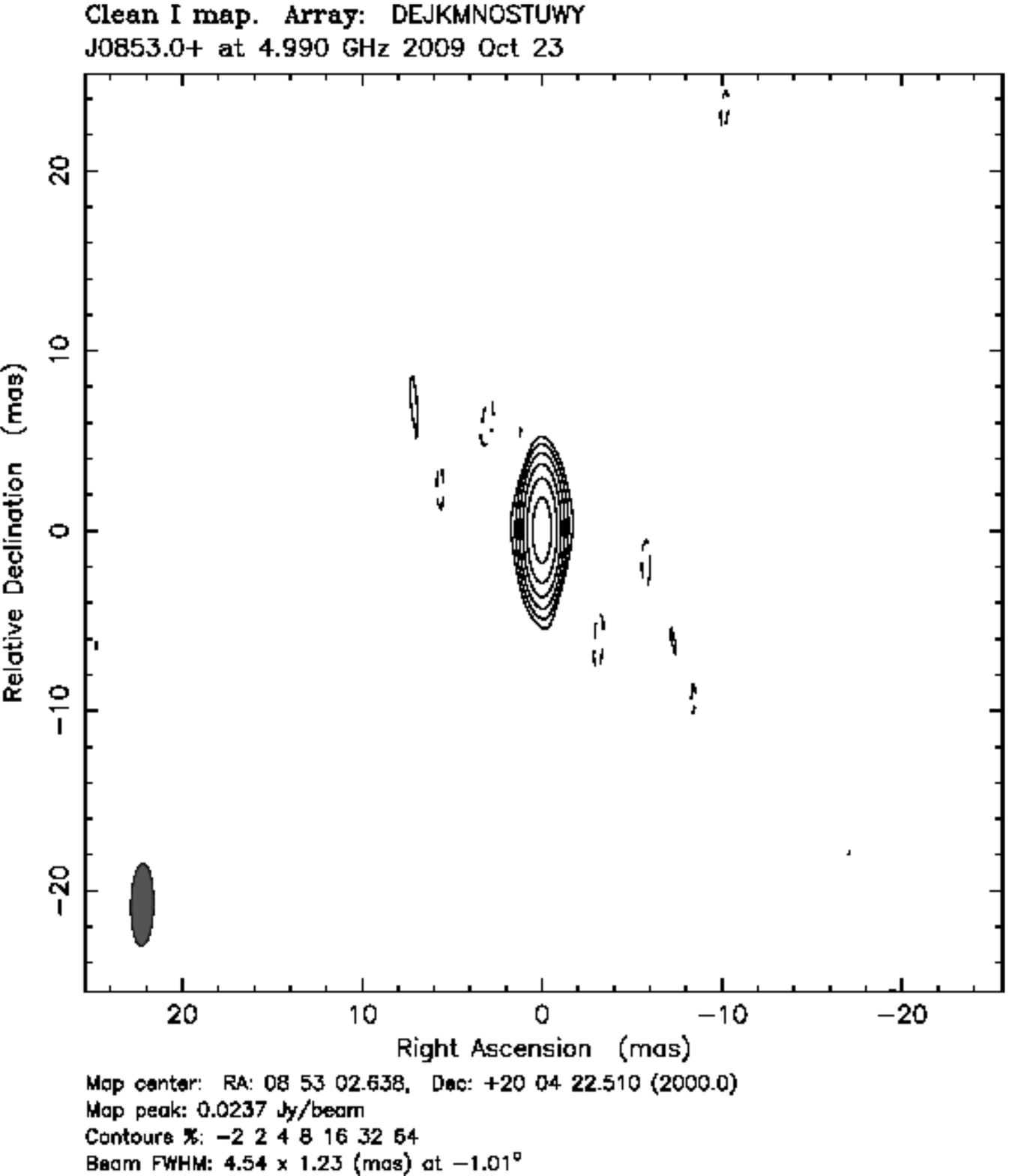}
\includegraphics[width=7cm,clip]{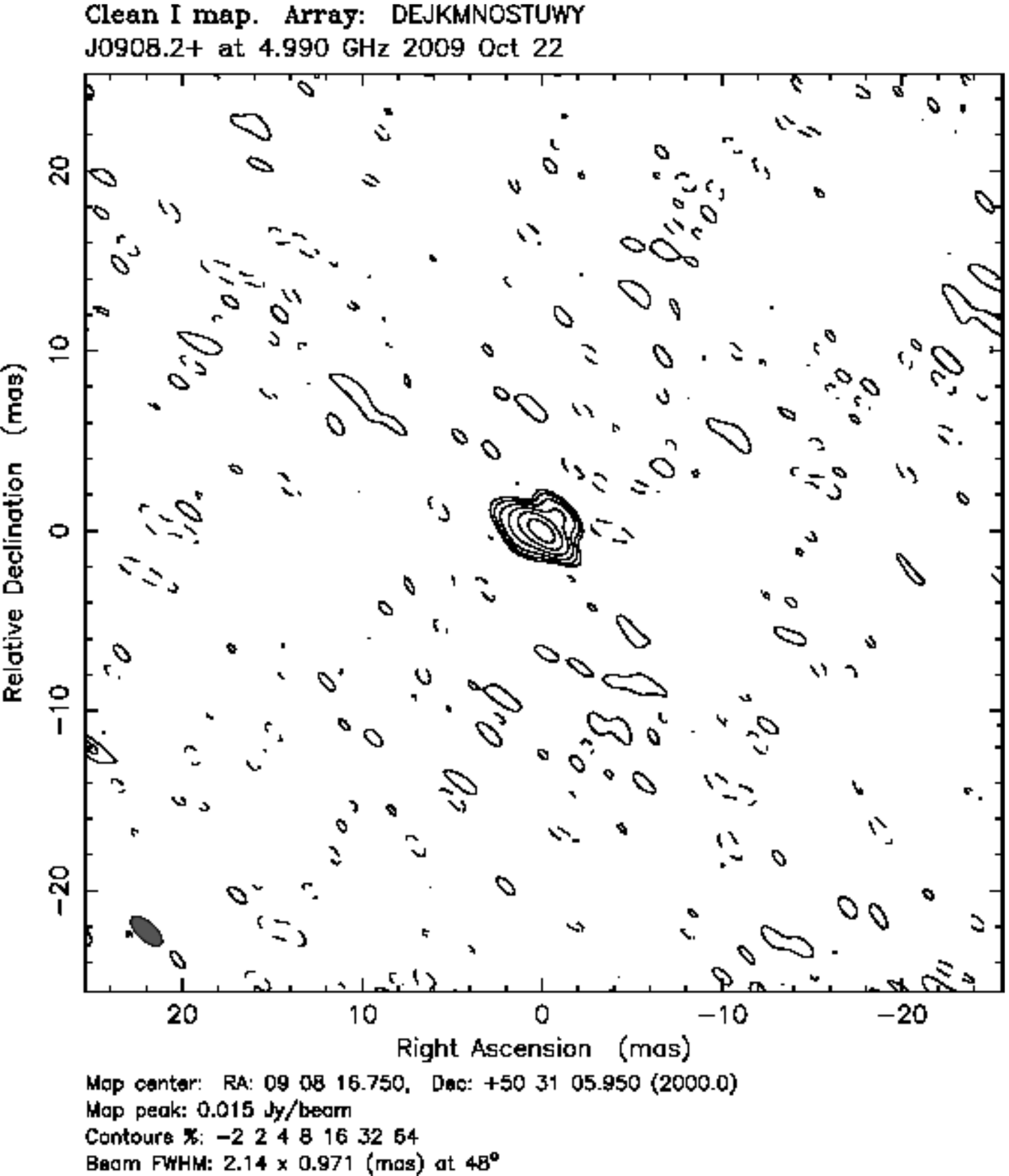} \\
\newline
\includegraphics[width=7cm,clip]{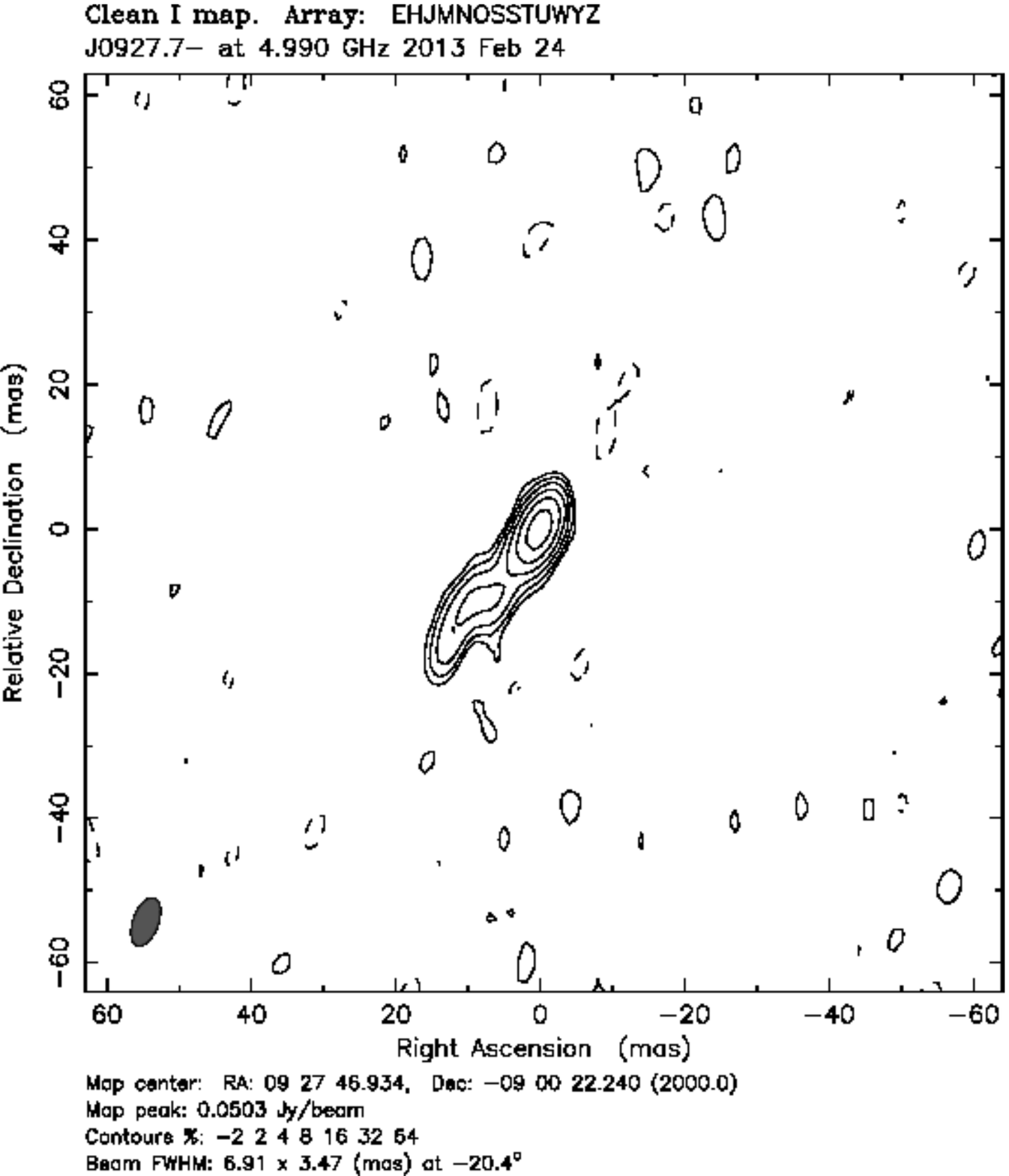}
\includegraphics[width=7cm,clip]{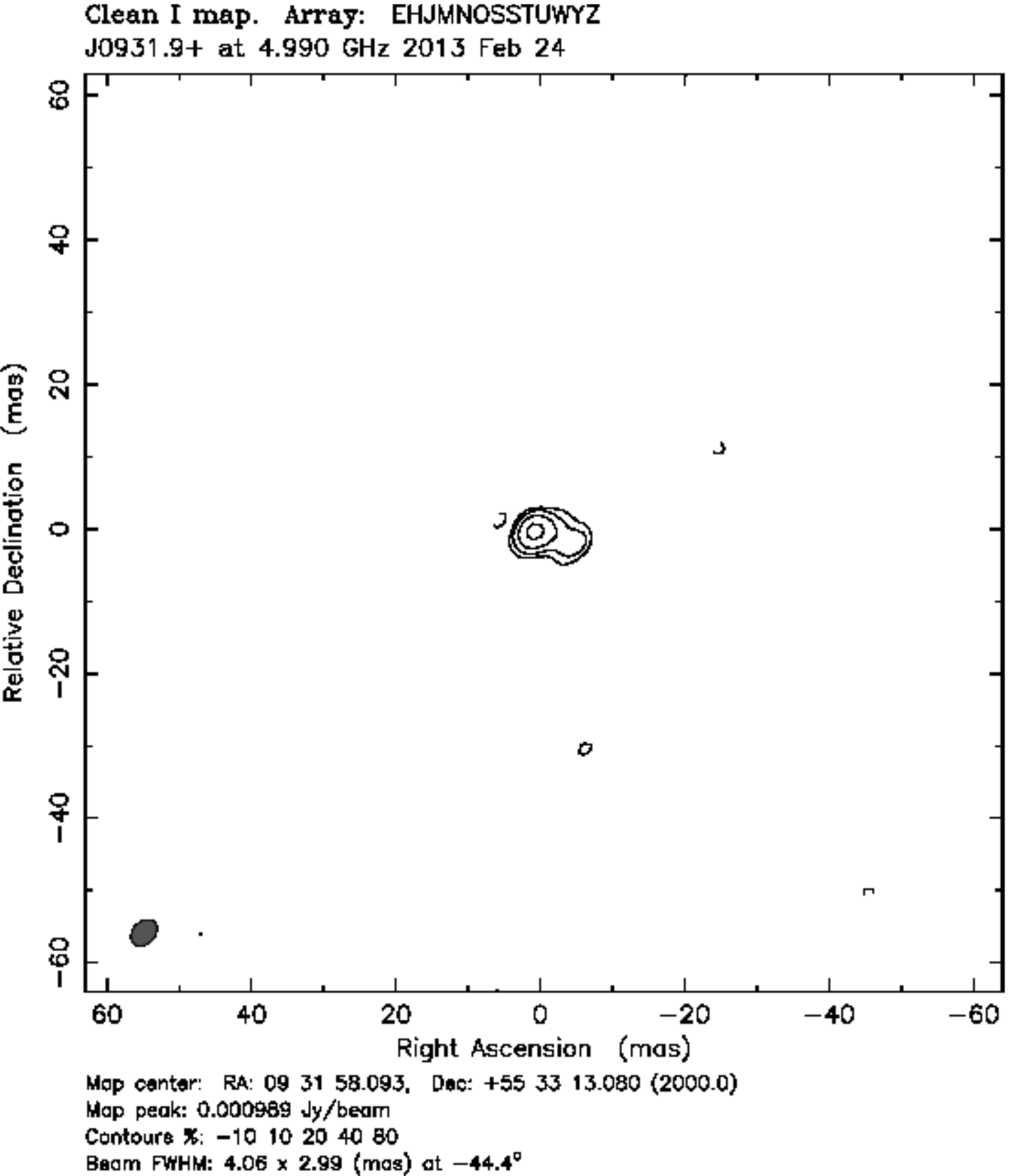} \\
\newline
\includegraphics[width=7cm,clip]{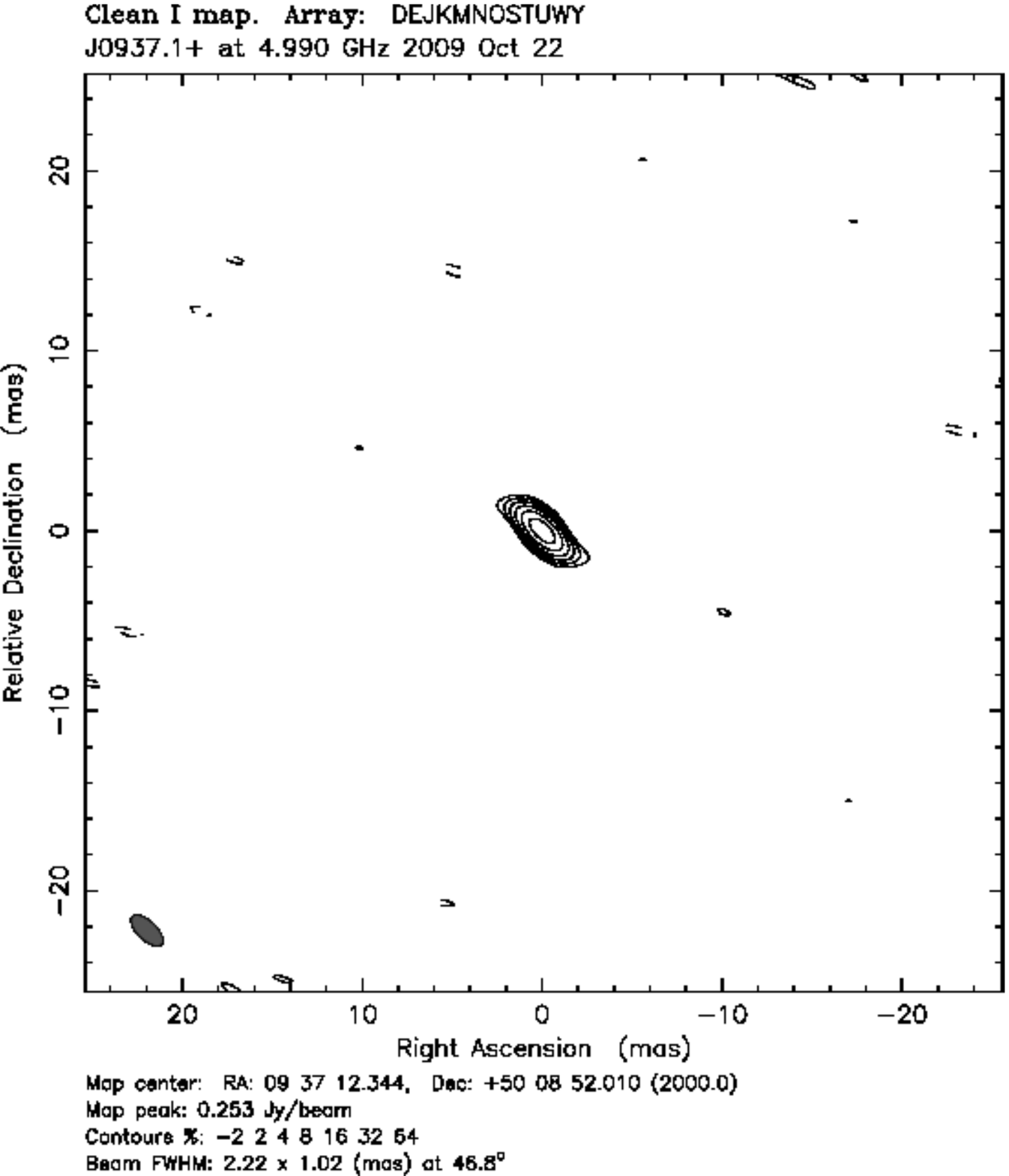}
\includegraphics[width=7cm,clip]{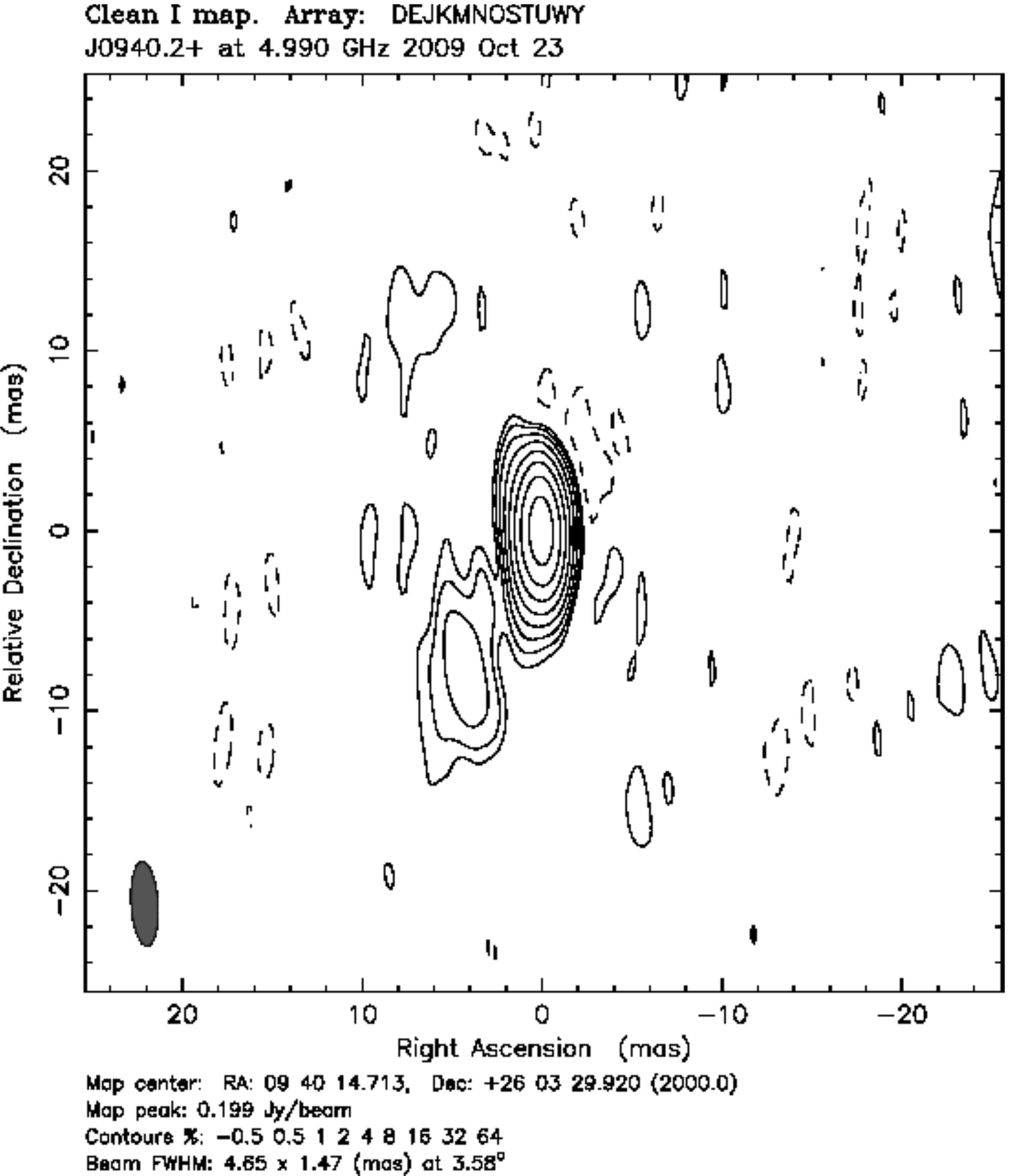}
\caption{EVN images at 5\,GHz. Continued}\label{appfig}
\end{figure*}
\begin{figure*}
\includegraphics[width=7cm,clip]{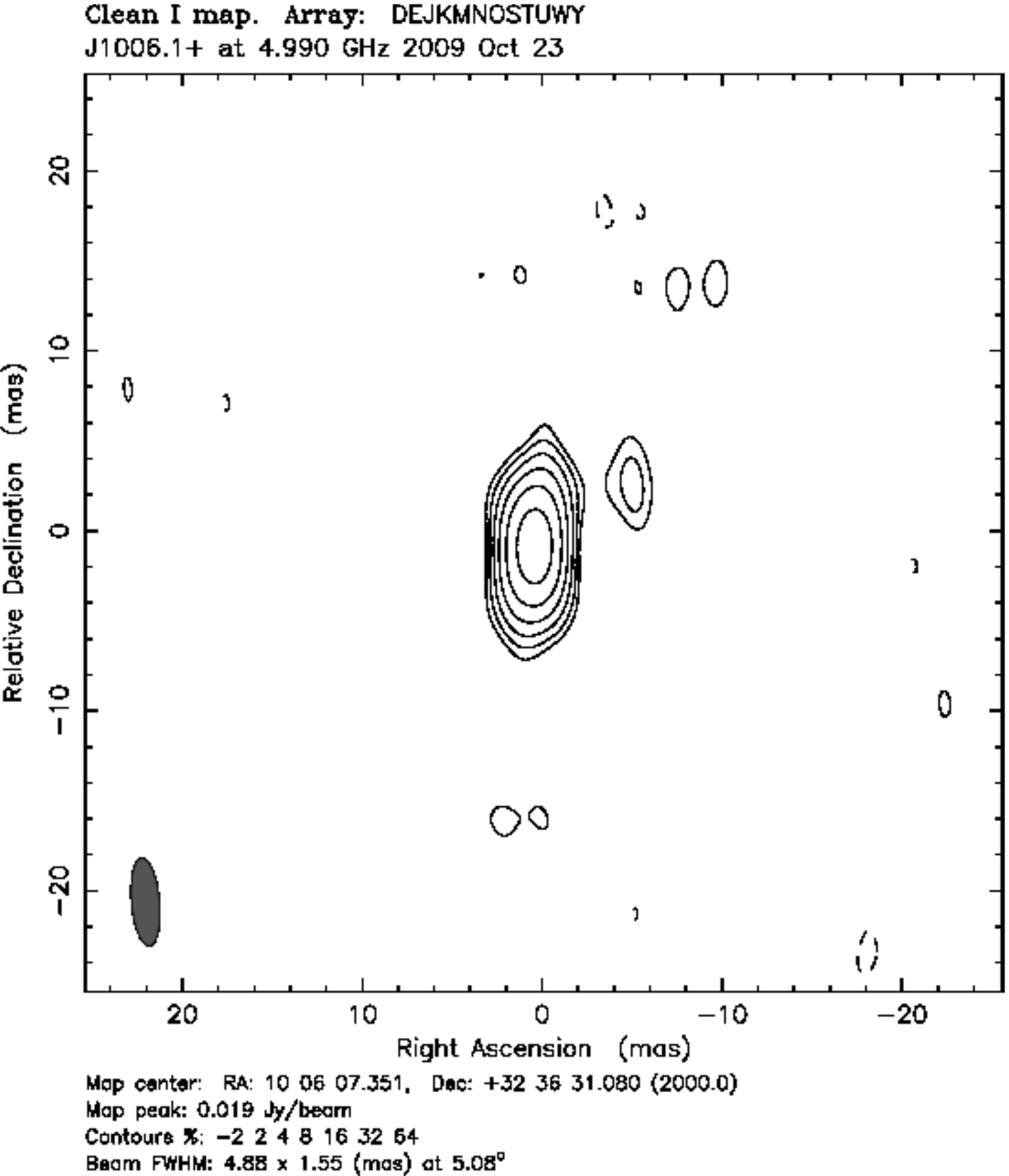}
\includegraphics[width=7cm,clip]{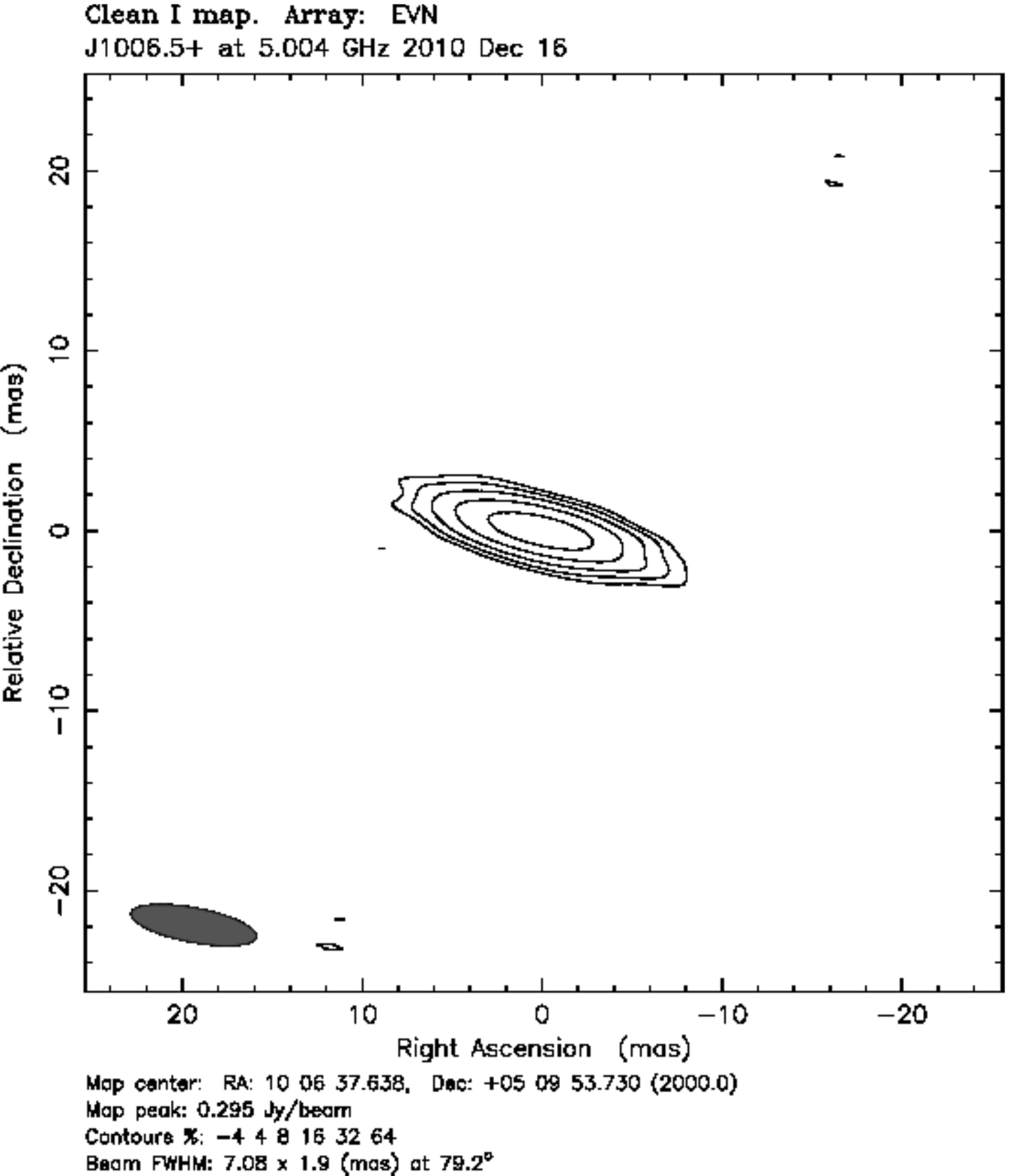} \\
\newline
\includegraphics[width=7cm,clip]{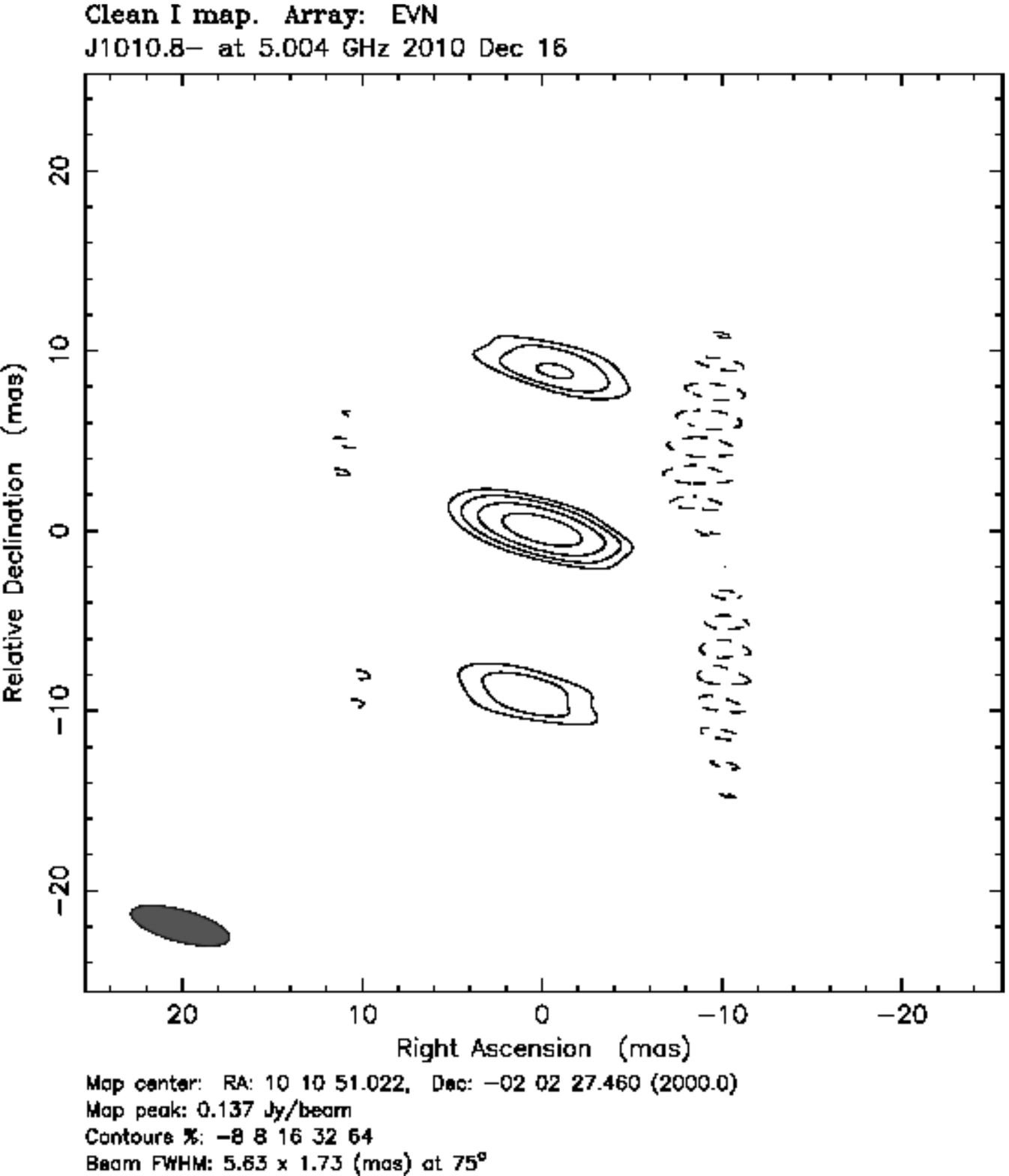}
\includegraphics[width=7cm,clip]{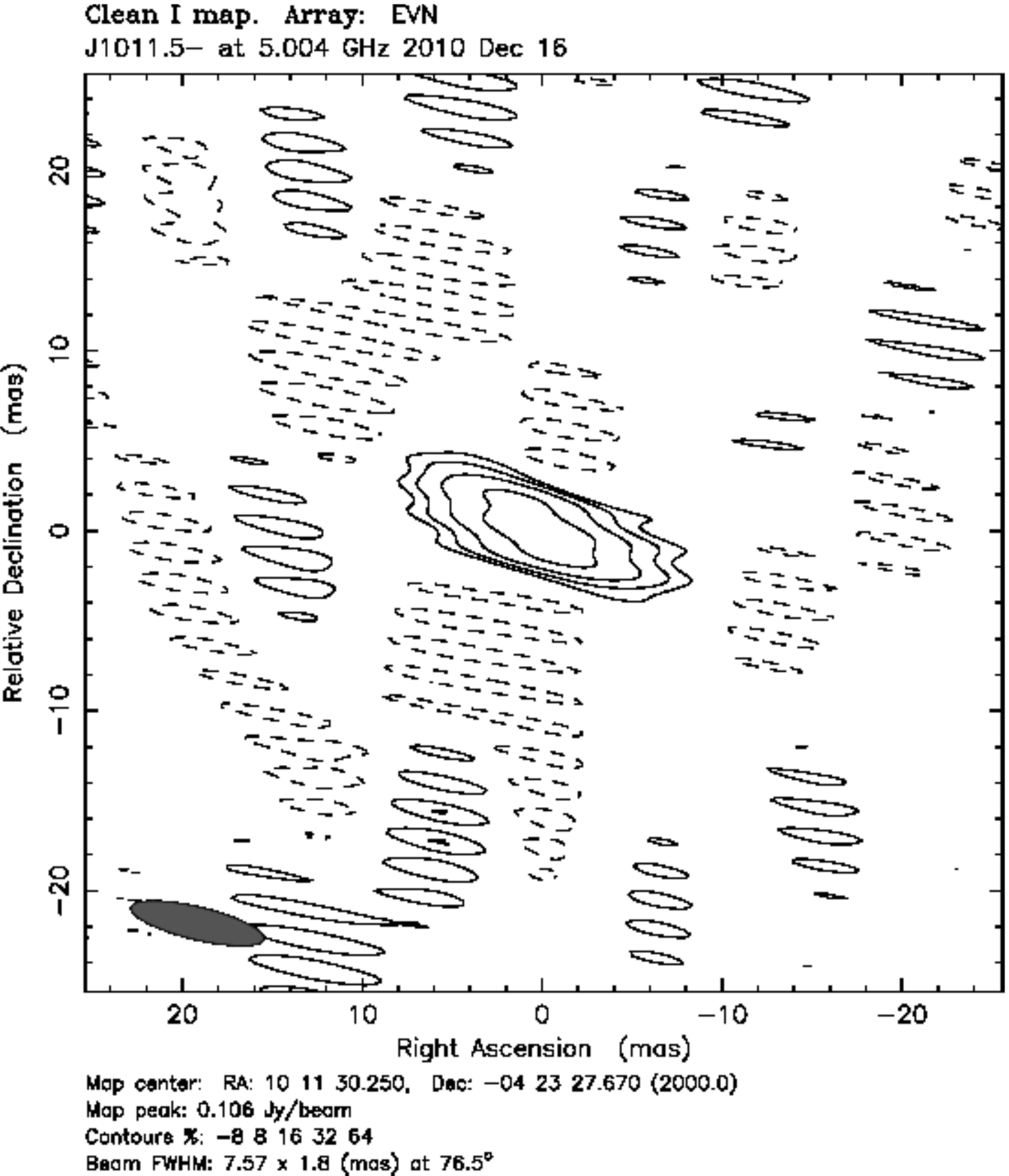} \\
\newline
\includegraphics[width=7cm,clip]{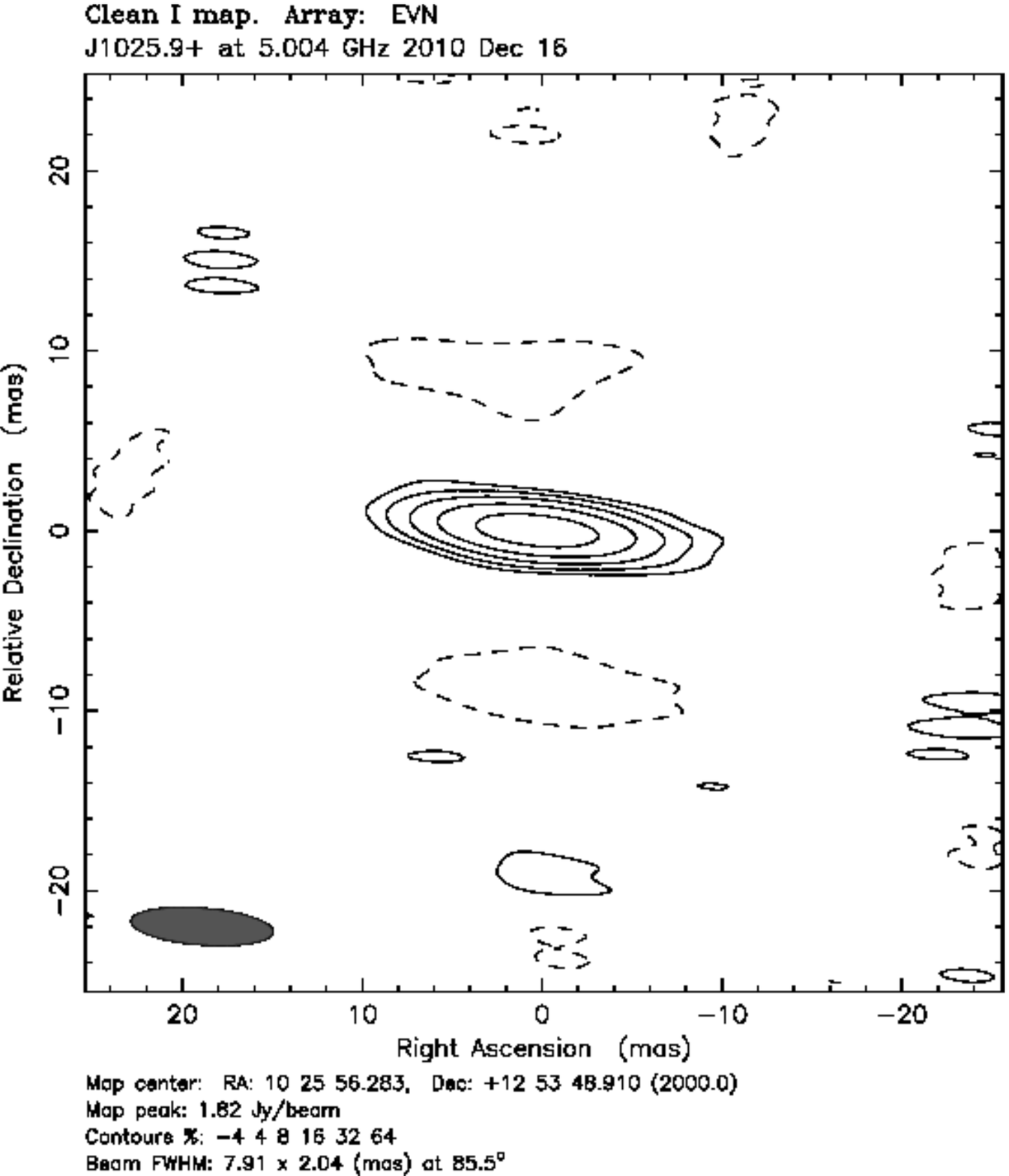}
\includegraphics[width=7cm,clip]{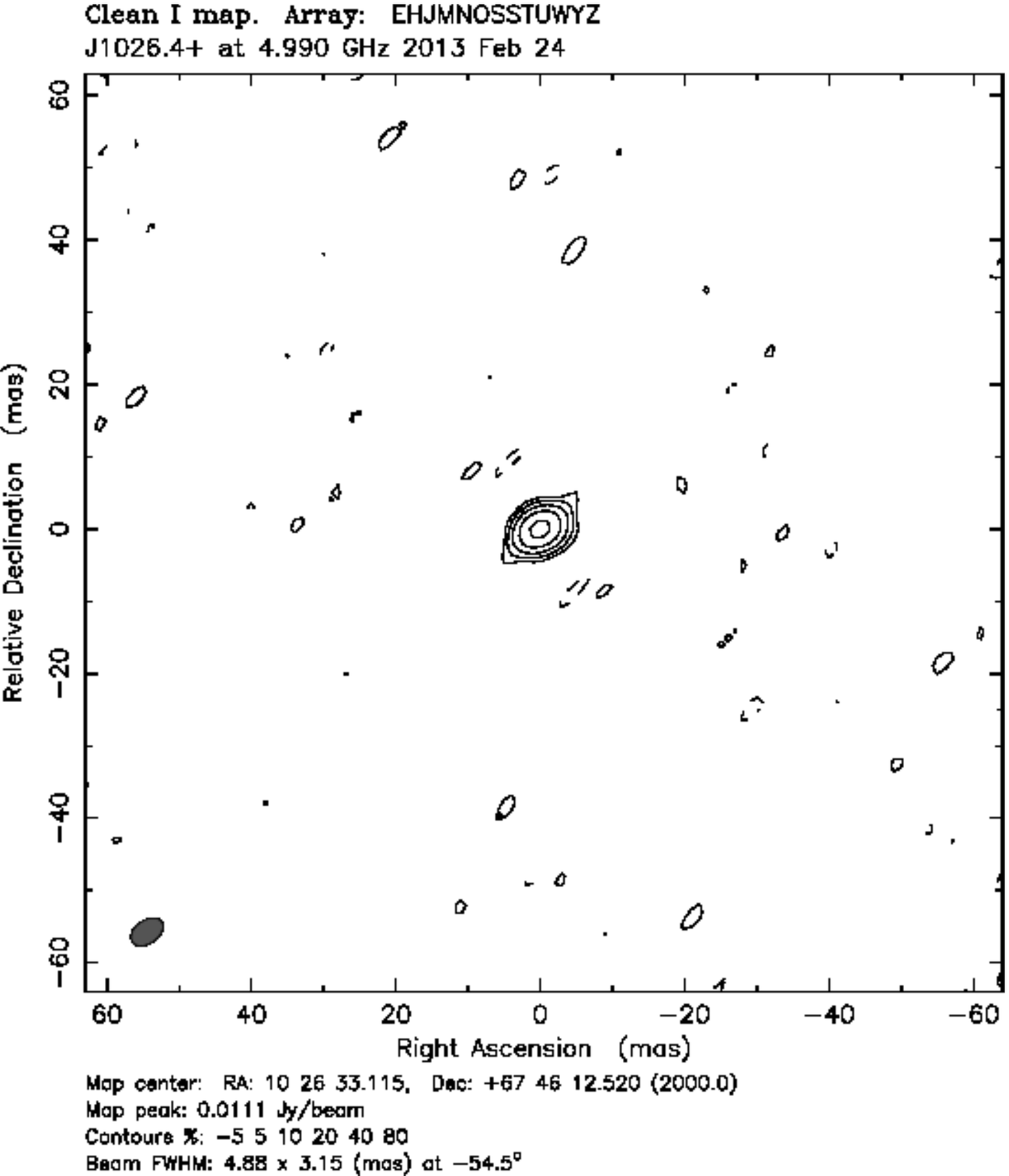}
\caption{EVN images at 5\,GHz.}\label{appfig}
\end{figure*}
\begin{figure*}
\includegraphics[width=7cm,clip]{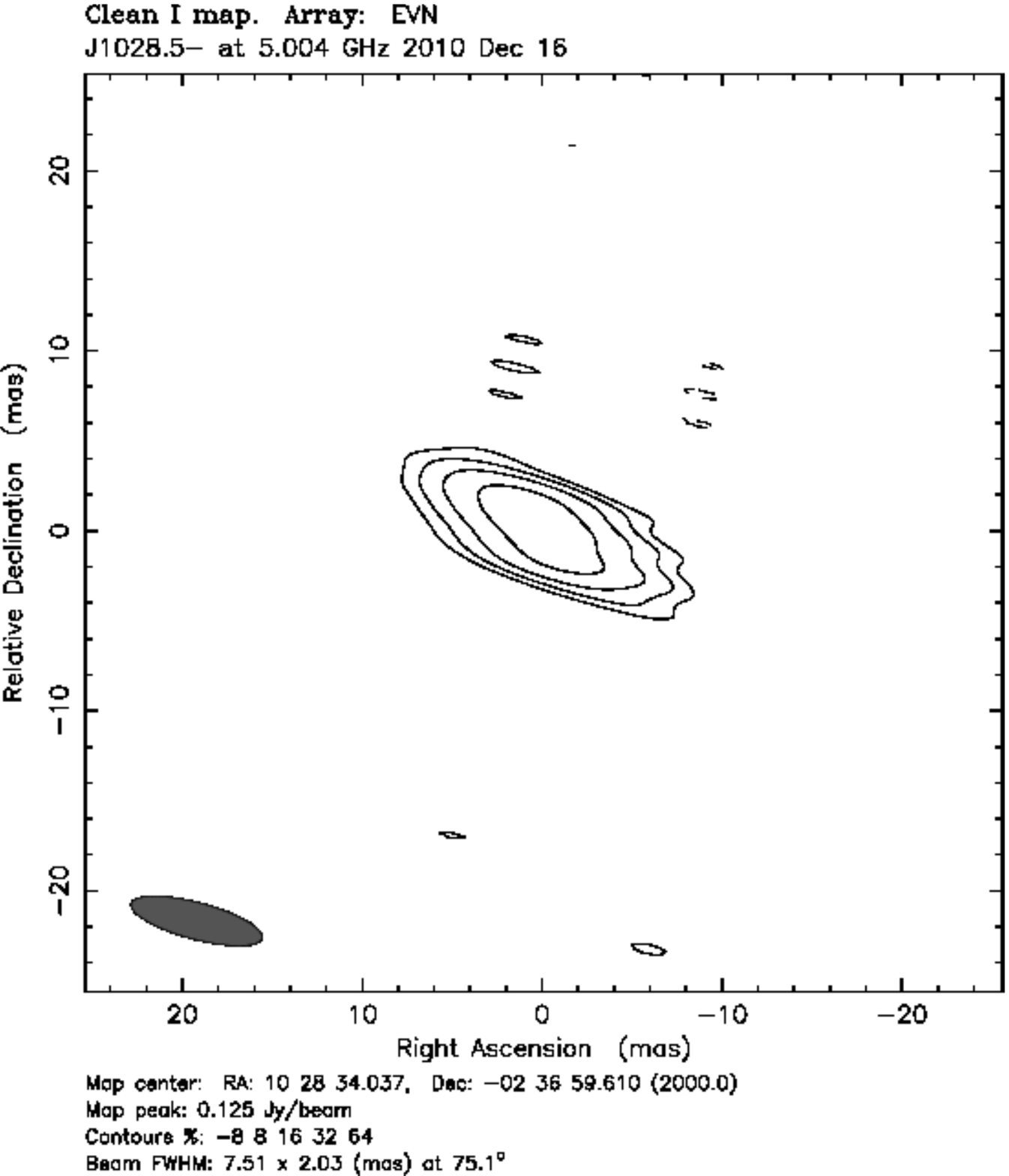}
\includegraphics[width=7cm,clip]{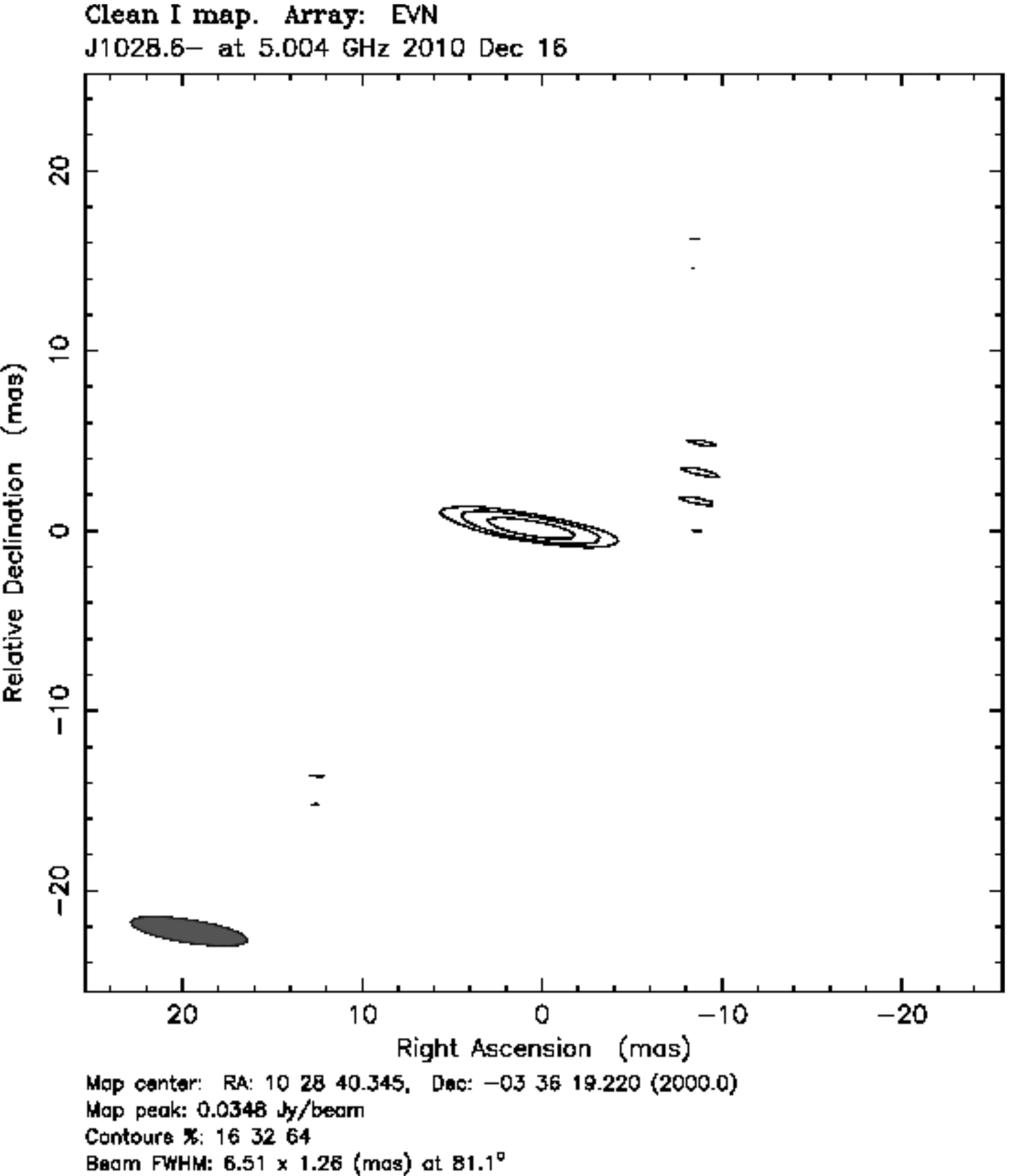} \\
\newline
\includegraphics[width=7cm,clip]{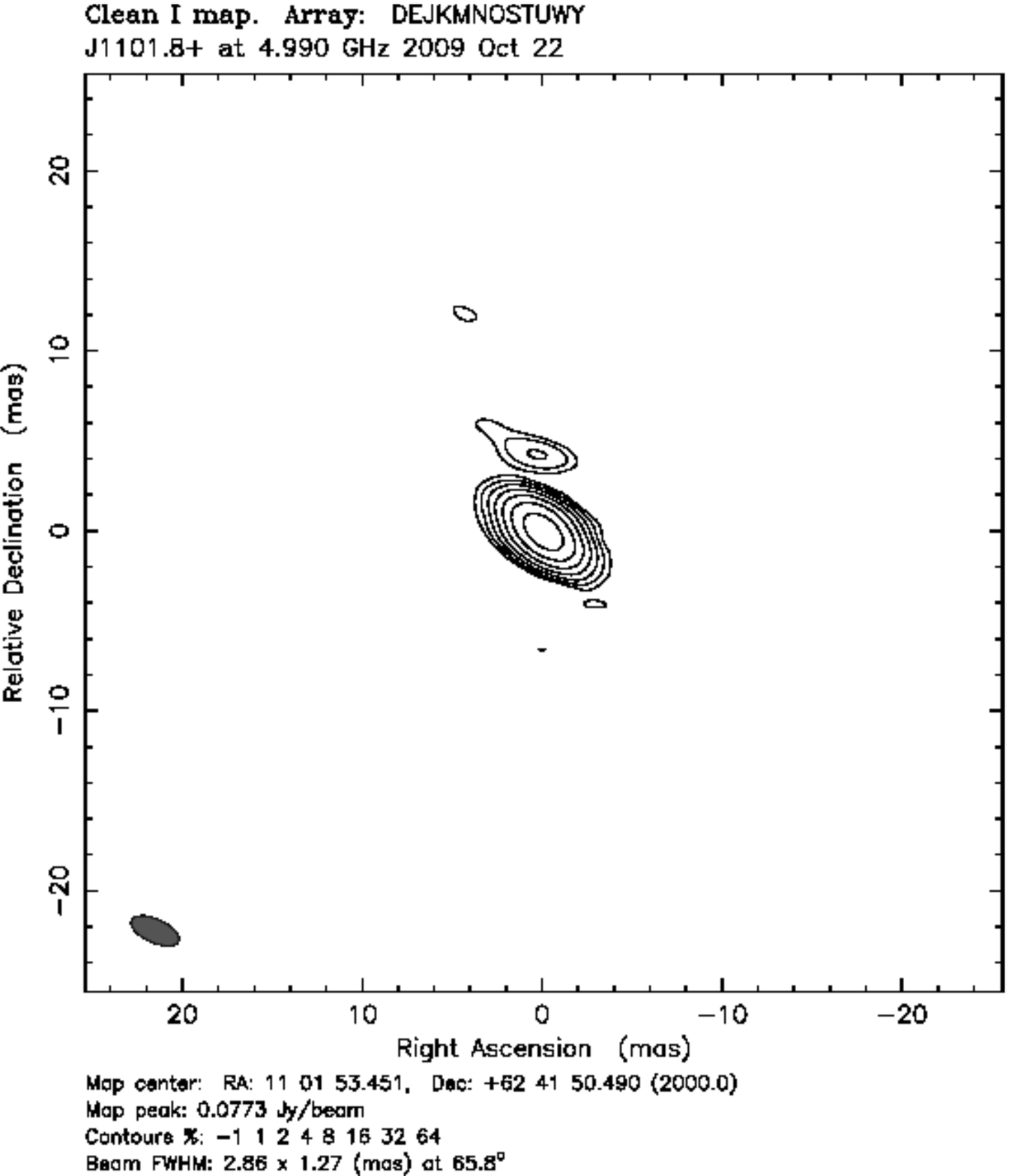}
\includegraphics[width=7cm,clip]{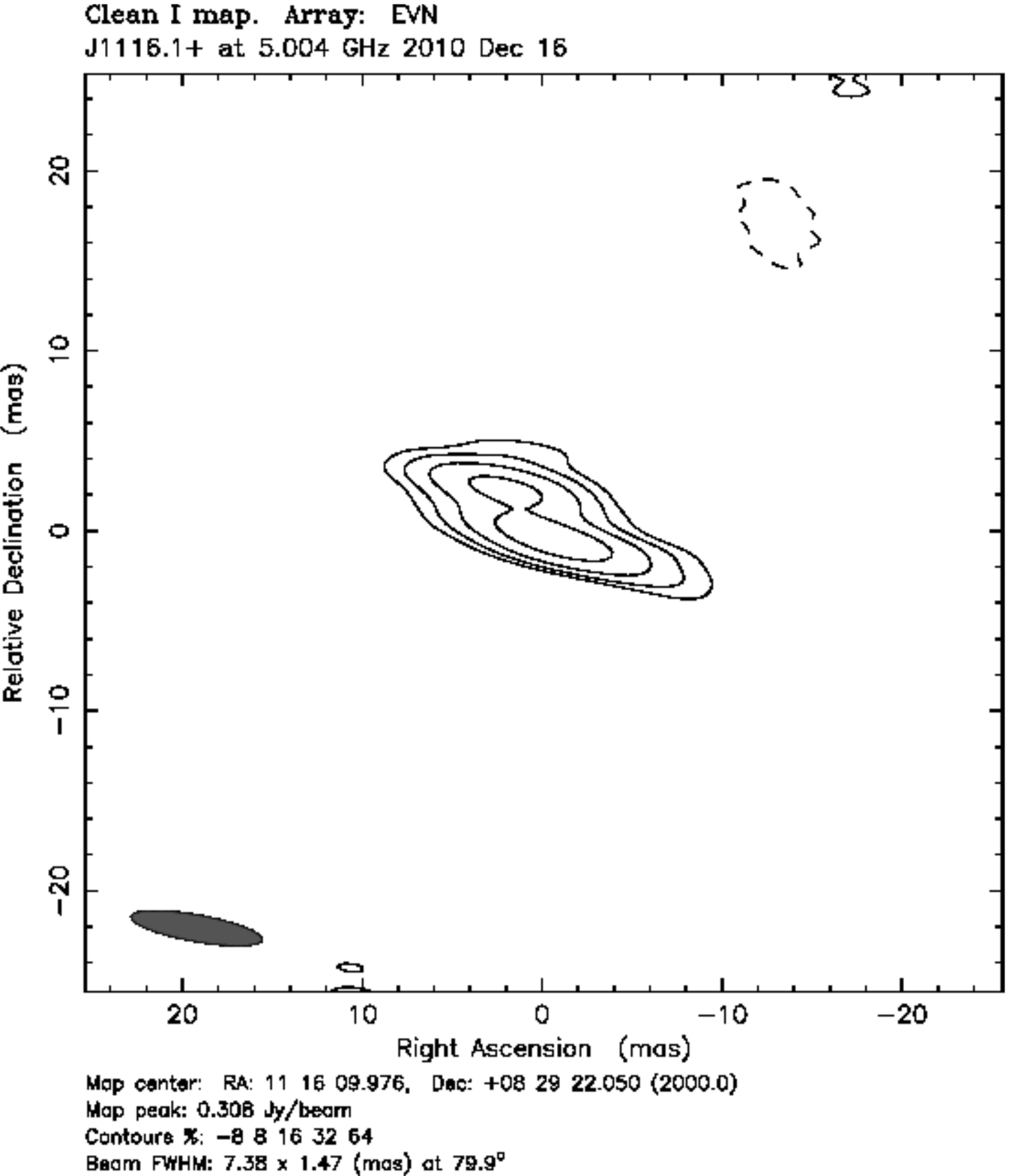} \\
\newline
\includegraphics[width=7cm,clip]{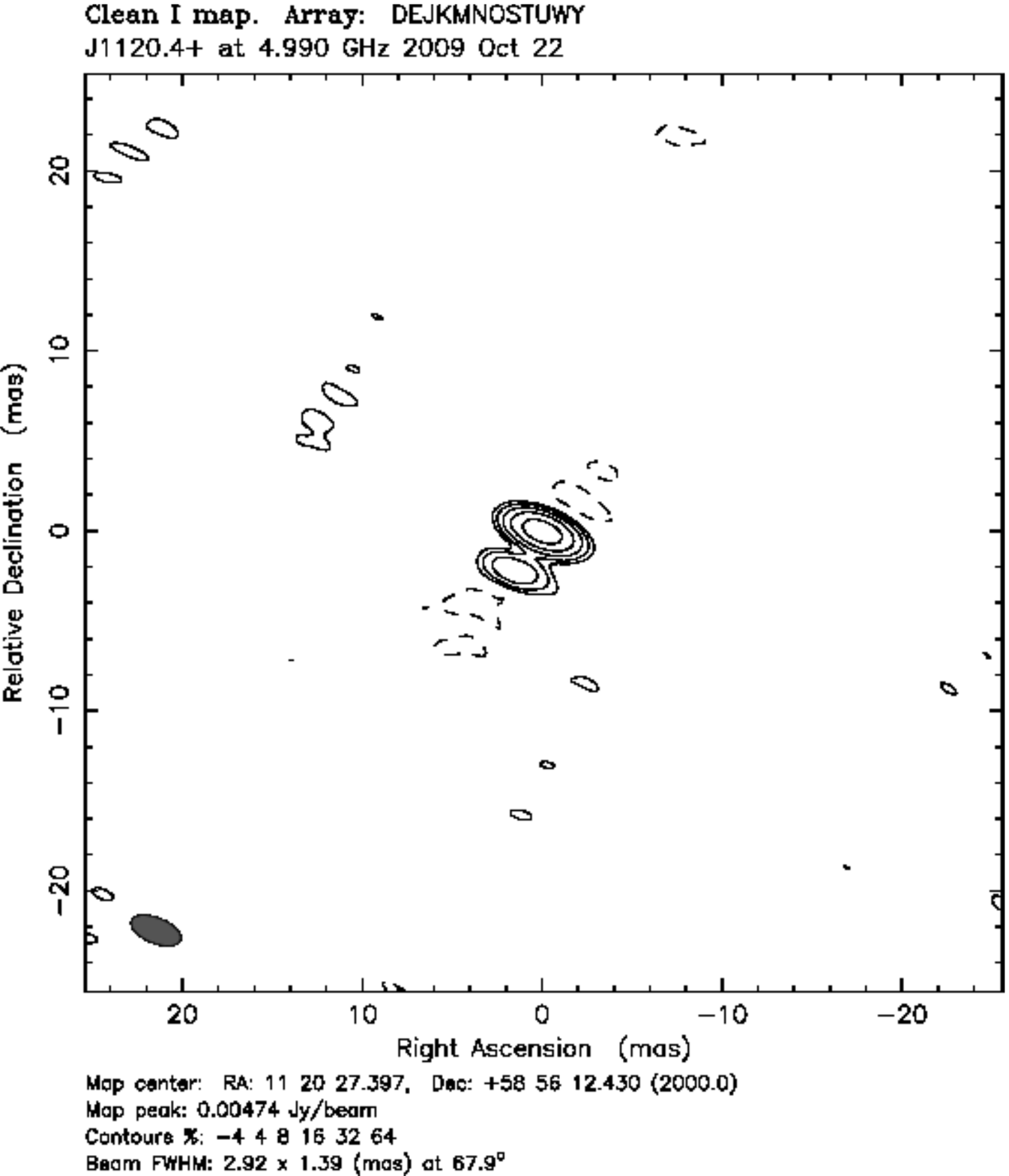}
\includegraphics[width=7cm,clip]{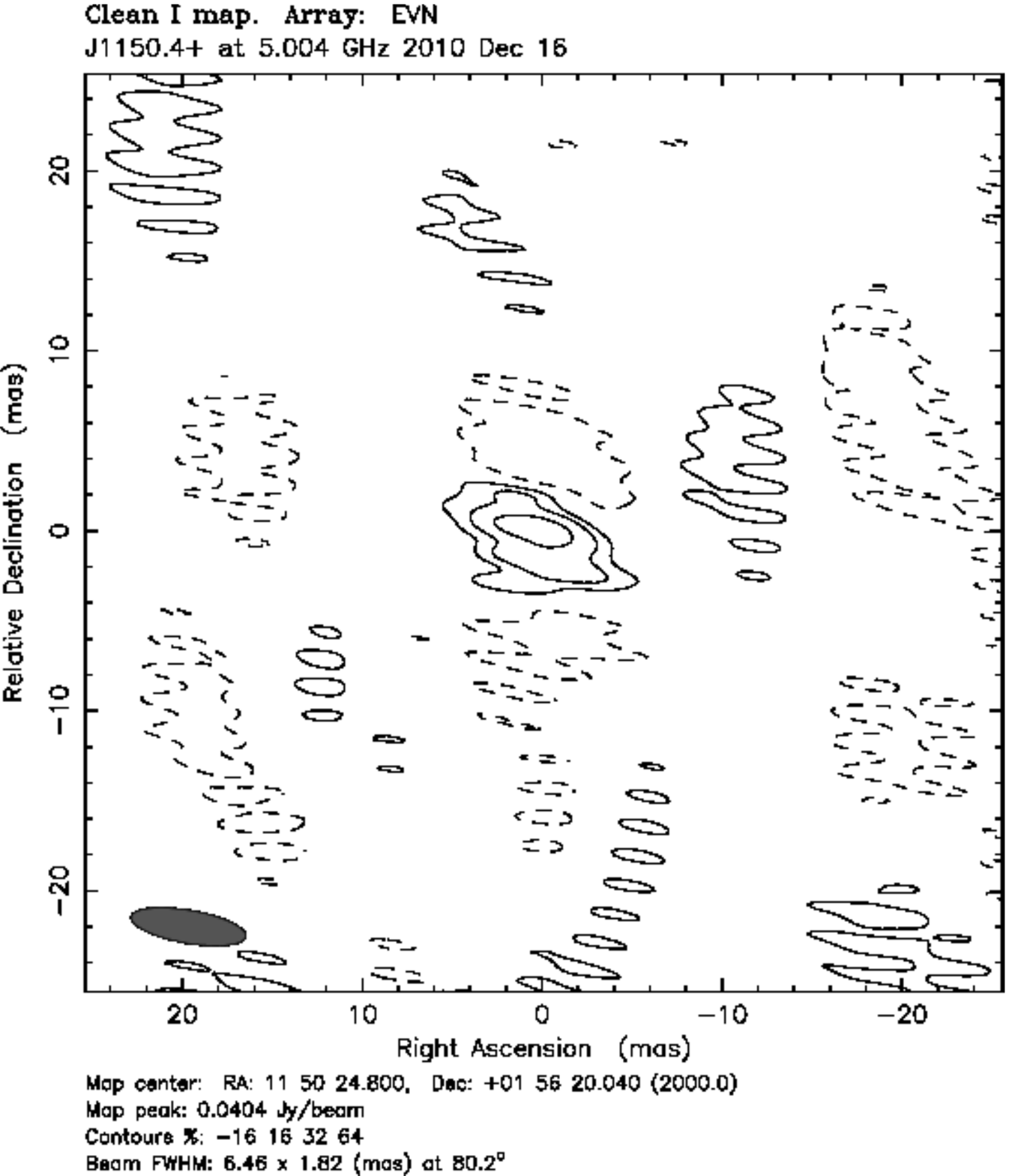}
\caption{EVN images at 5\,GHz. Continued}\label{appfig}
\end{figure*}
\begin{figure*}
\includegraphics[width=7cm,clip]{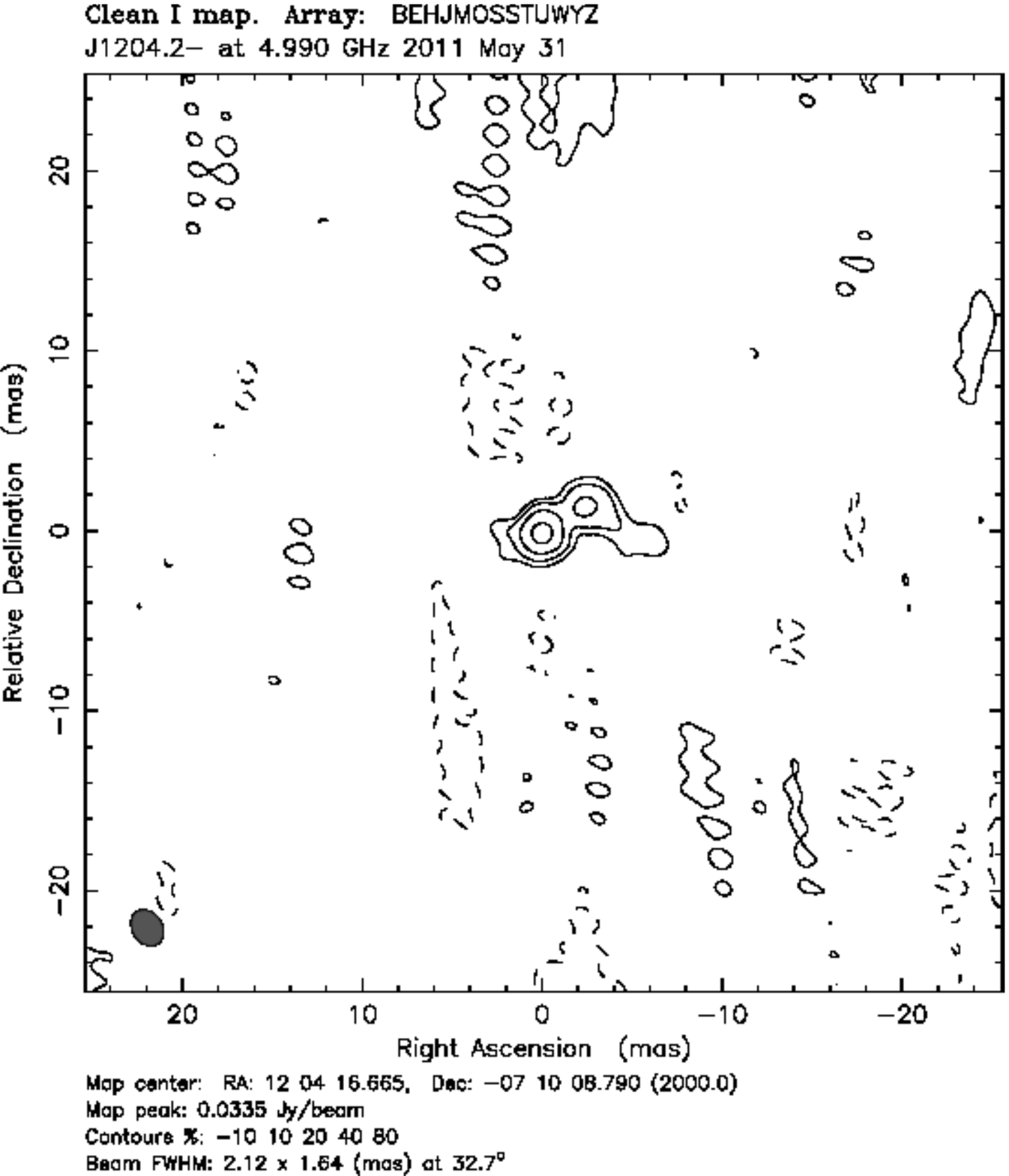}
\includegraphics[width=7cm,clip]{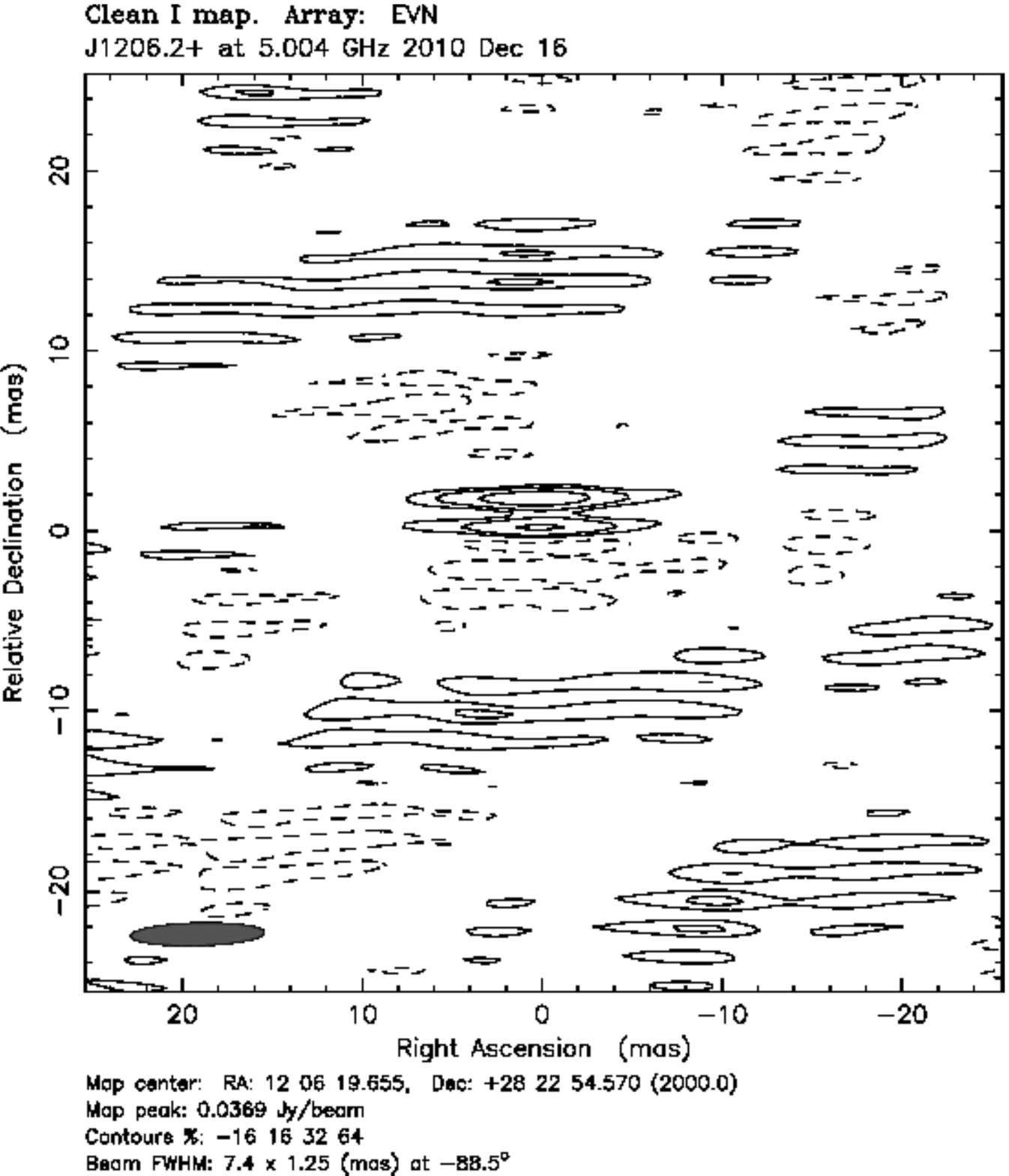} \\
\newline
\includegraphics[width=7cm,clip]{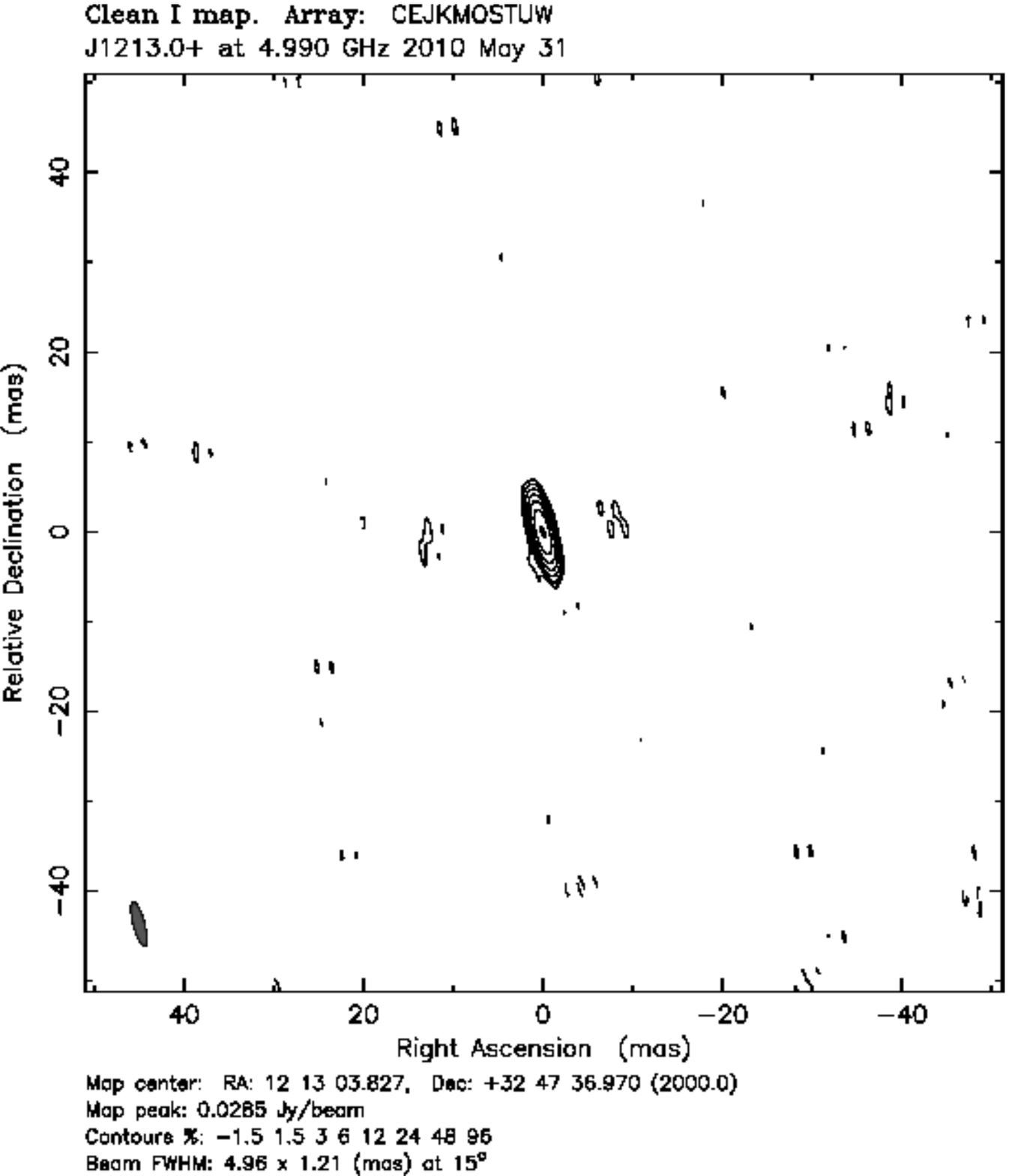}
\includegraphics[width=7cm,clip]{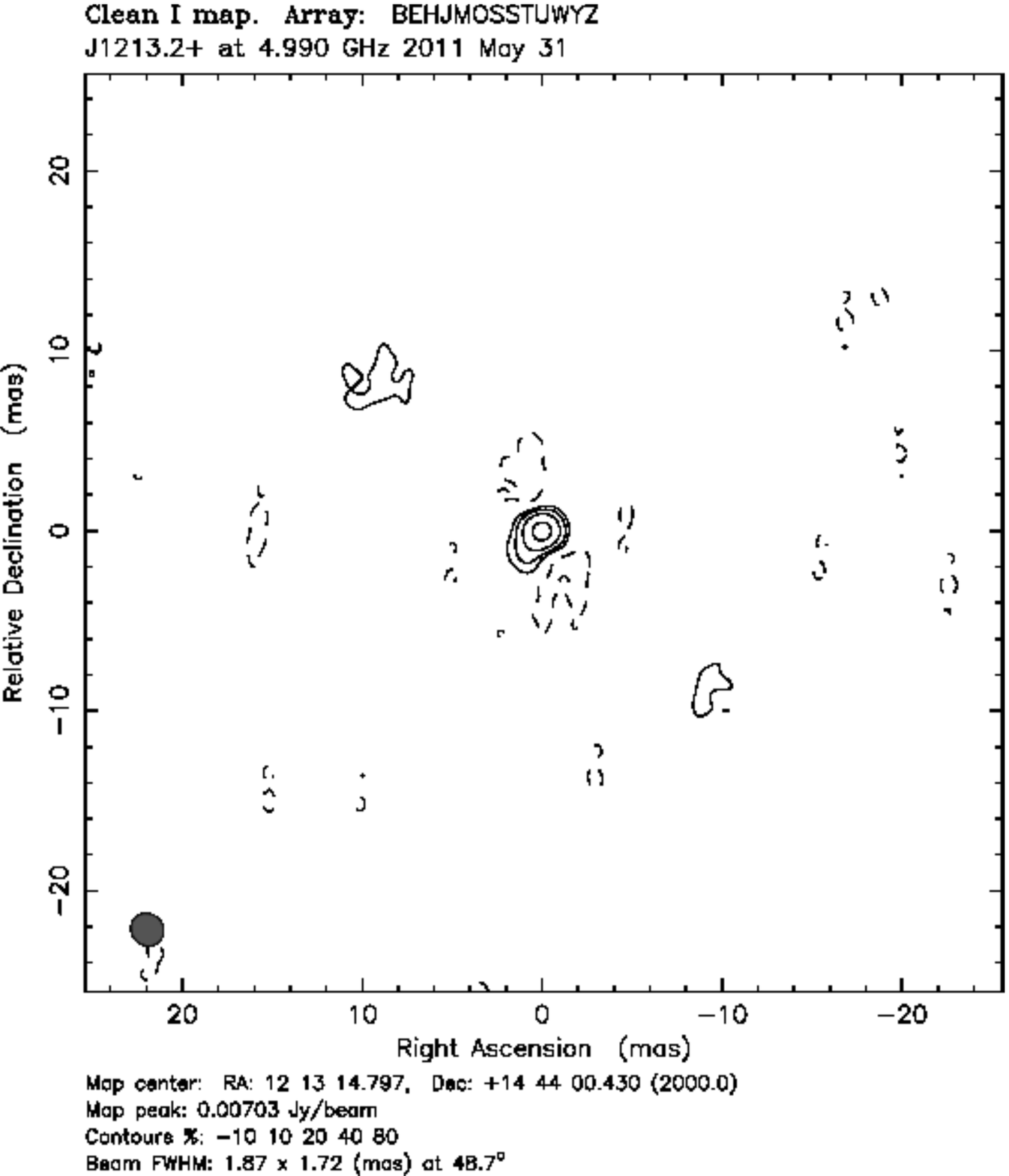} \\
\newline
\includegraphics[width=7cm,clip]{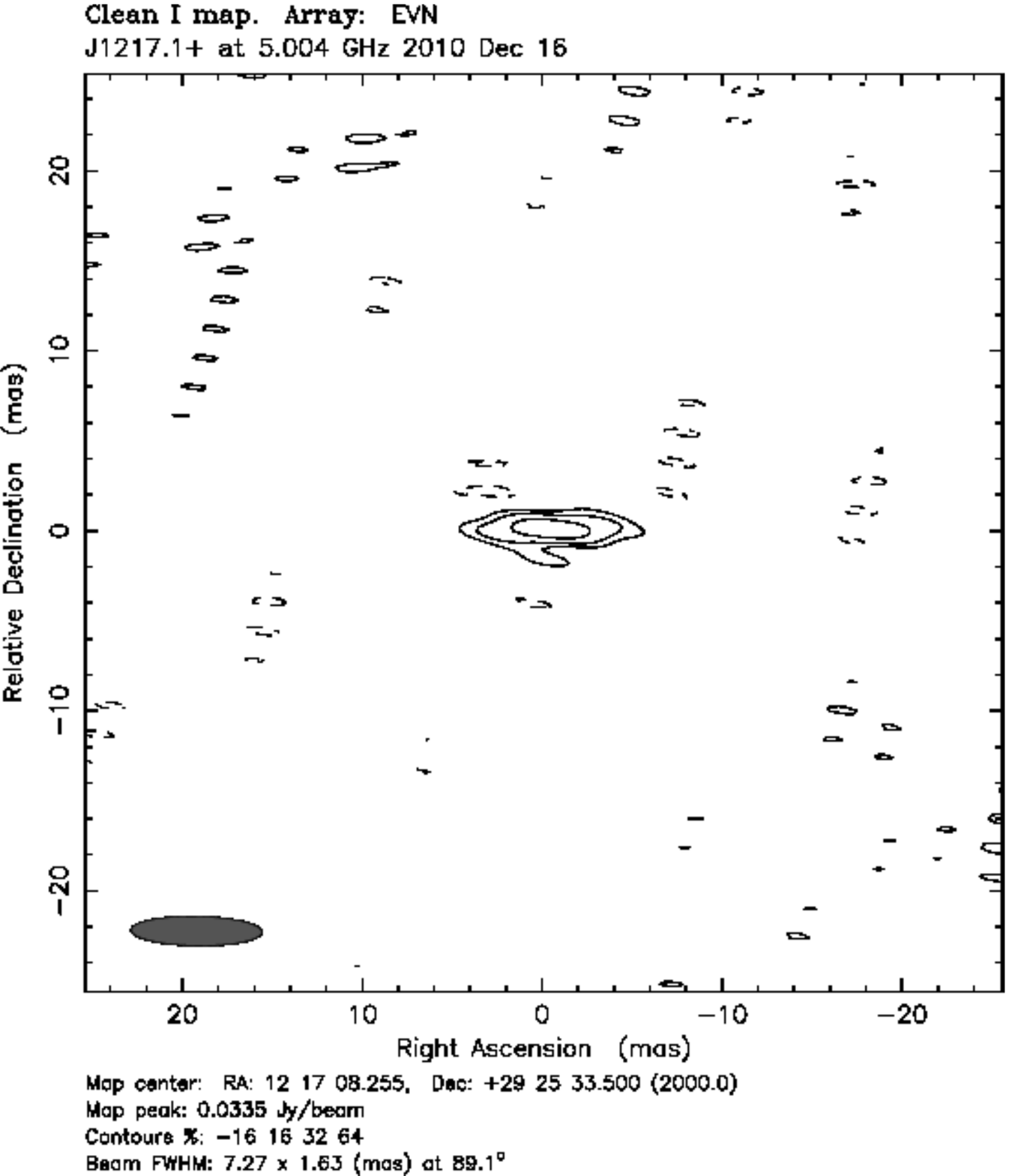}
\includegraphics[width=7cm,clip]{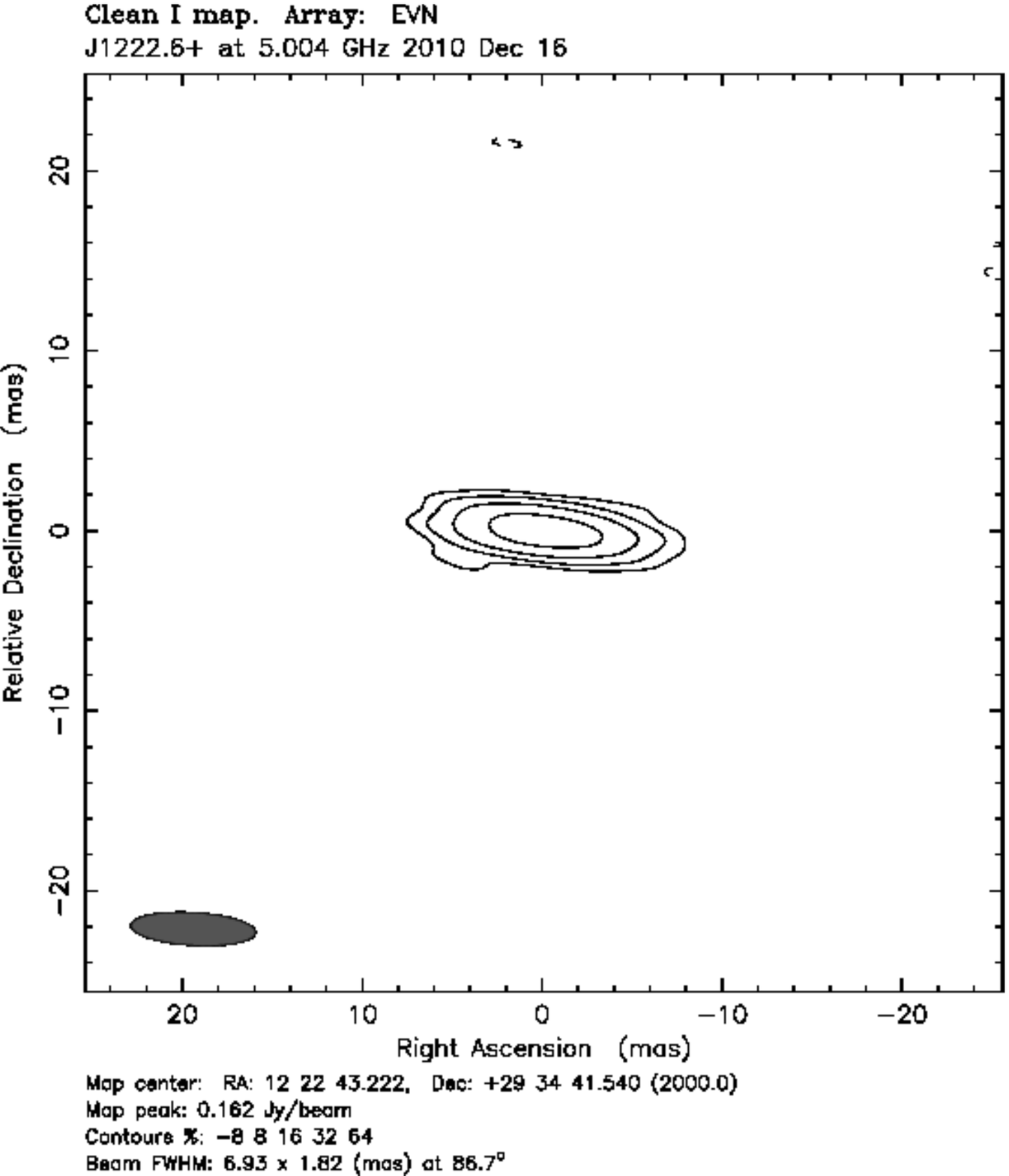}
\caption{EVN images at 5\,GHz.}\label{appfig}
\end{figure*}
\begin{figure*}
\includegraphics[width=7cm,clip]{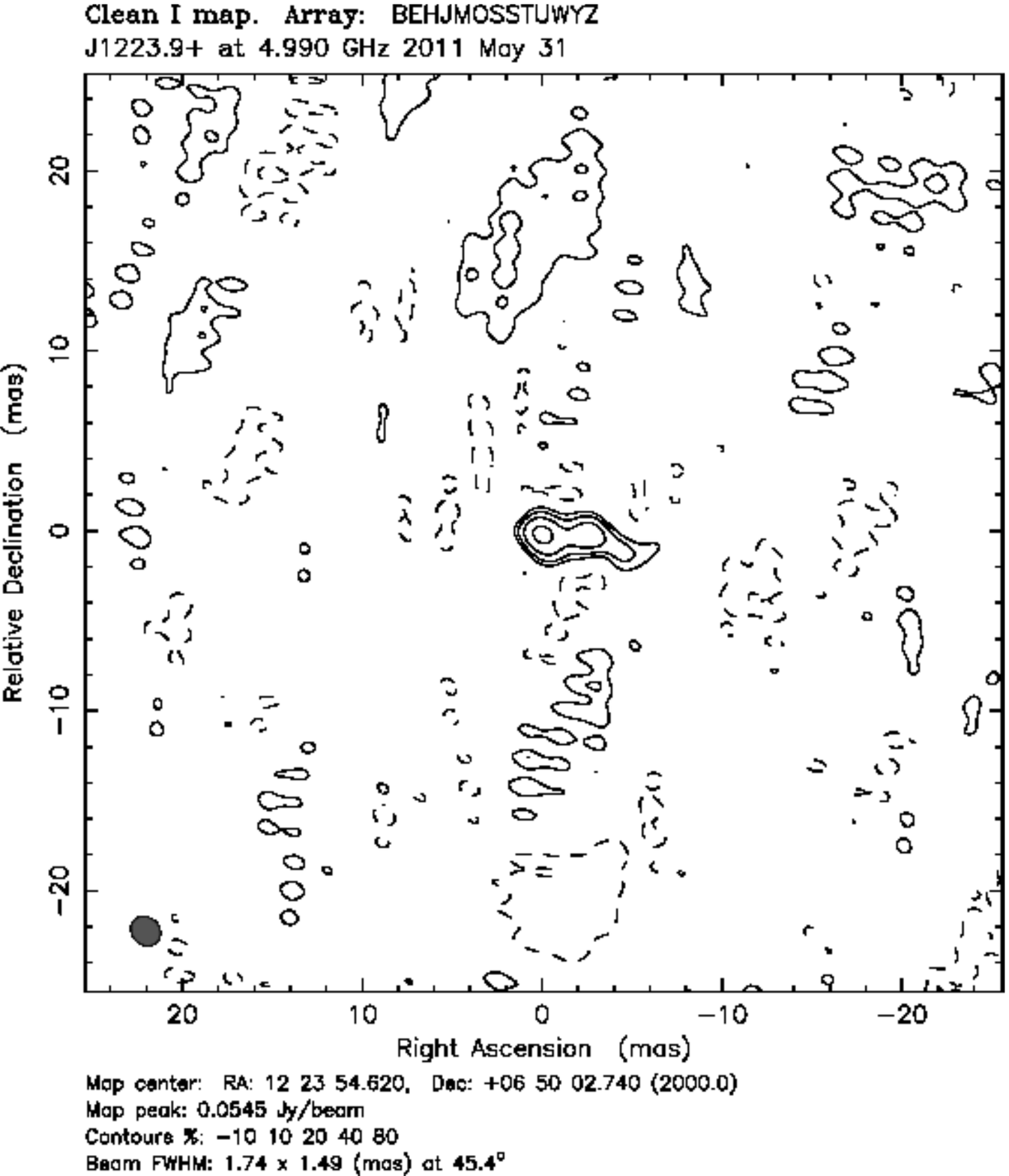}
\includegraphics[width=7cm,clip]{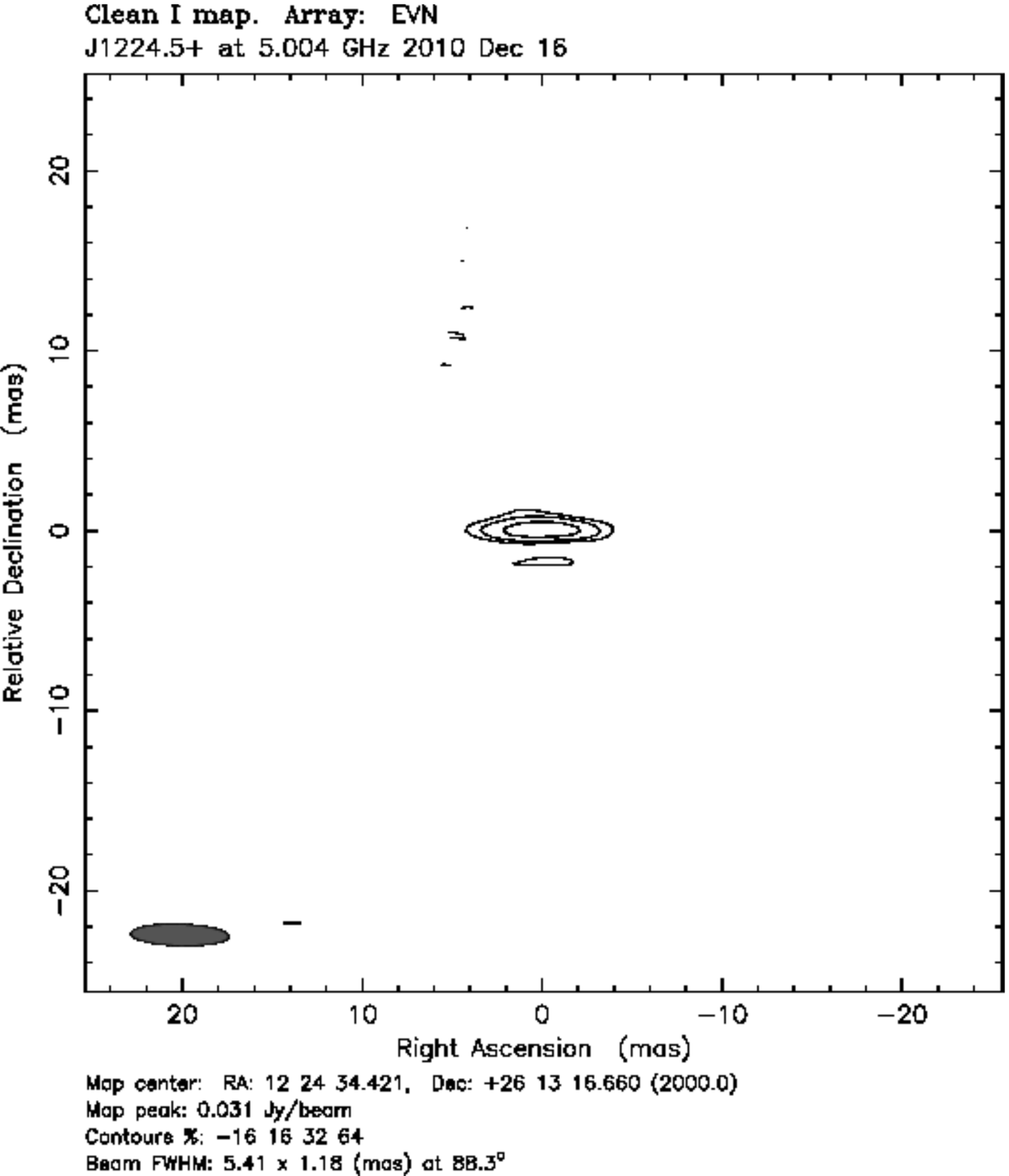} \\
\newline
\includegraphics[width=7cm,clip]{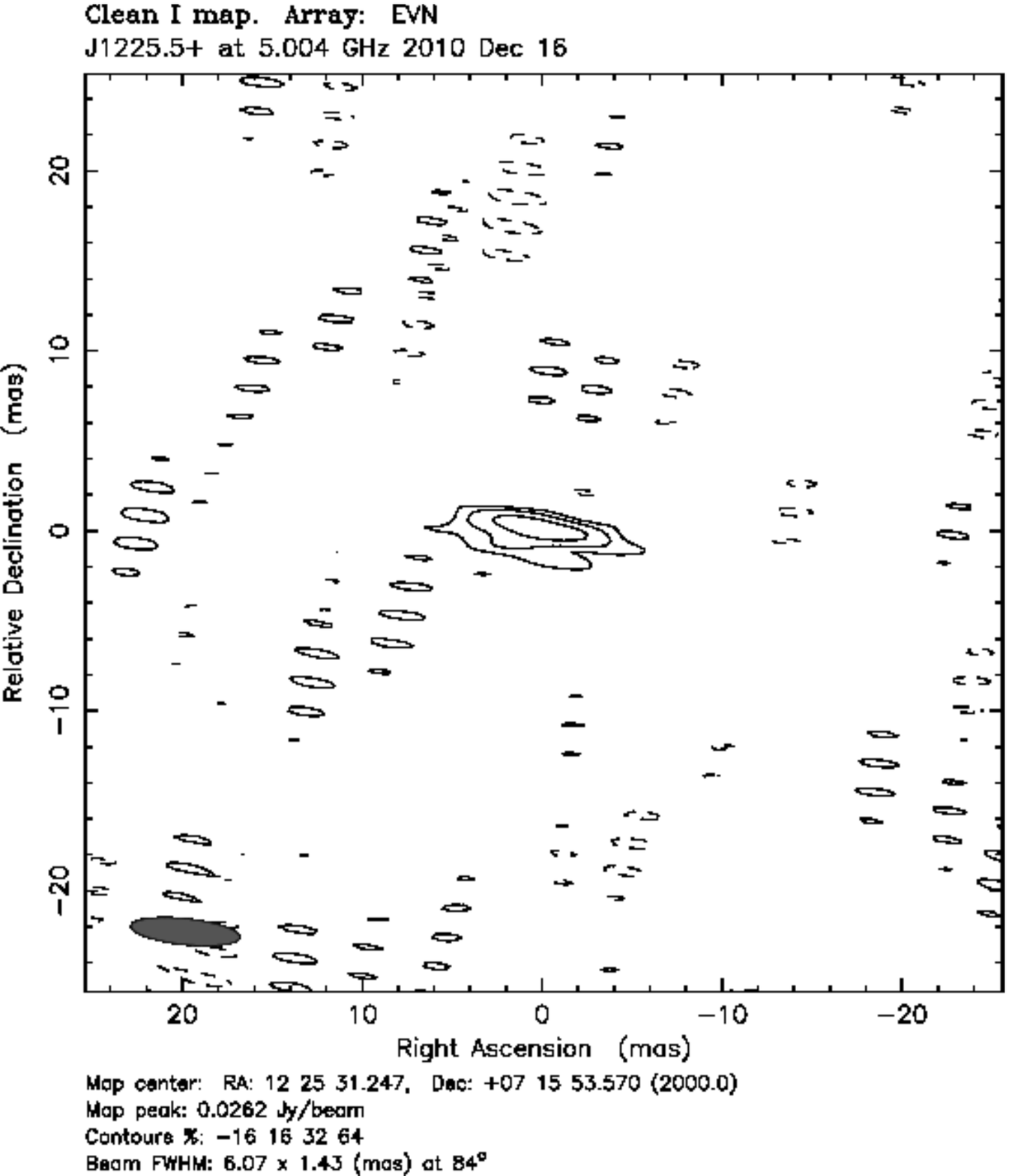}
\includegraphics[width=7cm,clip]{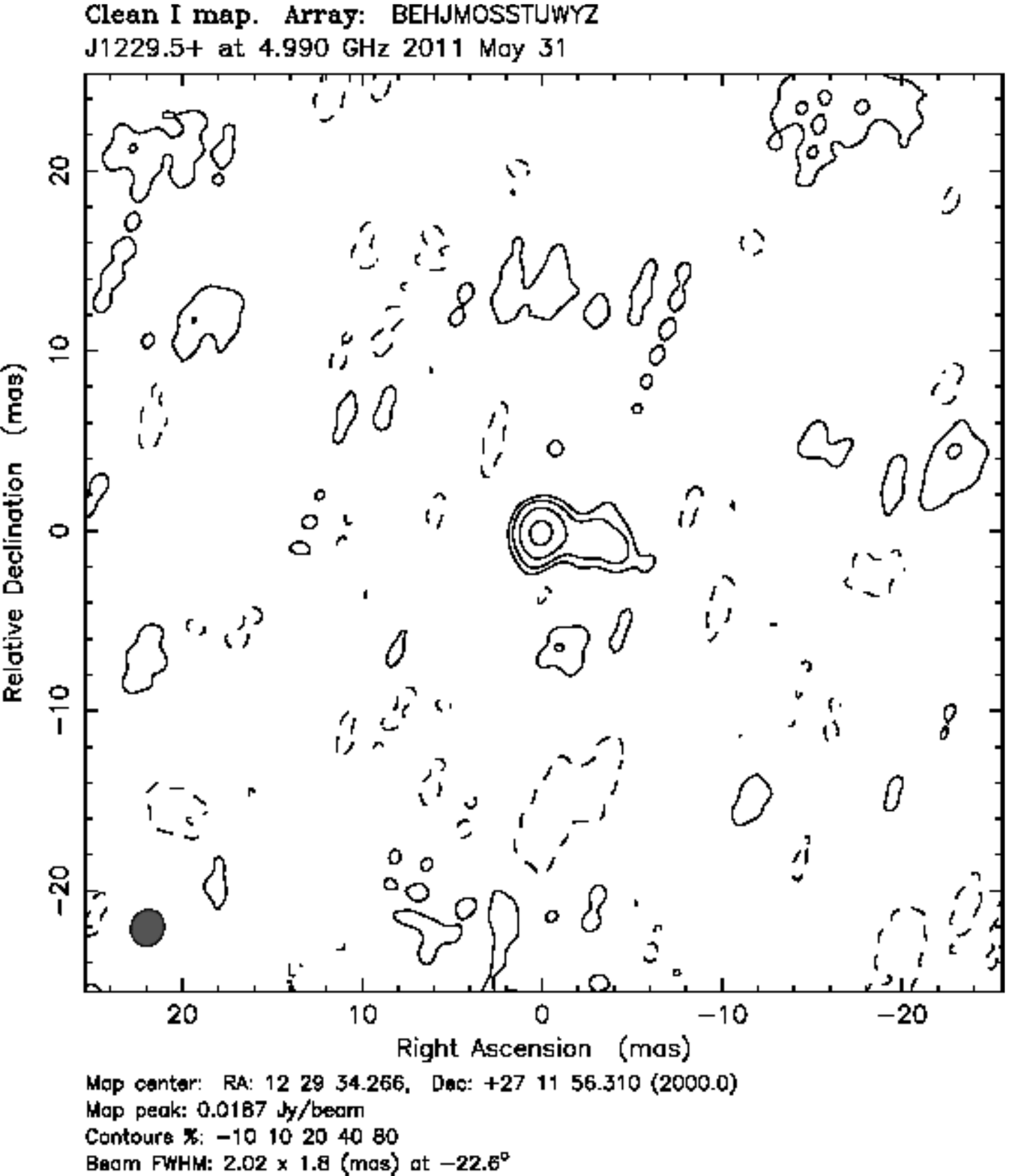} \\
\newline
\includegraphics[width=7cm,clip]{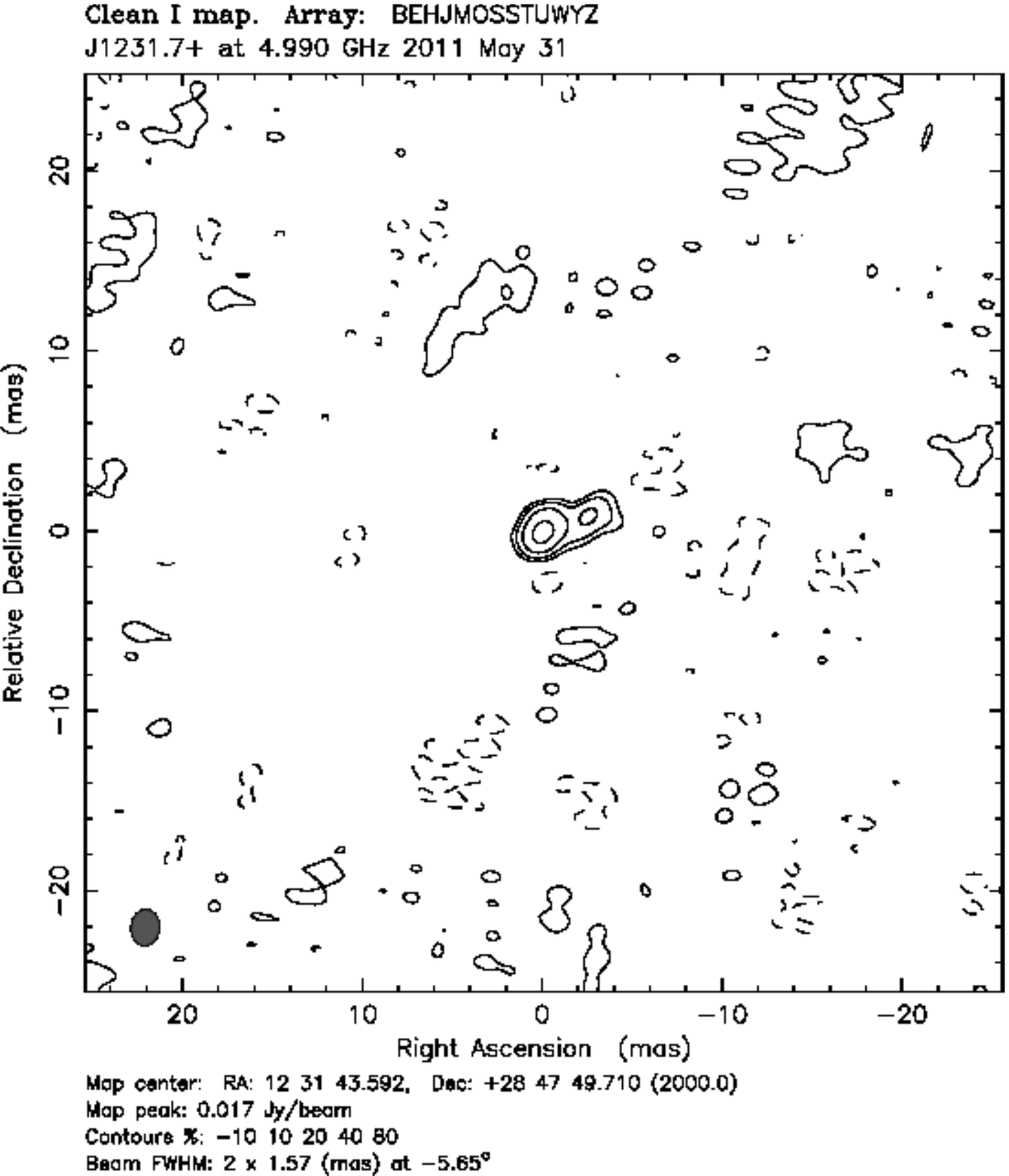}
\includegraphics[width=7cm,clip]{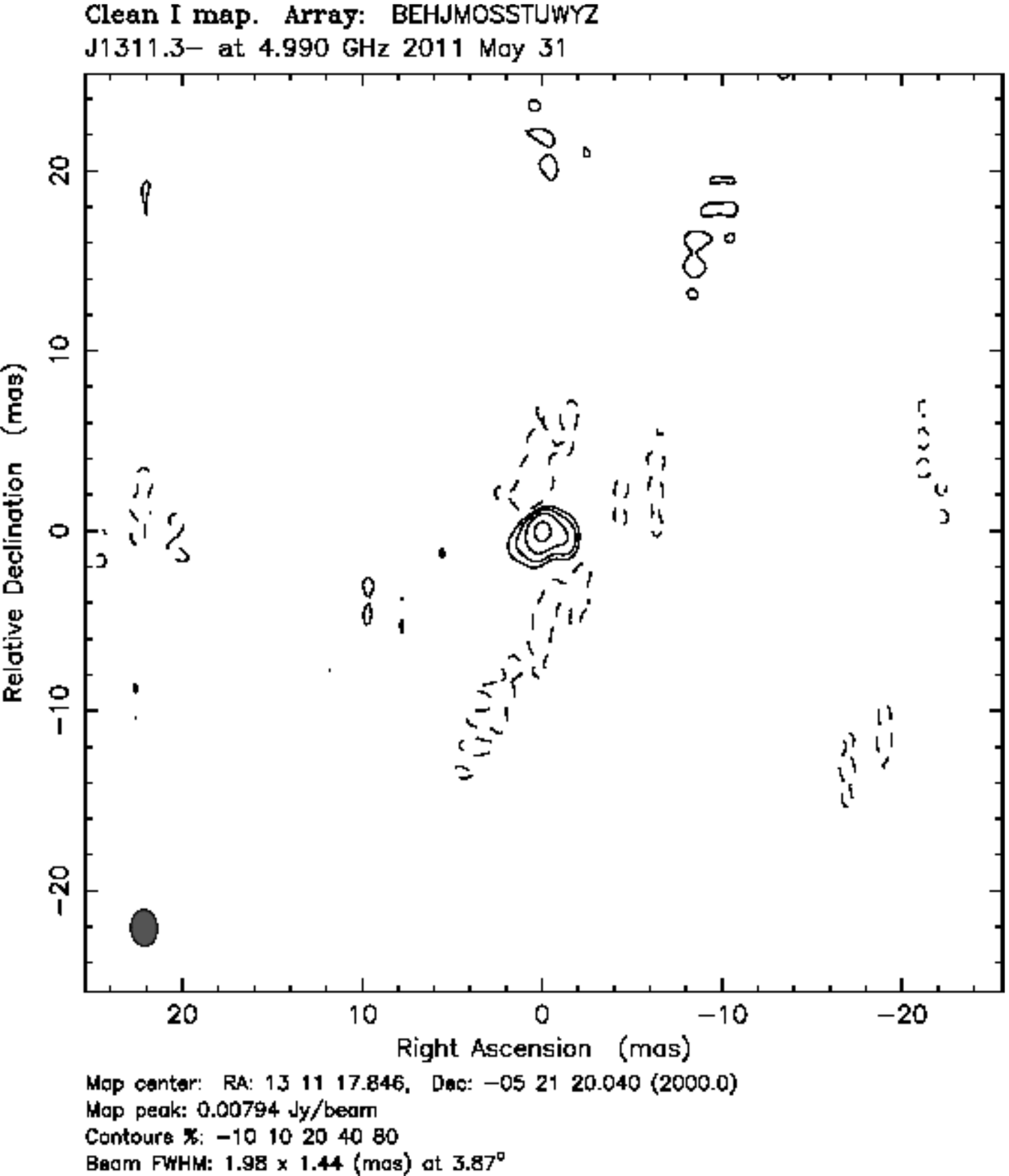}
\caption{EVN images at 5\,GHz. Continued}\label{appfig}
\end{figure*}
\begin{figure*}
\includegraphics[width=7cm,clip]{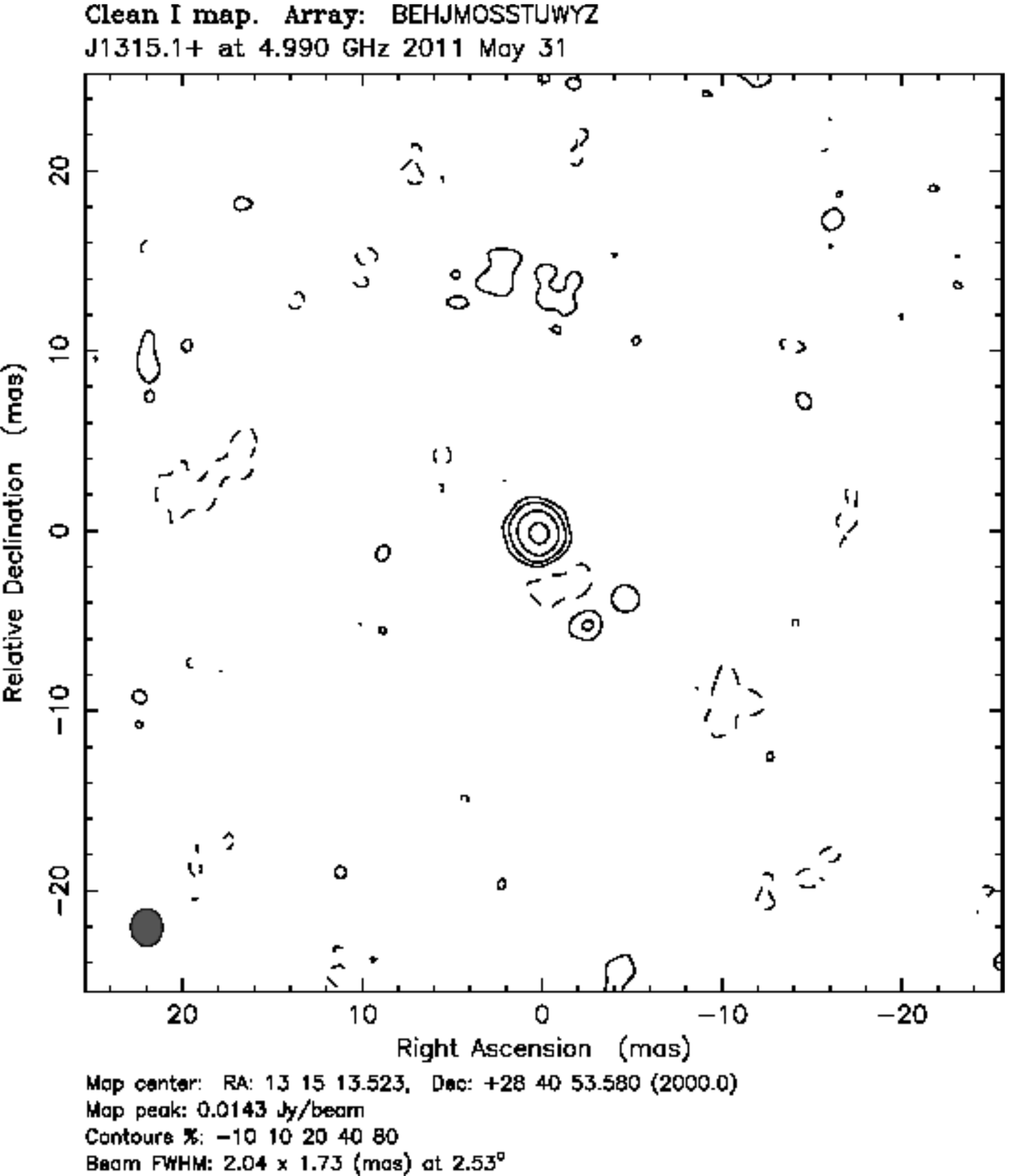}
\includegraphics[width=7cm,clip]{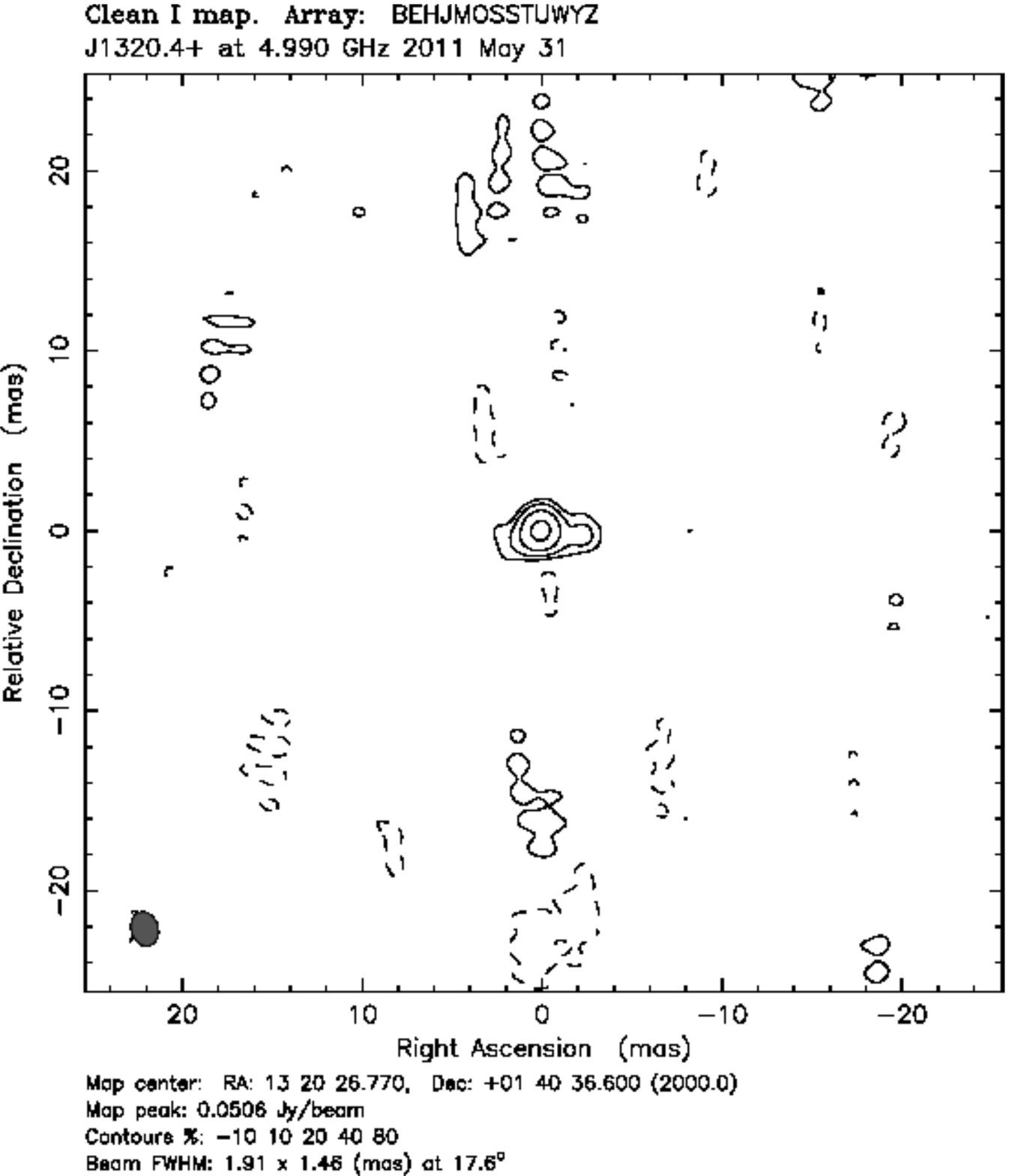} \\
\newline
\includegraphics[width=7cm,clip]{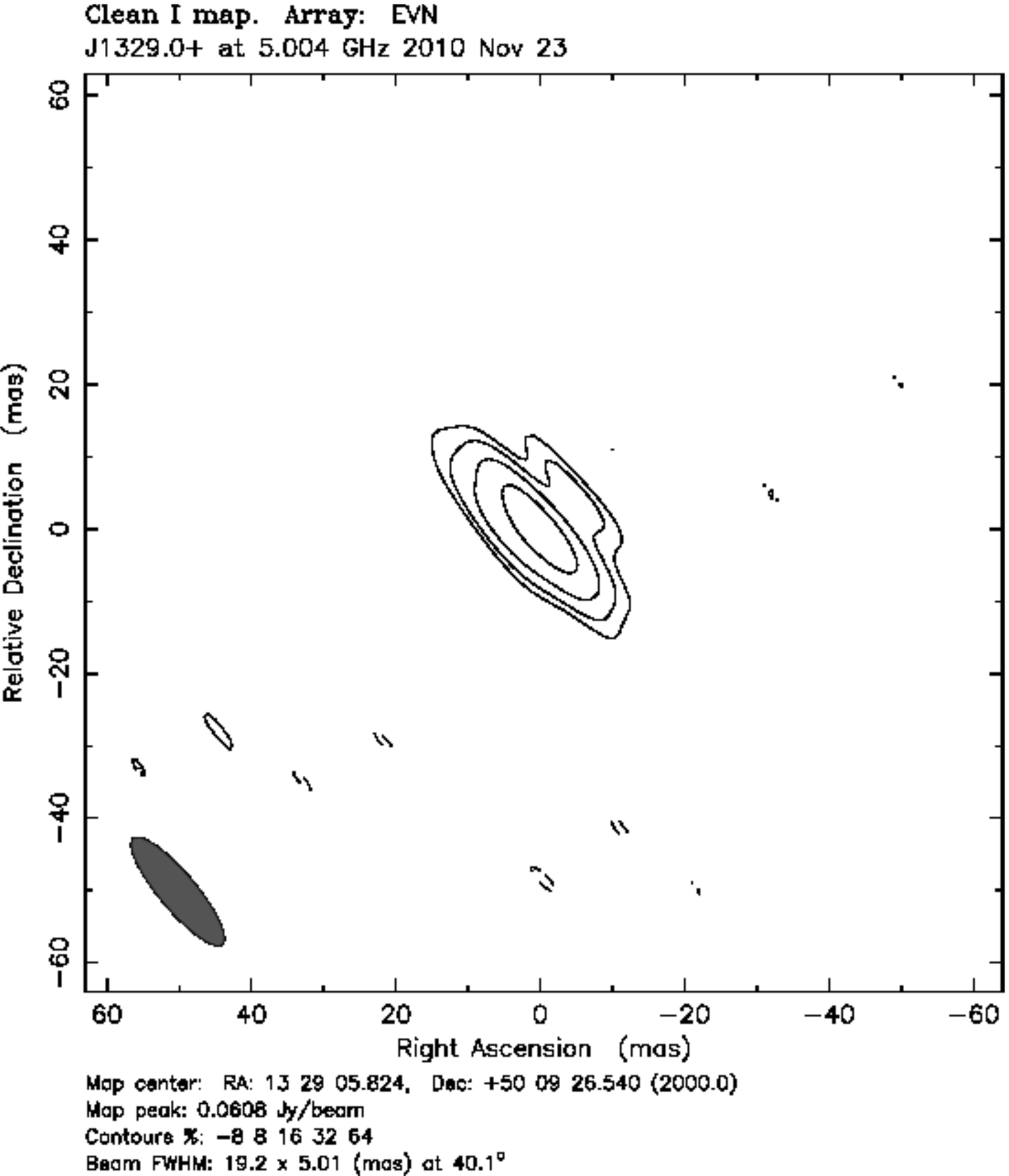}
\includegraphics[width=7cm,clip]{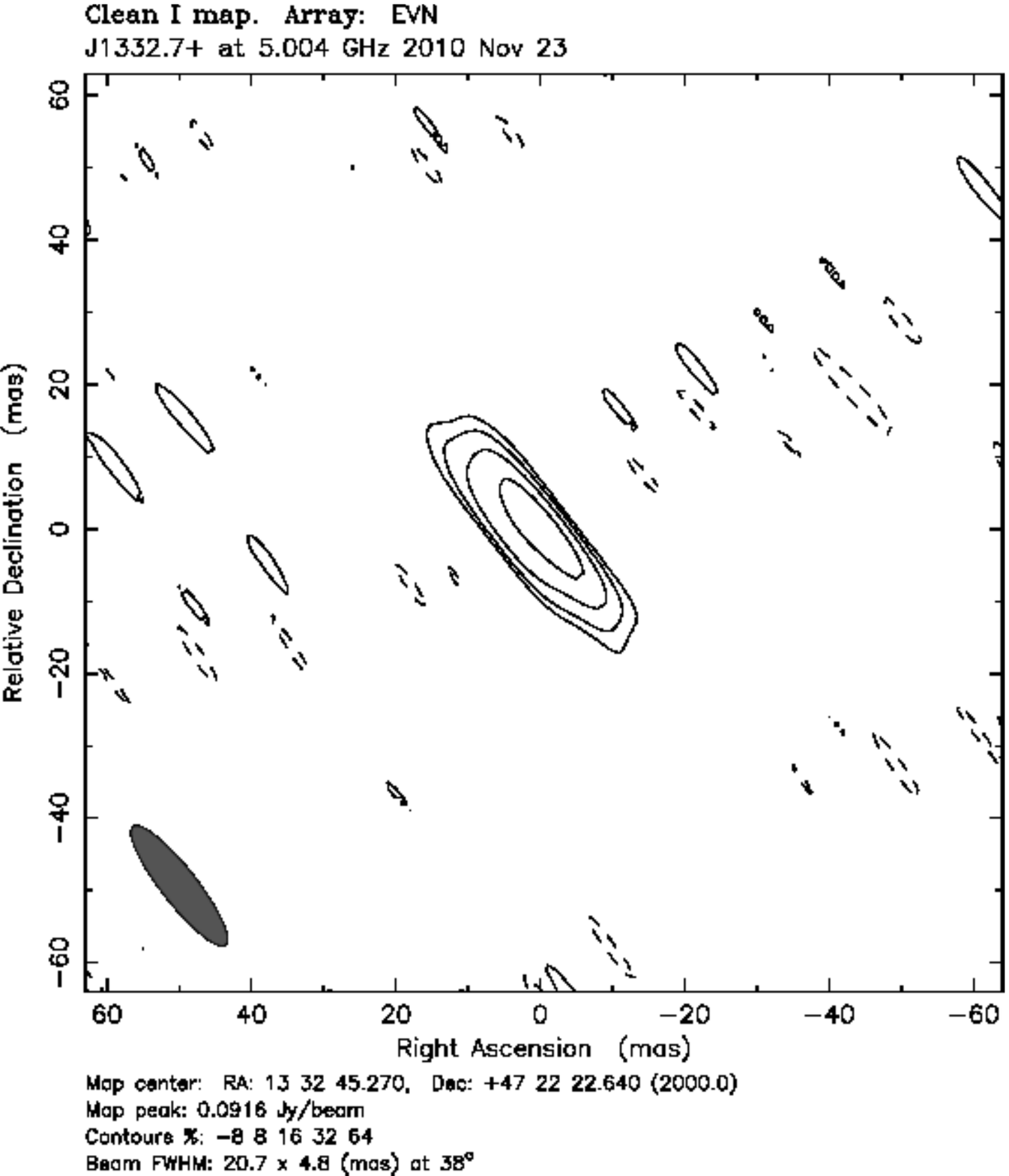} \\
\newline
\includegraphics[width=7cm,clip]{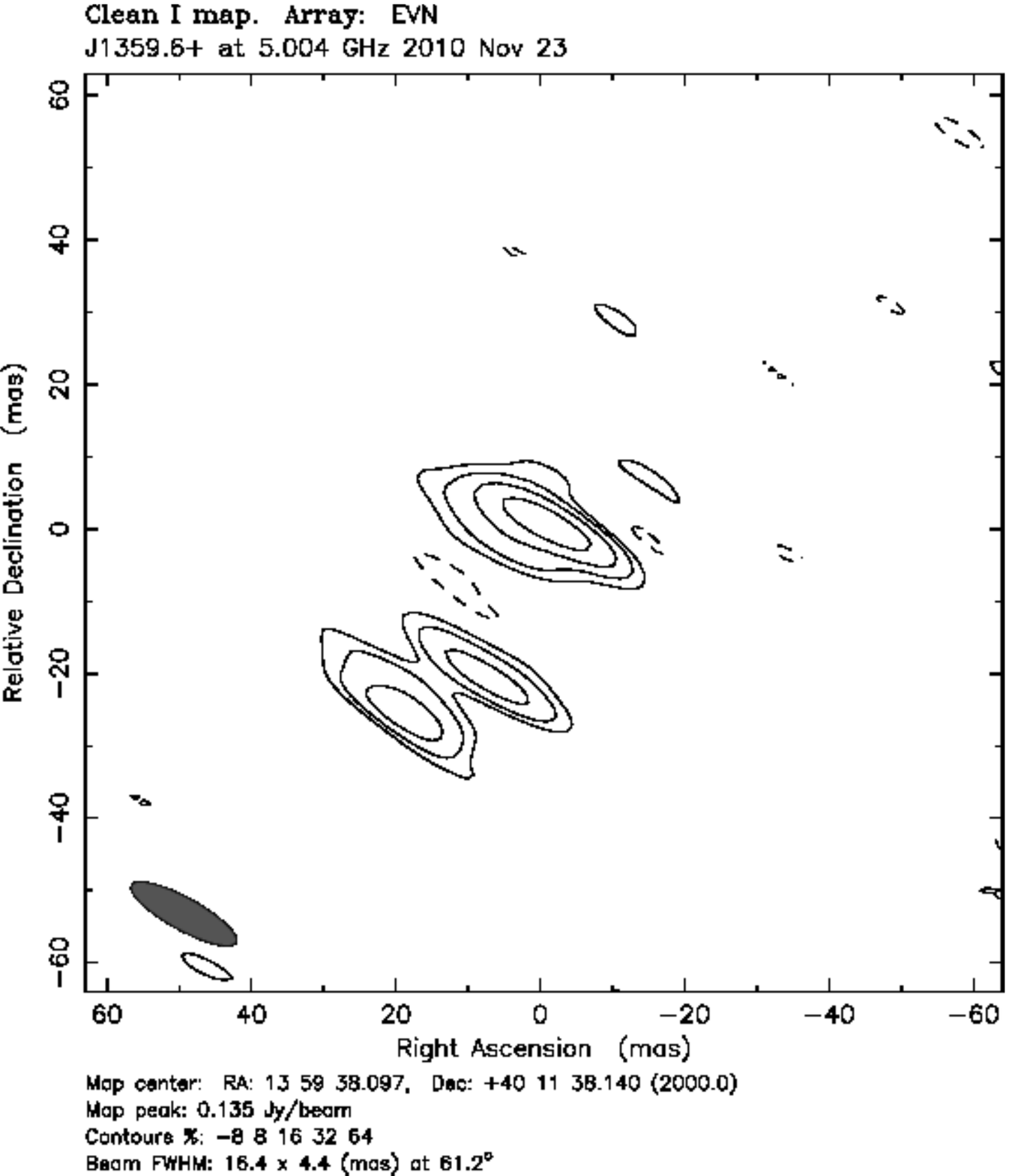} 
\includegraphics[width=7cm,clip]{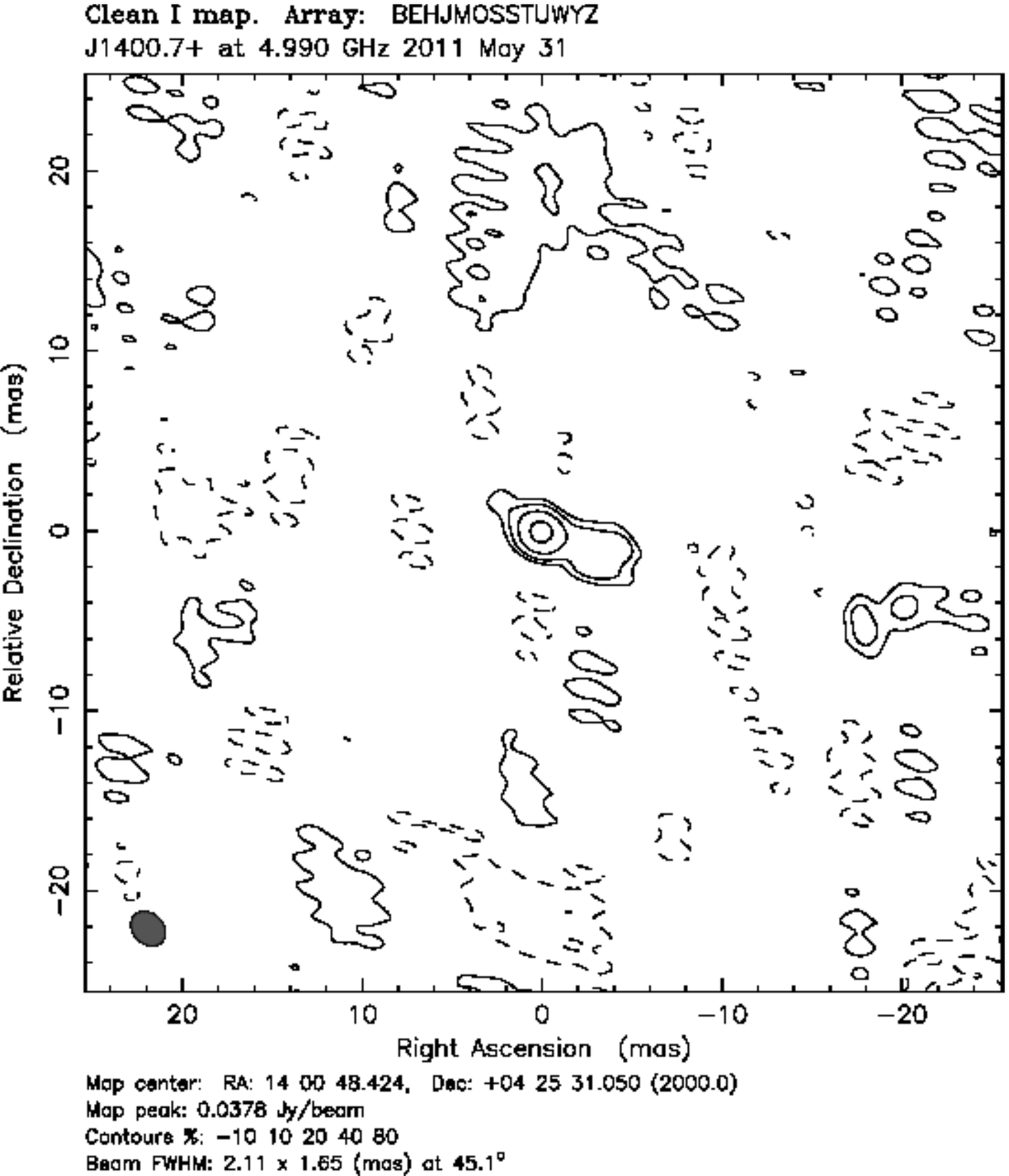}
\caption{EVN images at 5\,GHz. Continued}\label{appfig}
\end{figure*}
\begin{figure*}
\includegraphics[width=7cm,clip]{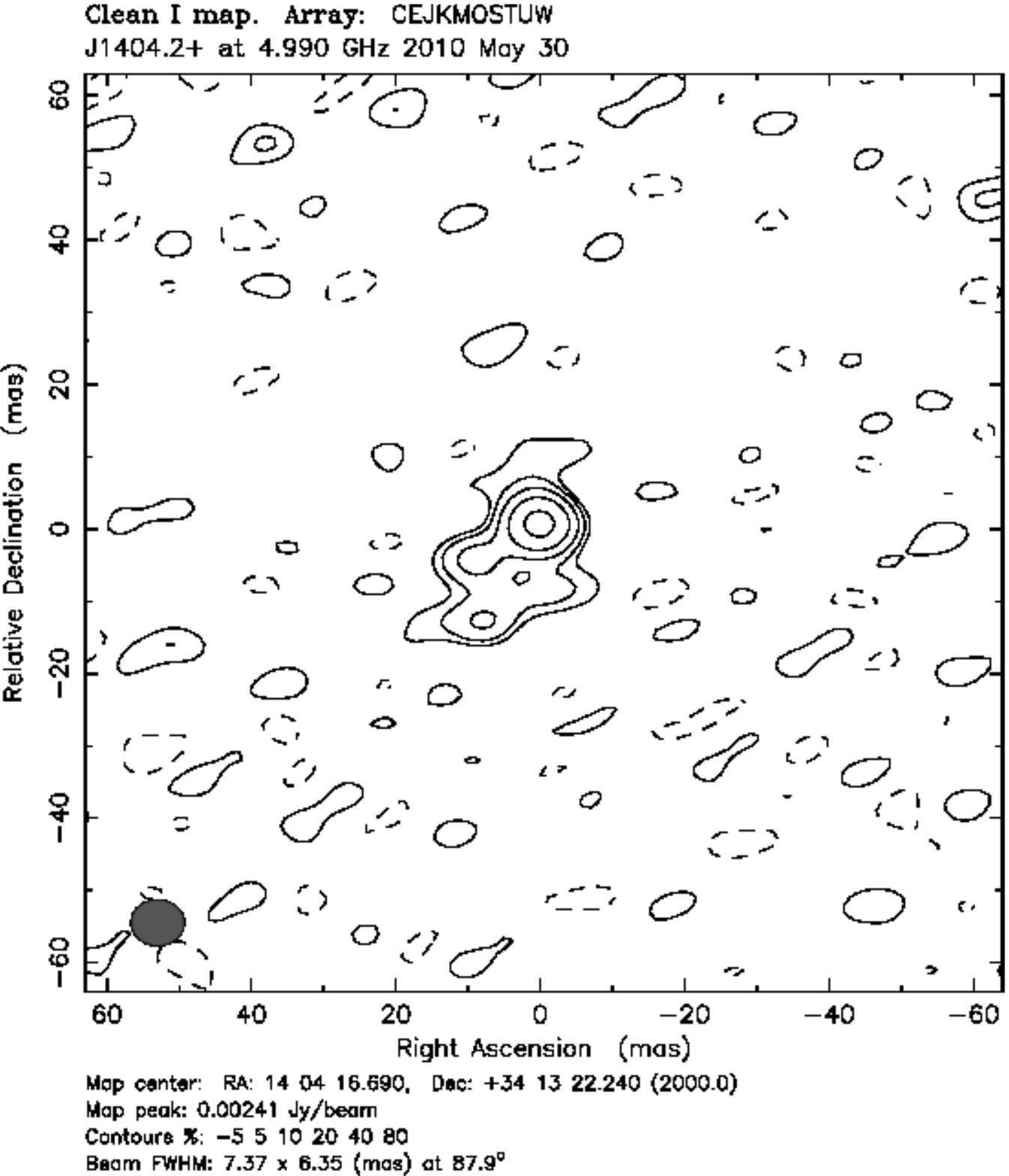}
\includegraphics[width=7cm,clip]{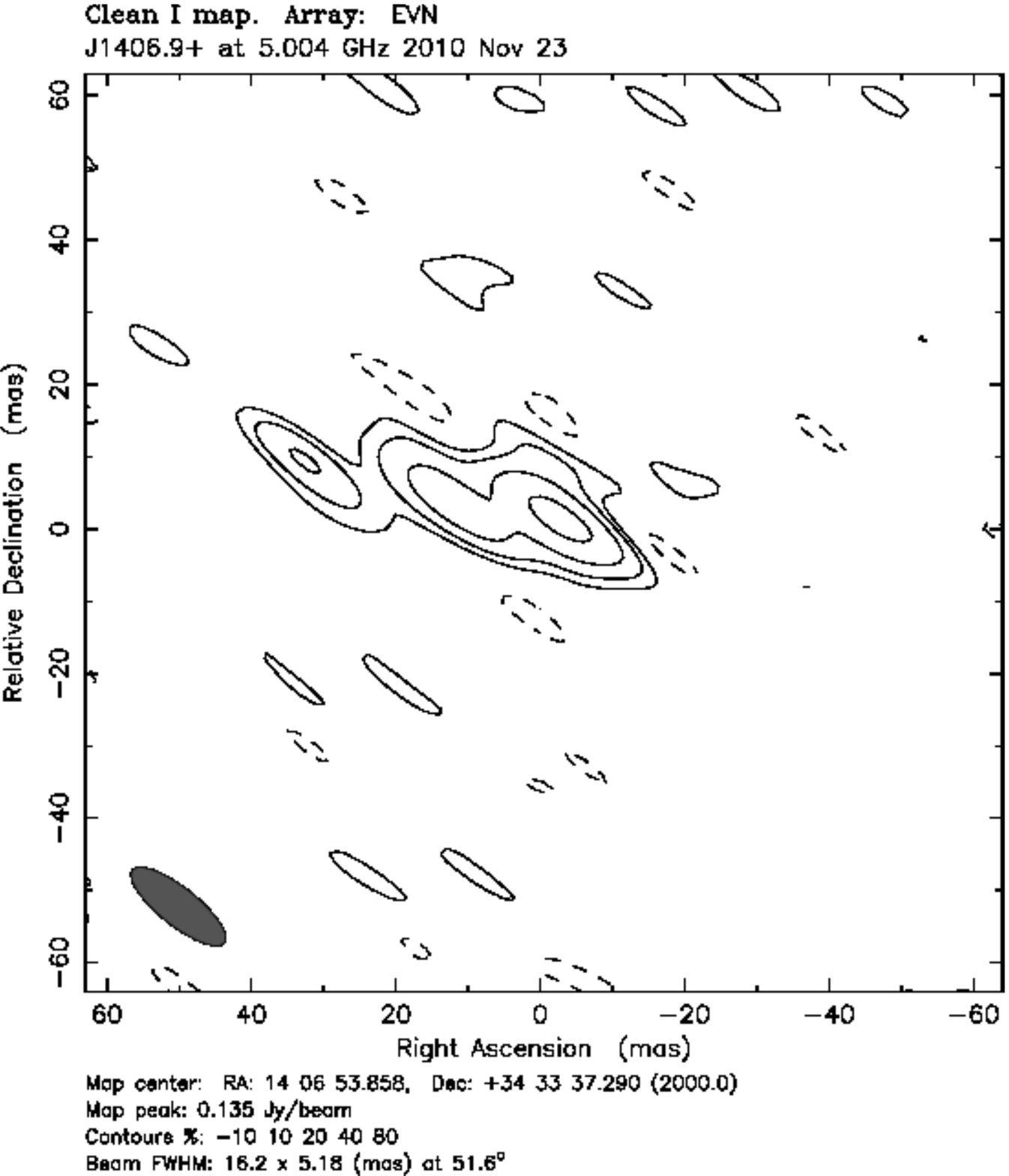} \\
\newline
\includegraphics[width=7cm,clip]{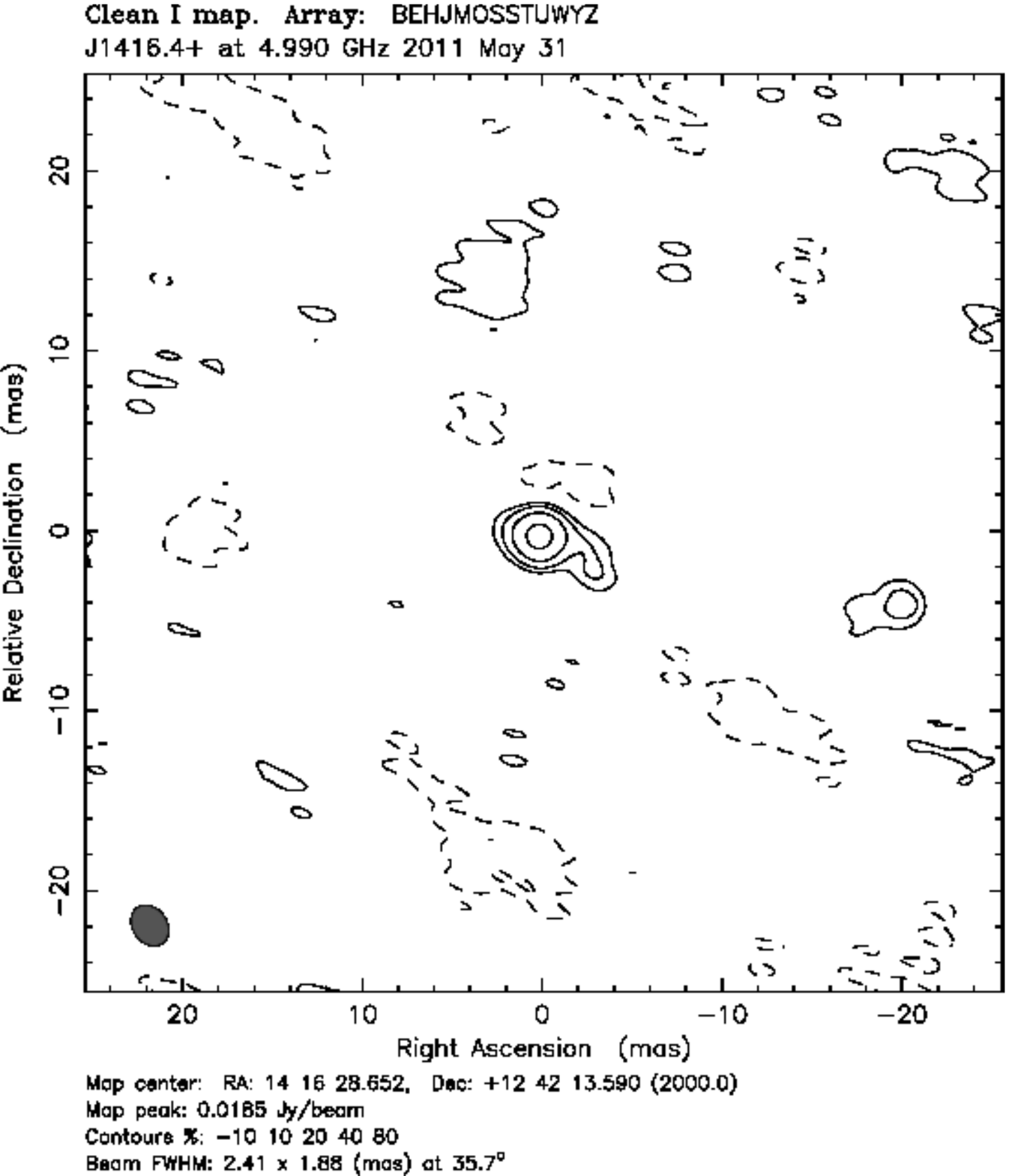} 
\includegraphics[width=7cm,clip]{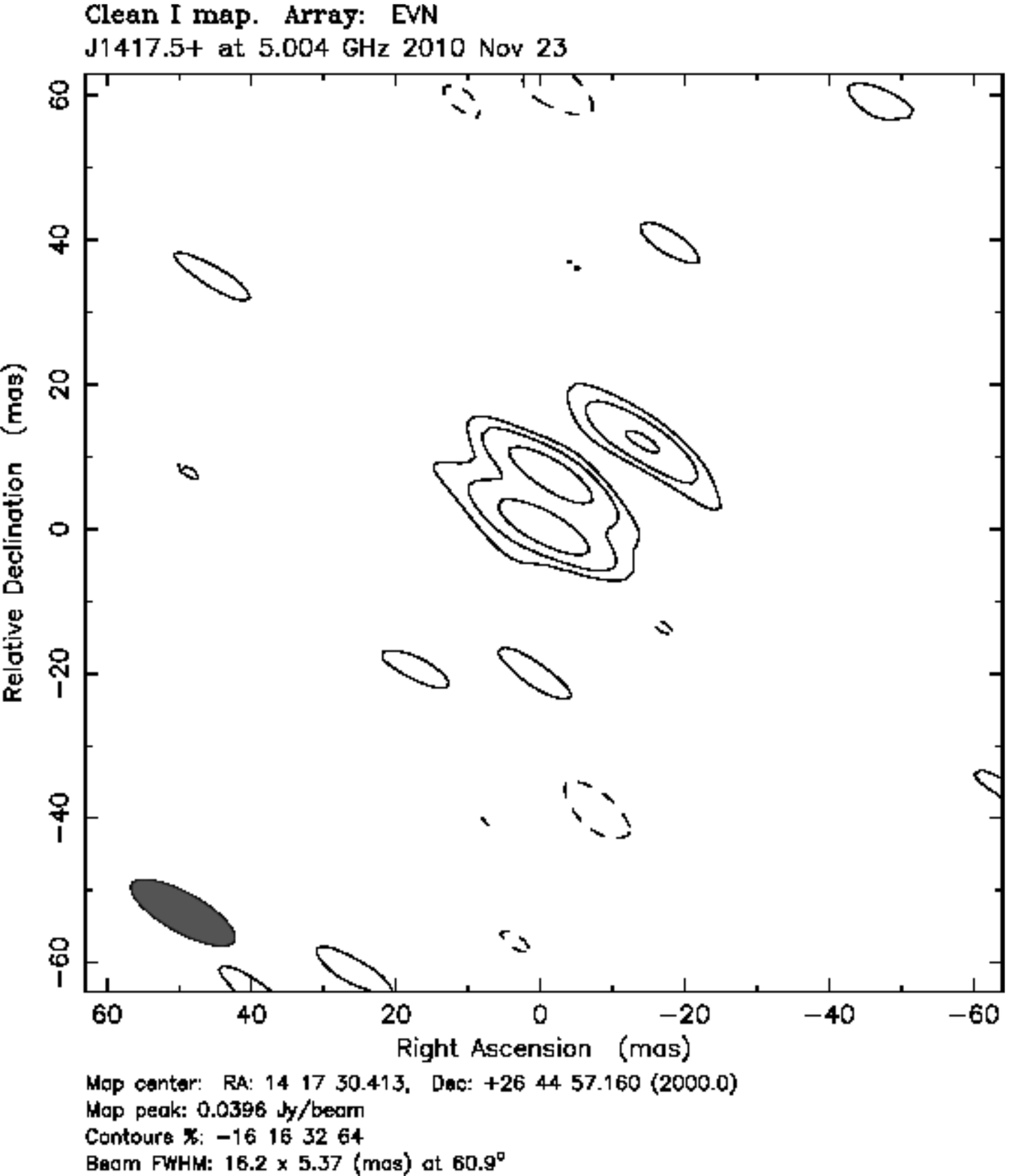} \\
\newline
\includegraphics[width=7cm,clip]{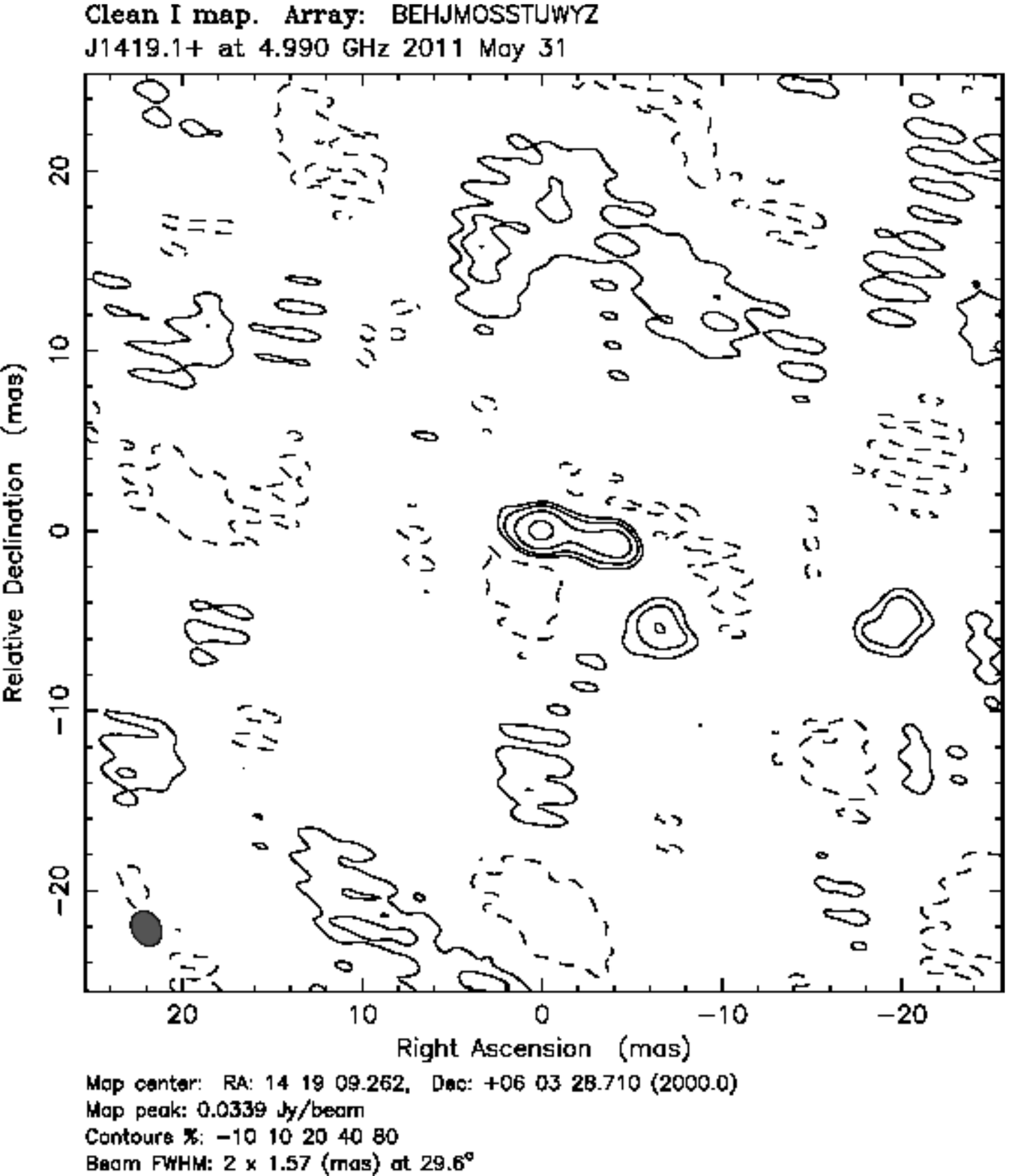}
\includegraphics[width=7cm,clip]{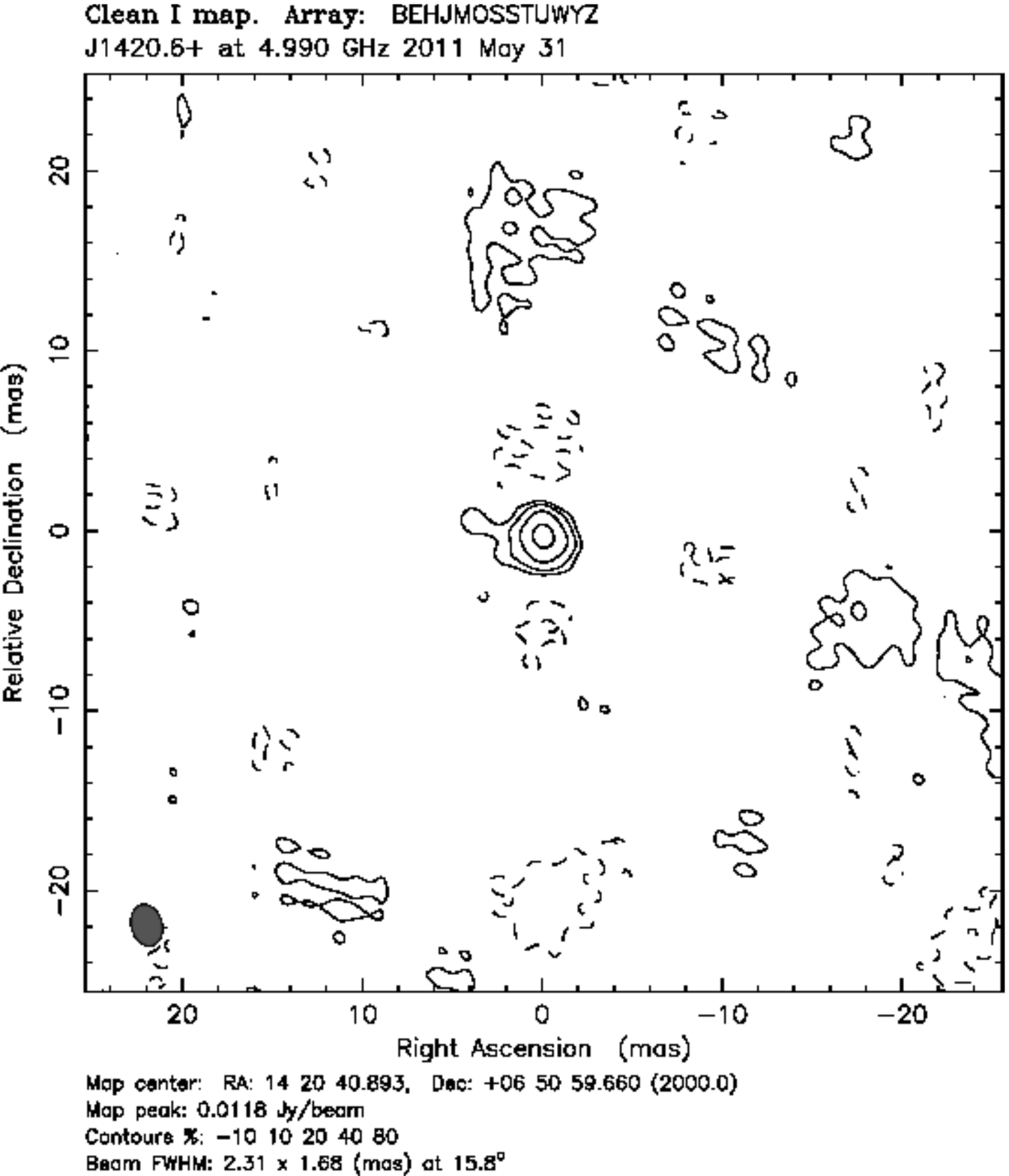}
\caption{EVN images at 5\,GHz. Continued}\label{appfig}
\end{figure*}
\begin{figure*}
\includegraphics[width=7cm,clip]{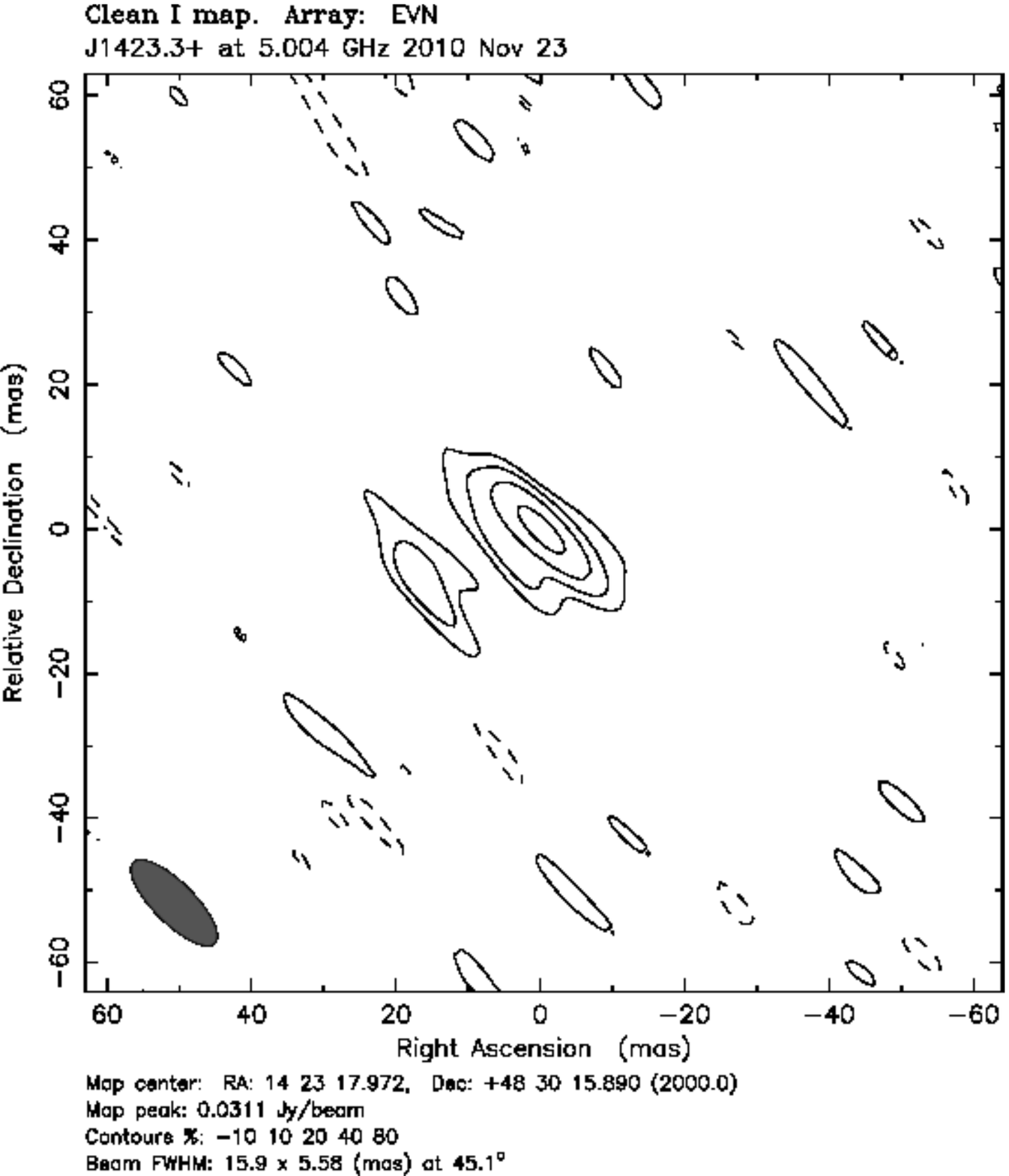}
\includegraphics[width=7cm,clip]{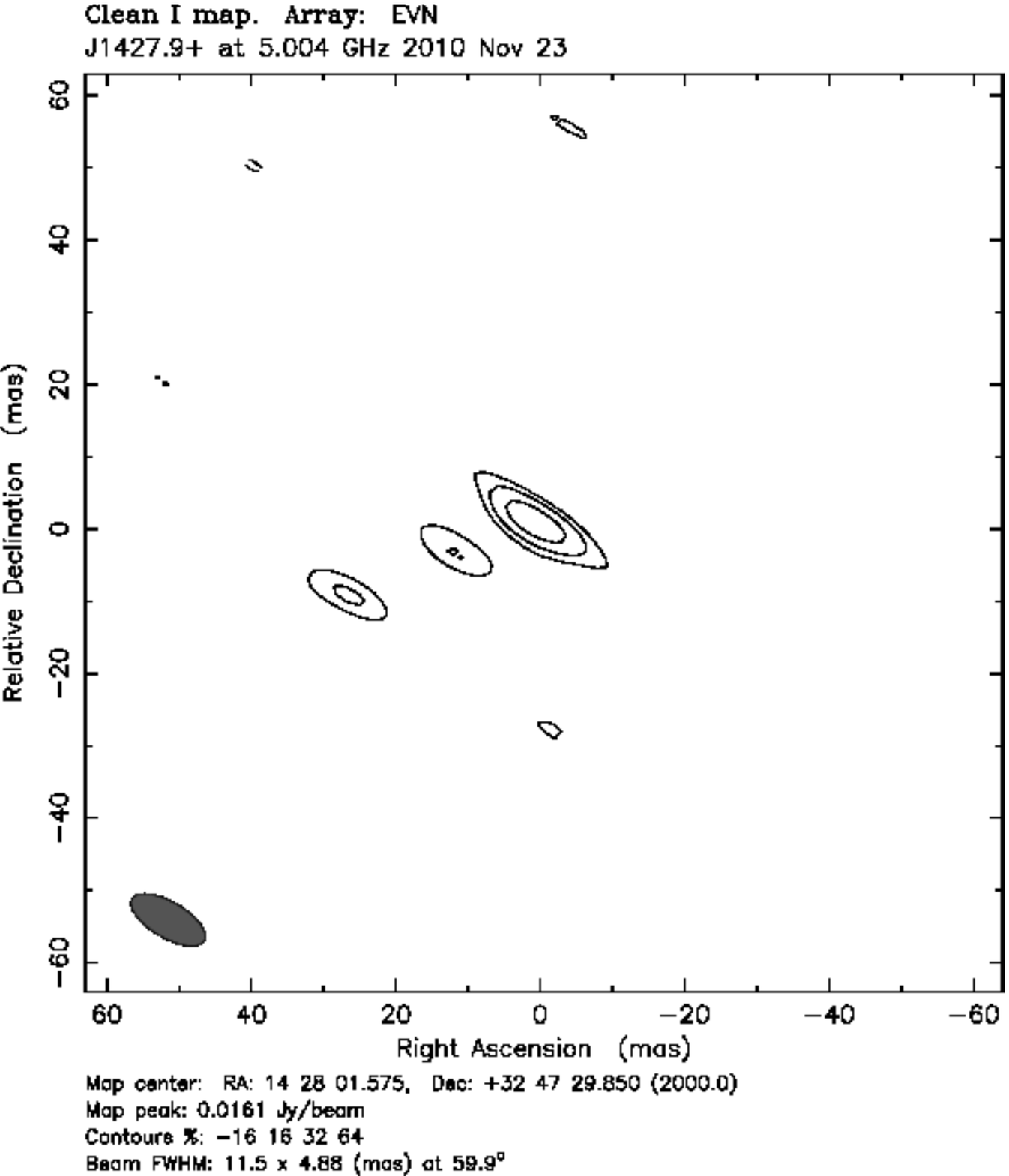} \\
\newline
\includegraphics[width=7cm,clip]{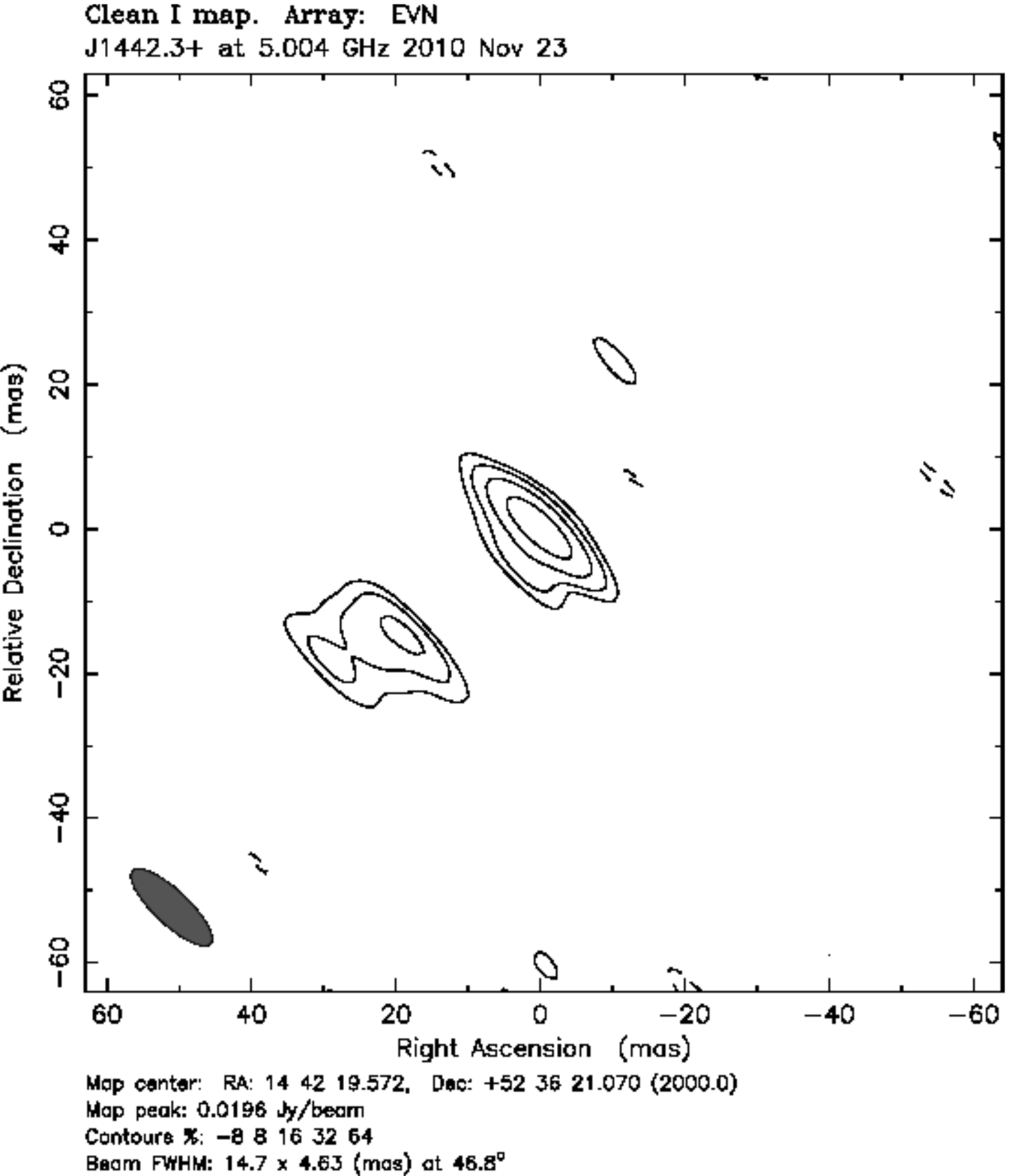}
\includegraphics[width=7cm,clip]{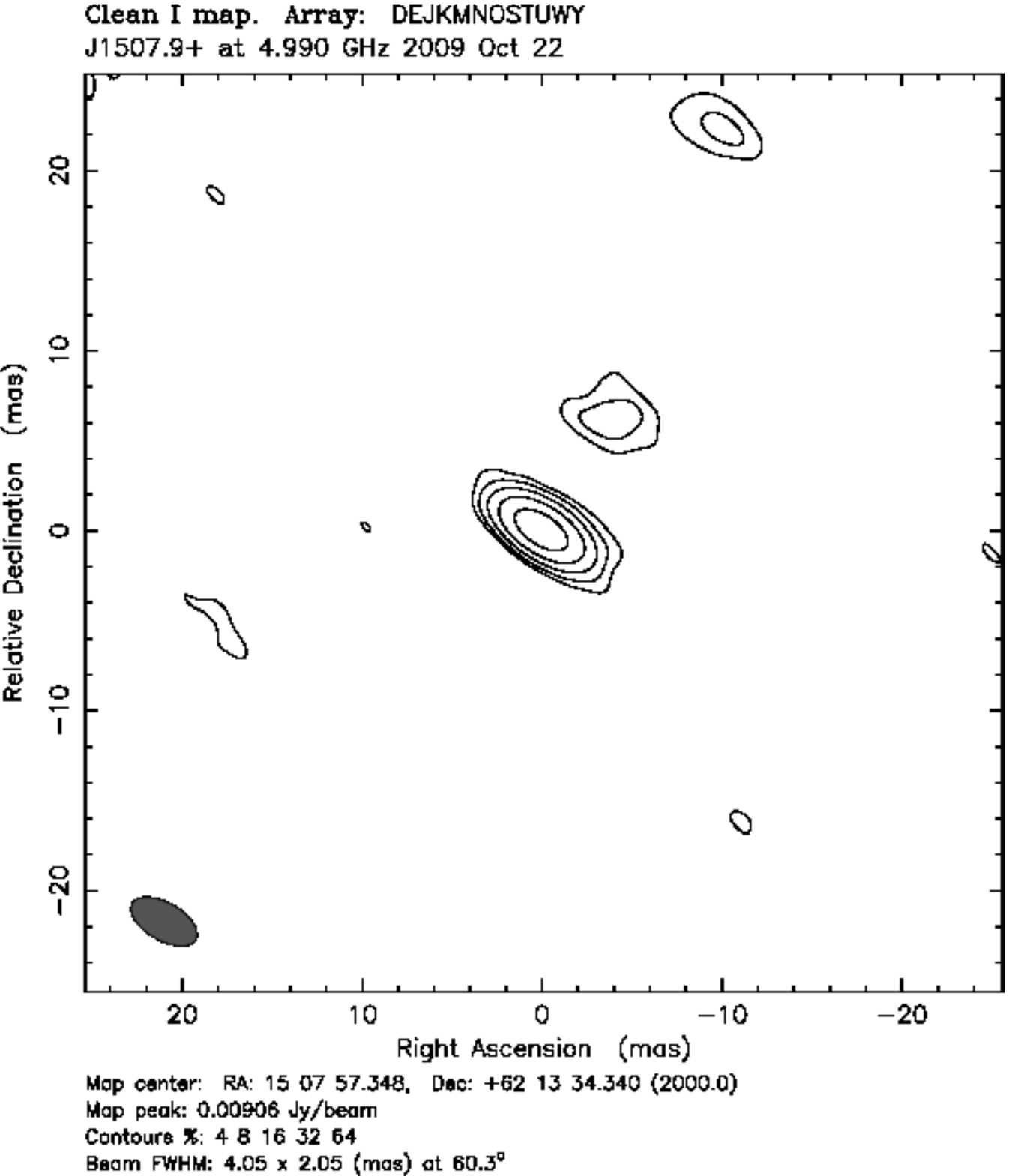} \\
\newline
\includegraphics[width=7cm,clip]{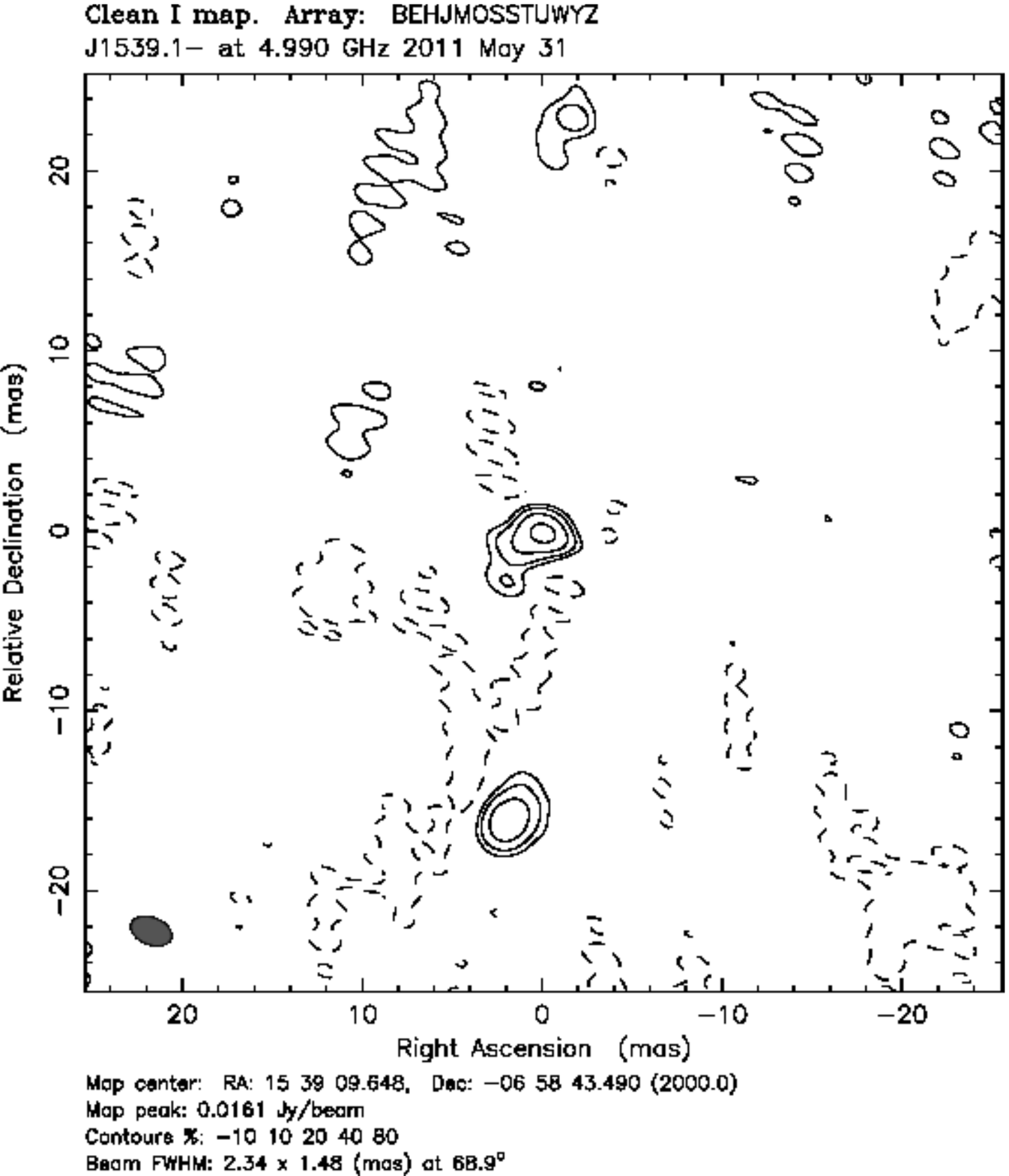}
\includegraphics[width=7cm,clip]{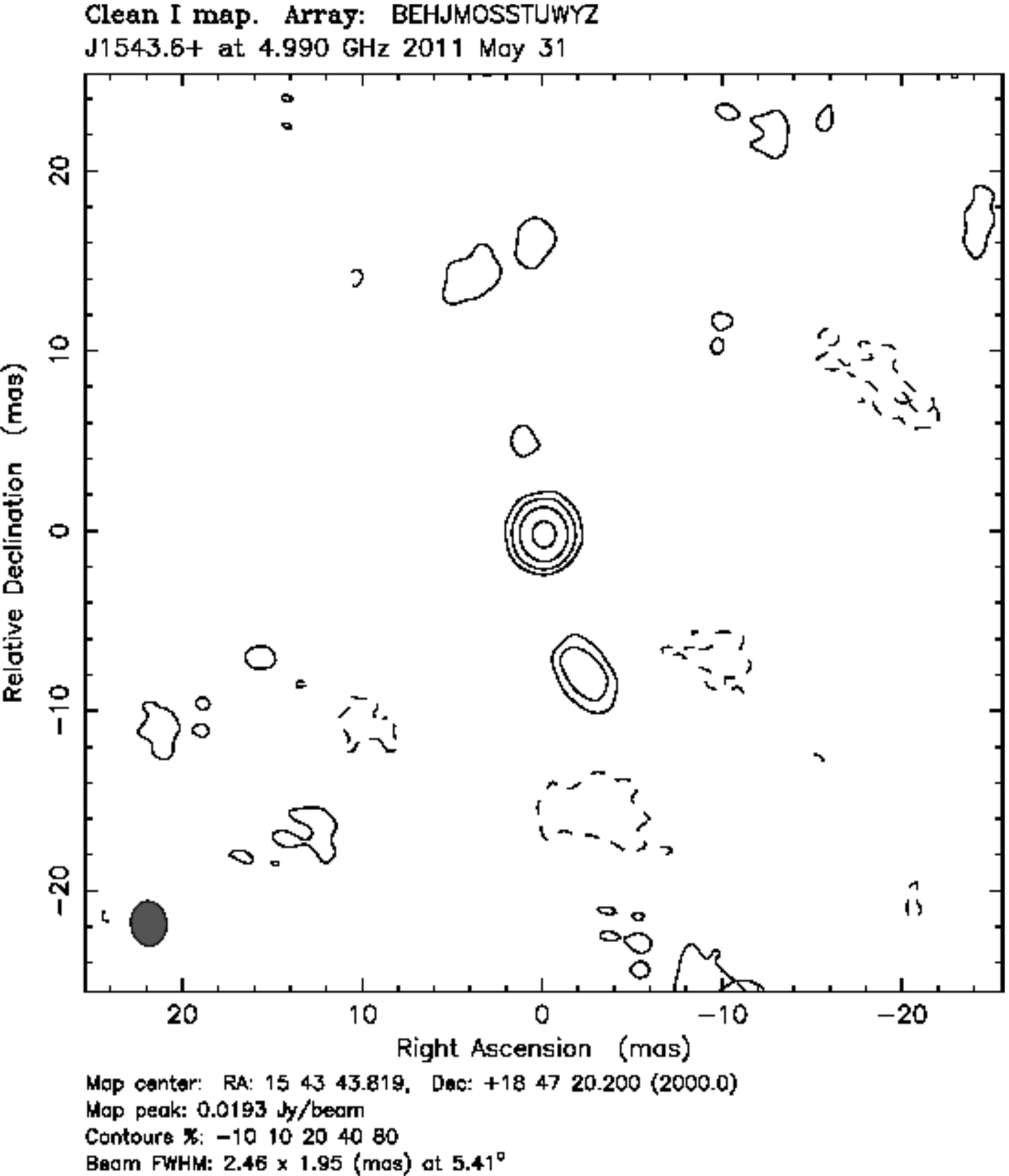}
\caption{EVN images at 5\,GHz.}\label{appfig}
\end{figure*}
\begin{figure*}
\includegraphics[width=7cm,clip]{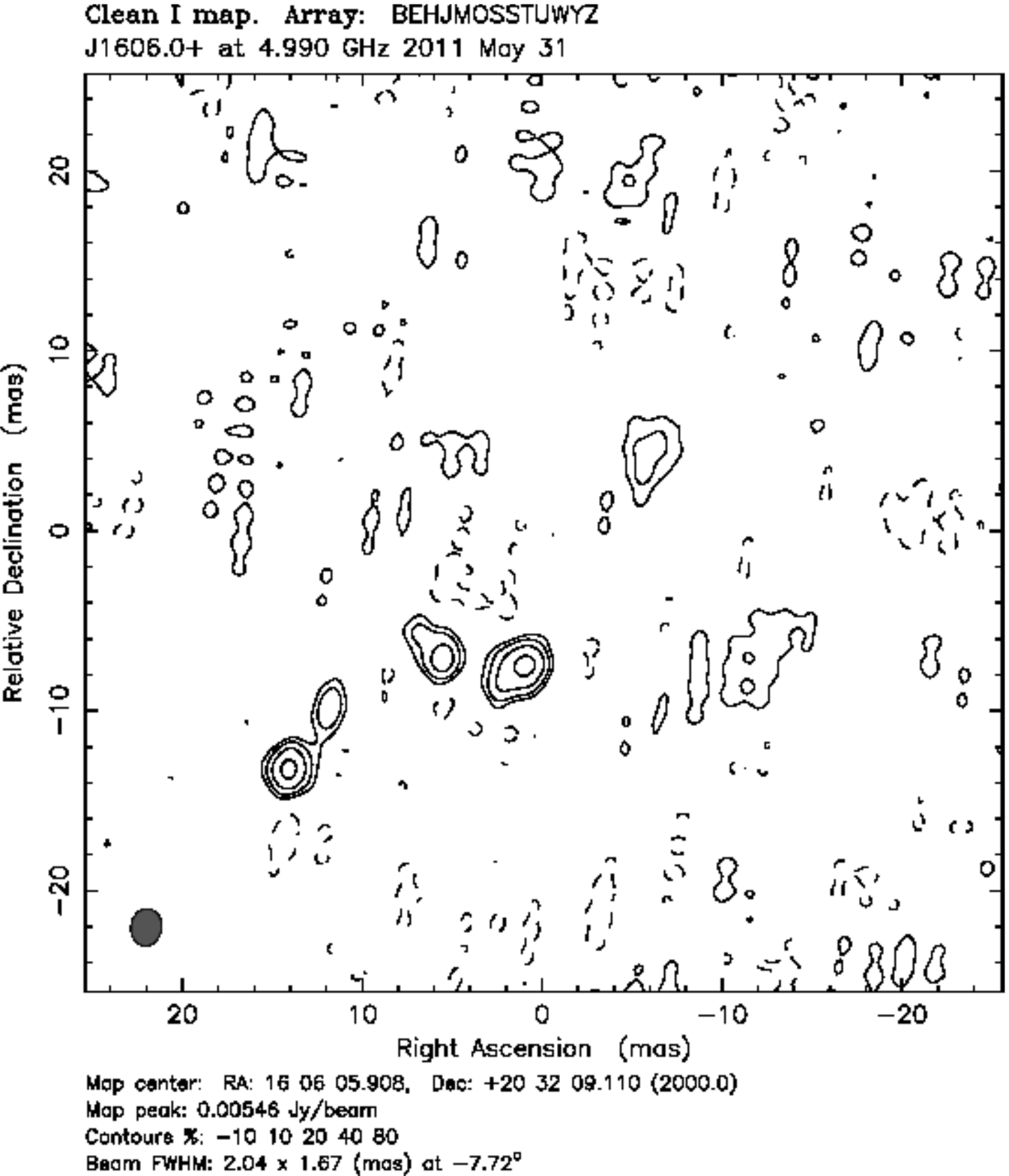}
\includegraphics[width=7cm,clip]{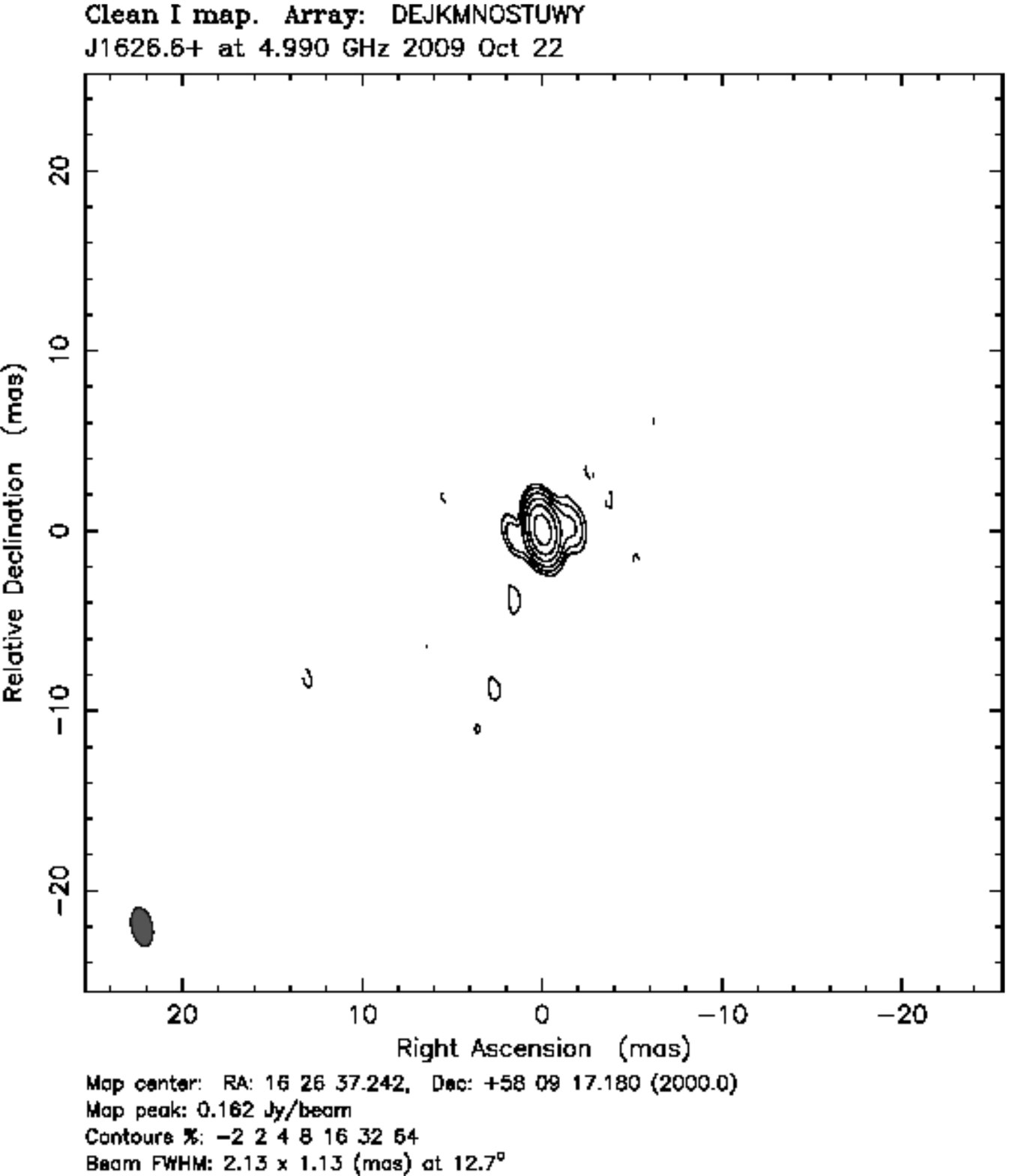} \\
\newline
\includegraphics[width=7cm,clip]{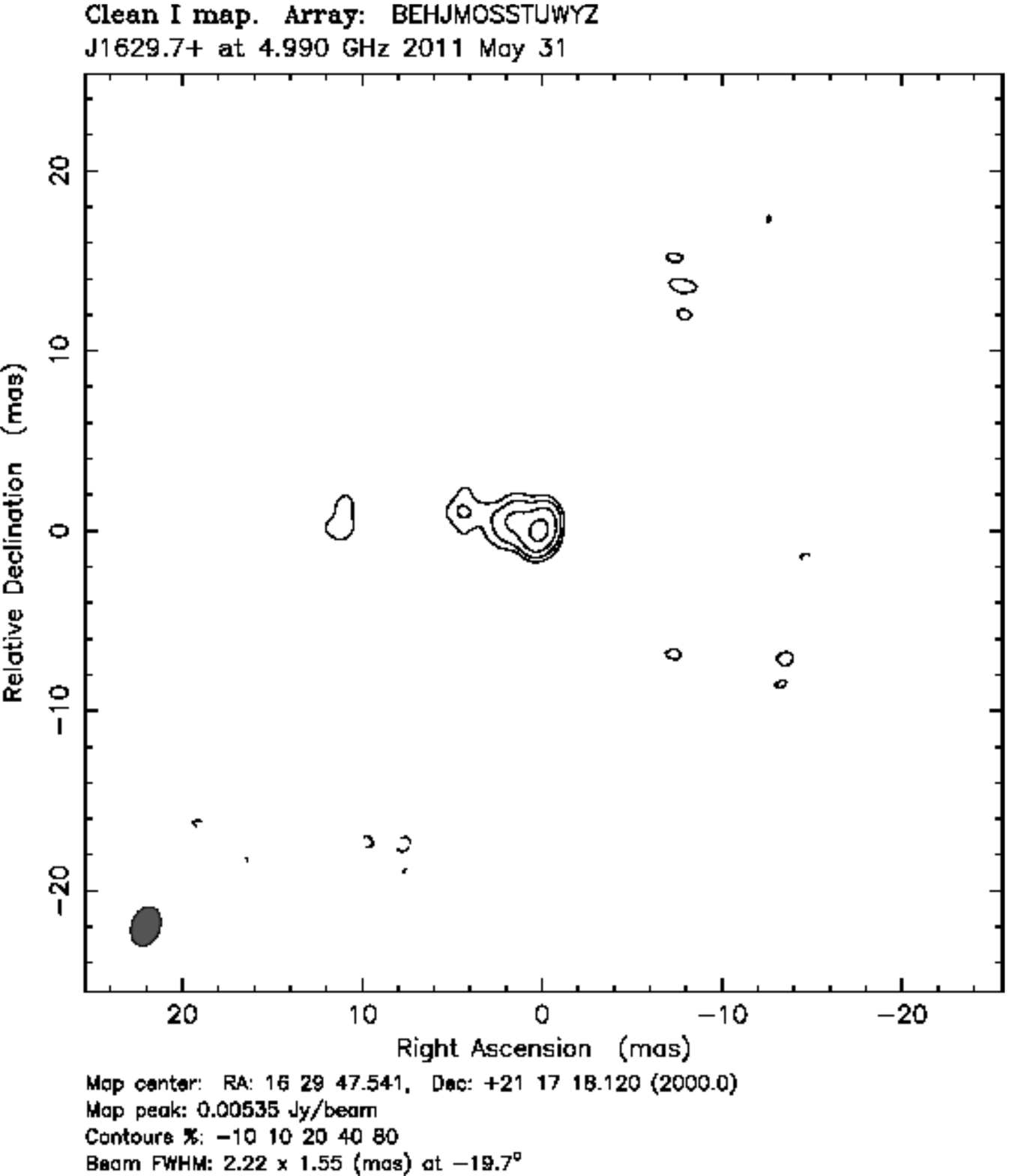}
\includegraphics[width=7cm,clip]{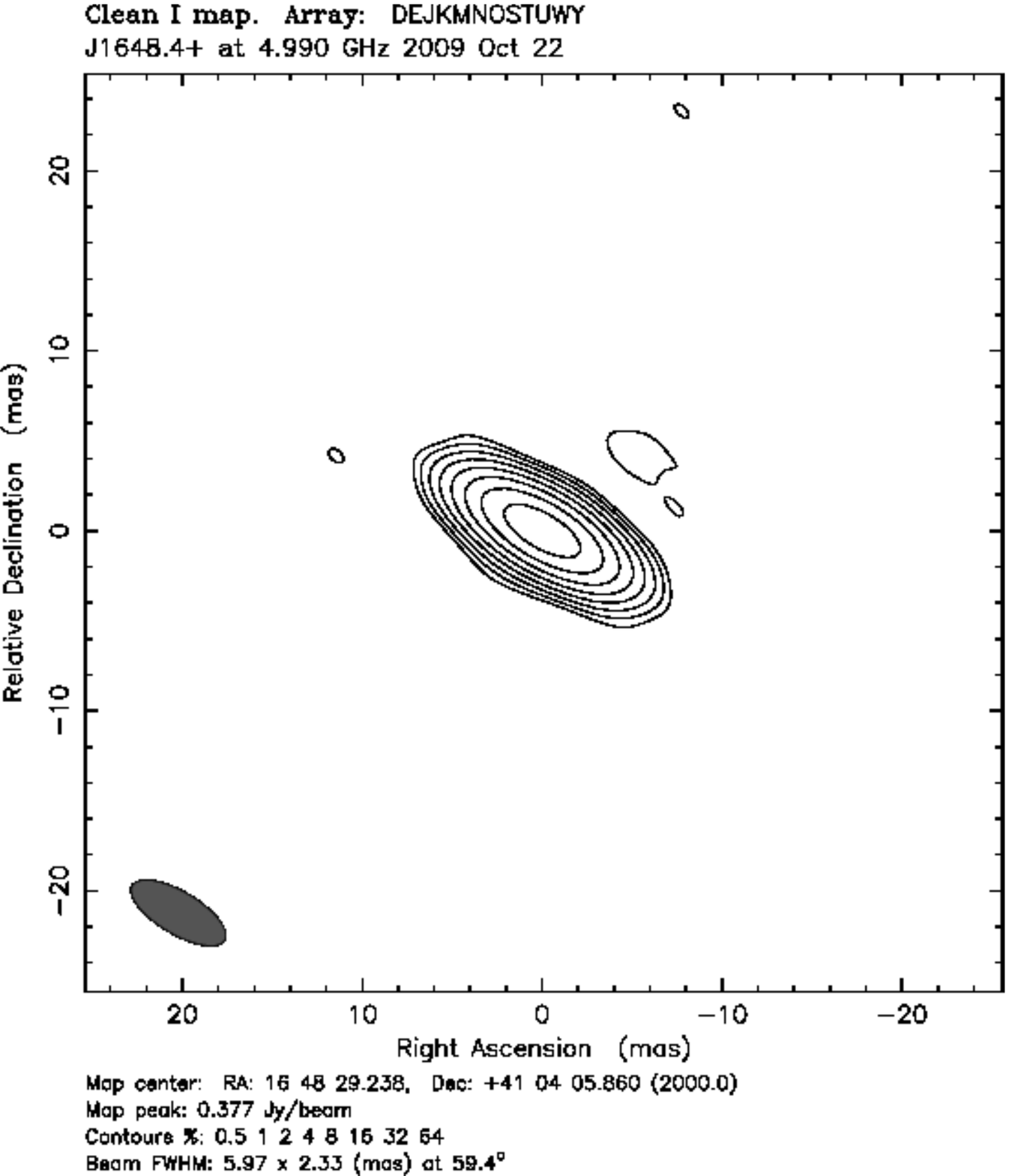} \\
\newline
\includegraphics[width=7cm,clip]{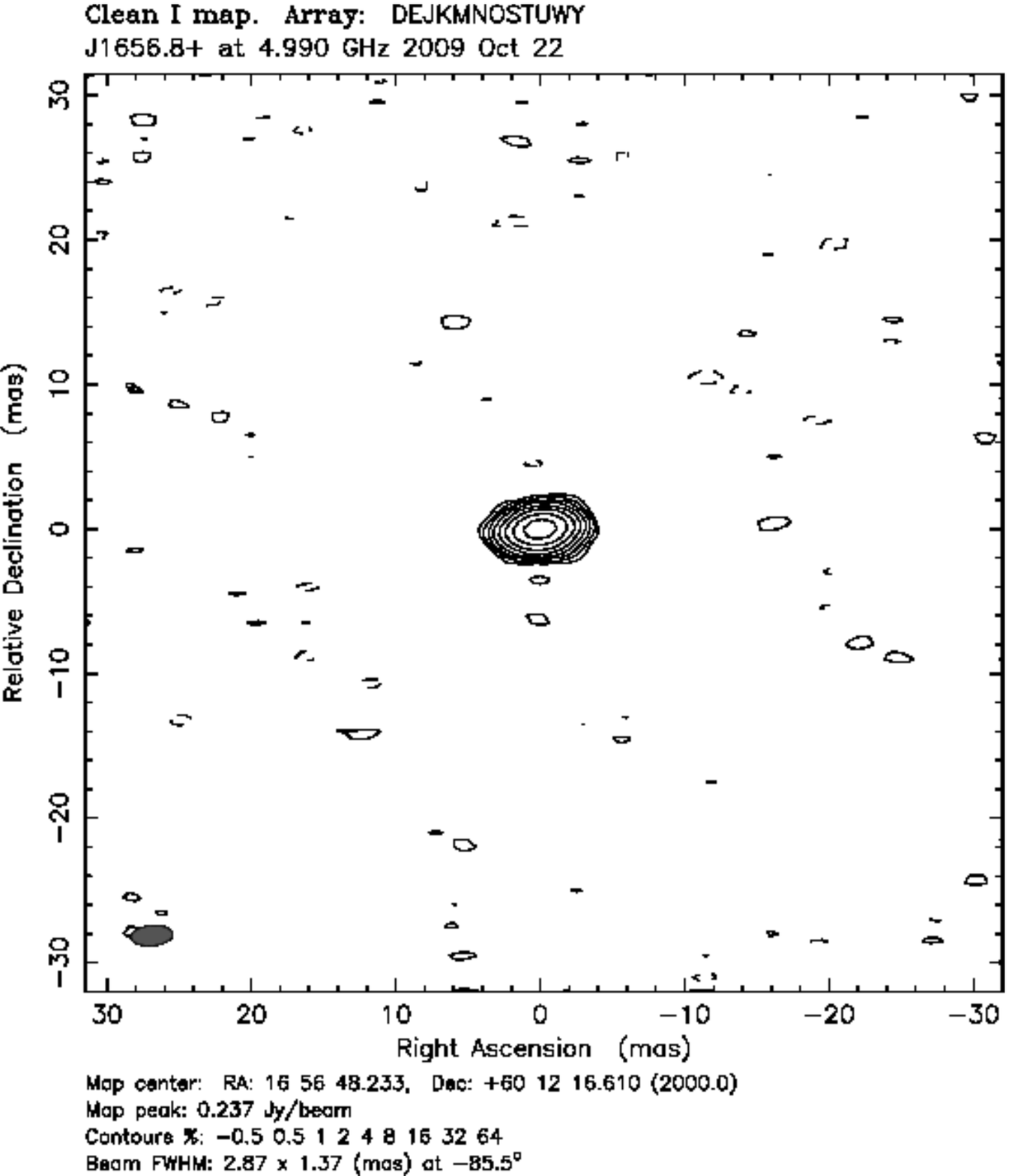}
\includegraphics[width=7cm,clip]{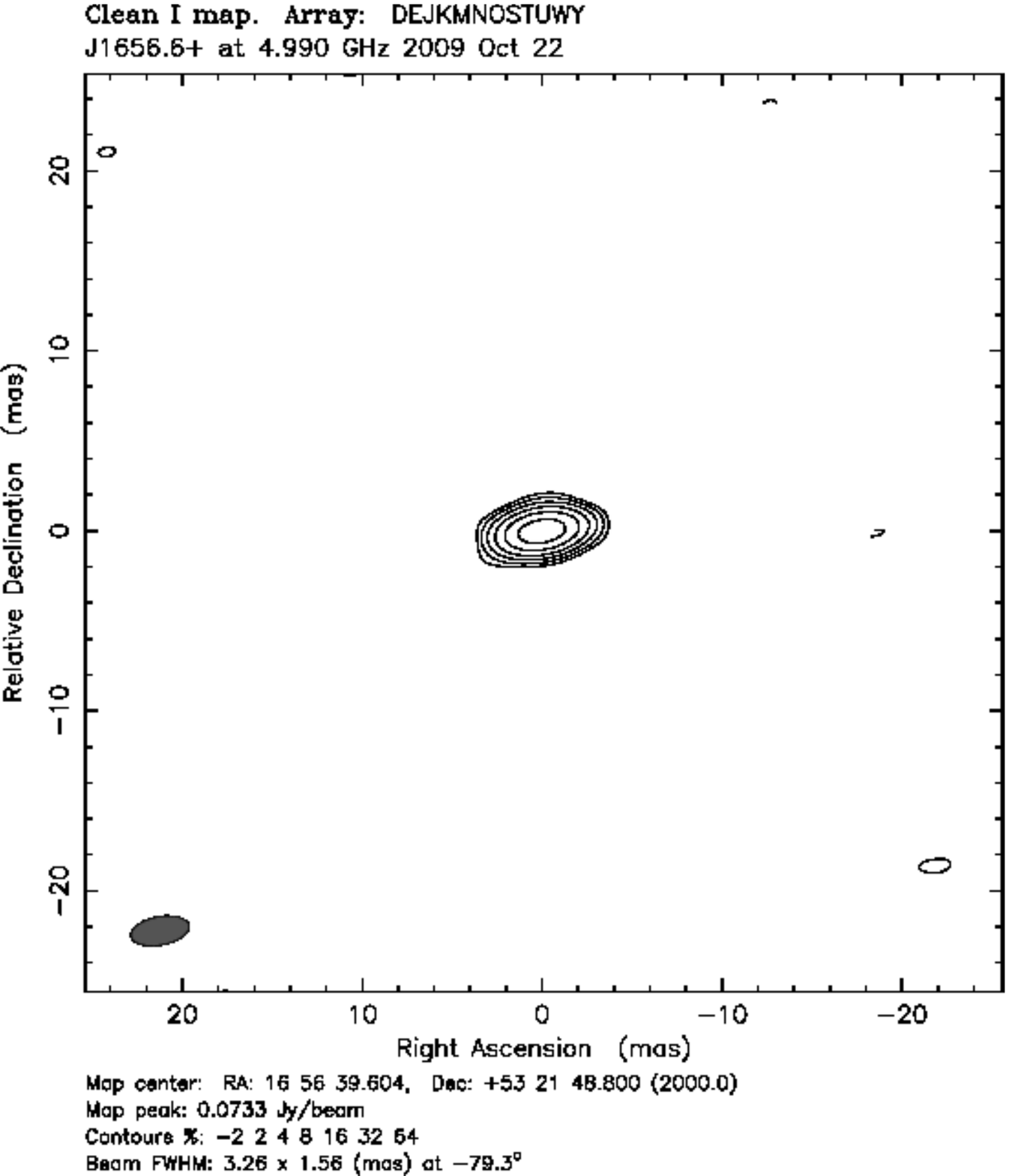}
\caption{EVN images at 5\,GHz.}\label{appfig}
\end{figure*}
\begin{figure*}
\includegraphics[width=7cm,clip]{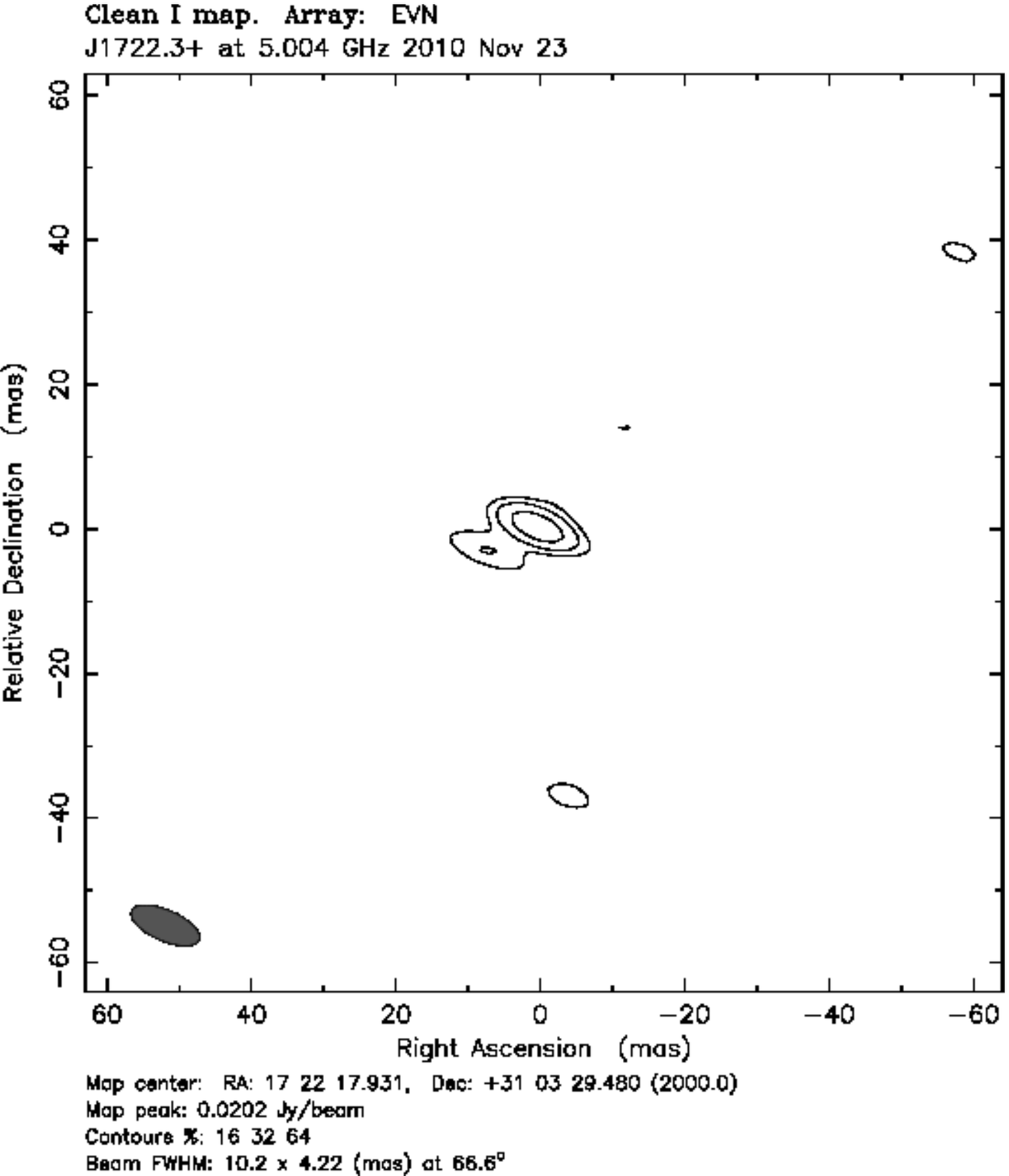}
\includegraphics[width=7cm,clip]{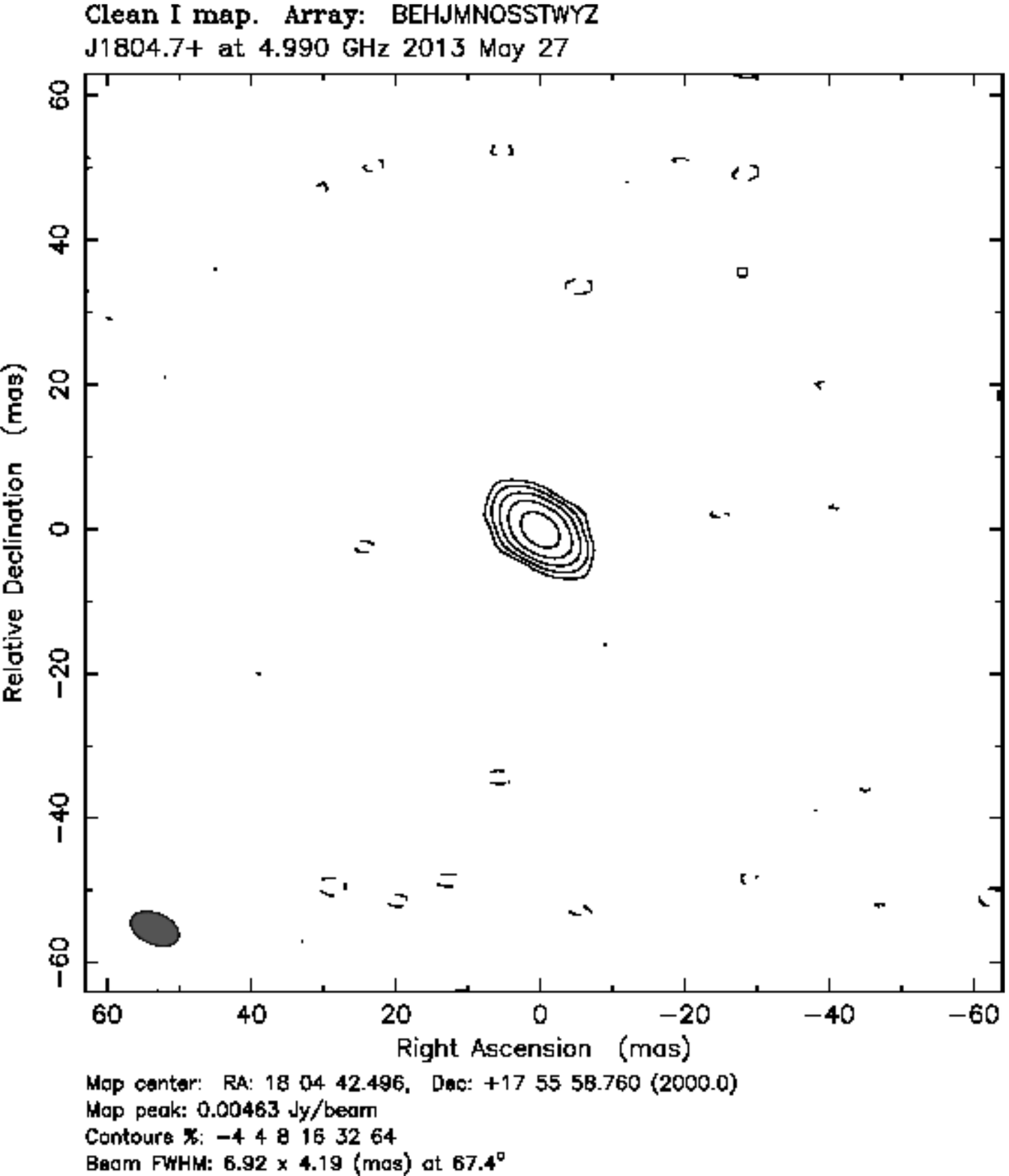} \\
\newline
\includegraphics[width=7cm,clip]{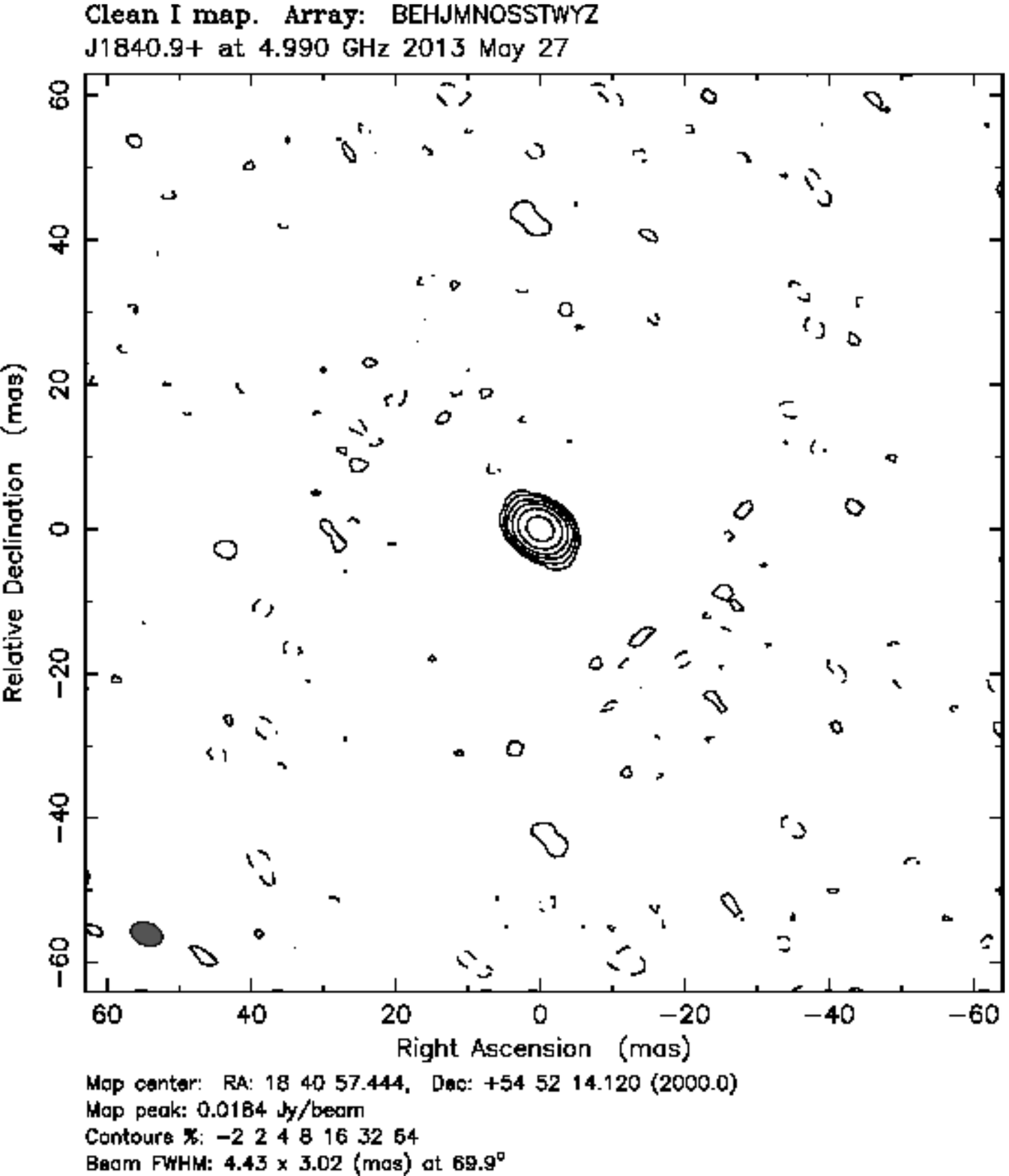}
\includegraphics[width=7cm,clip]{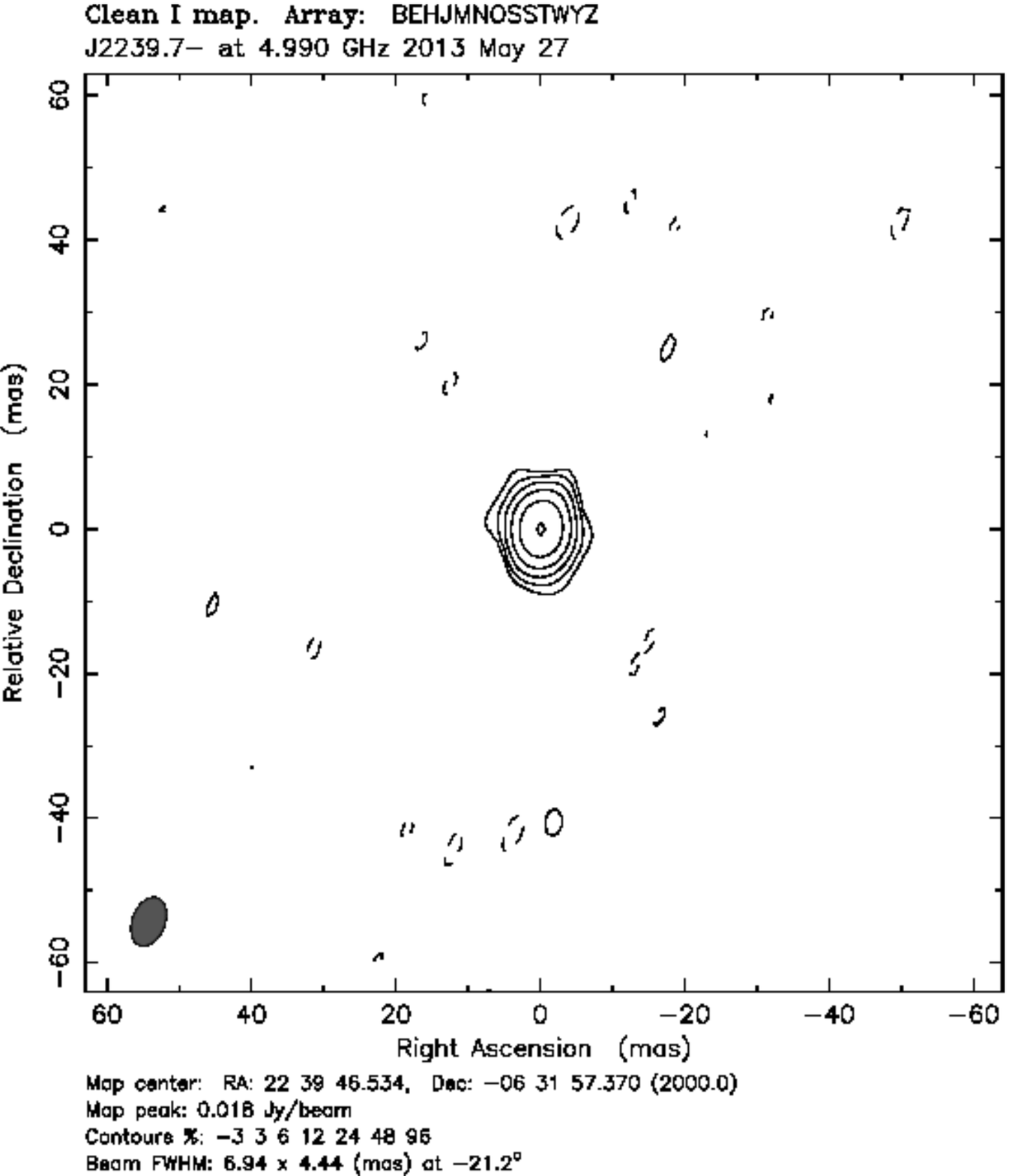} \\
\newline
\includegraphics[width=7cm,clip]{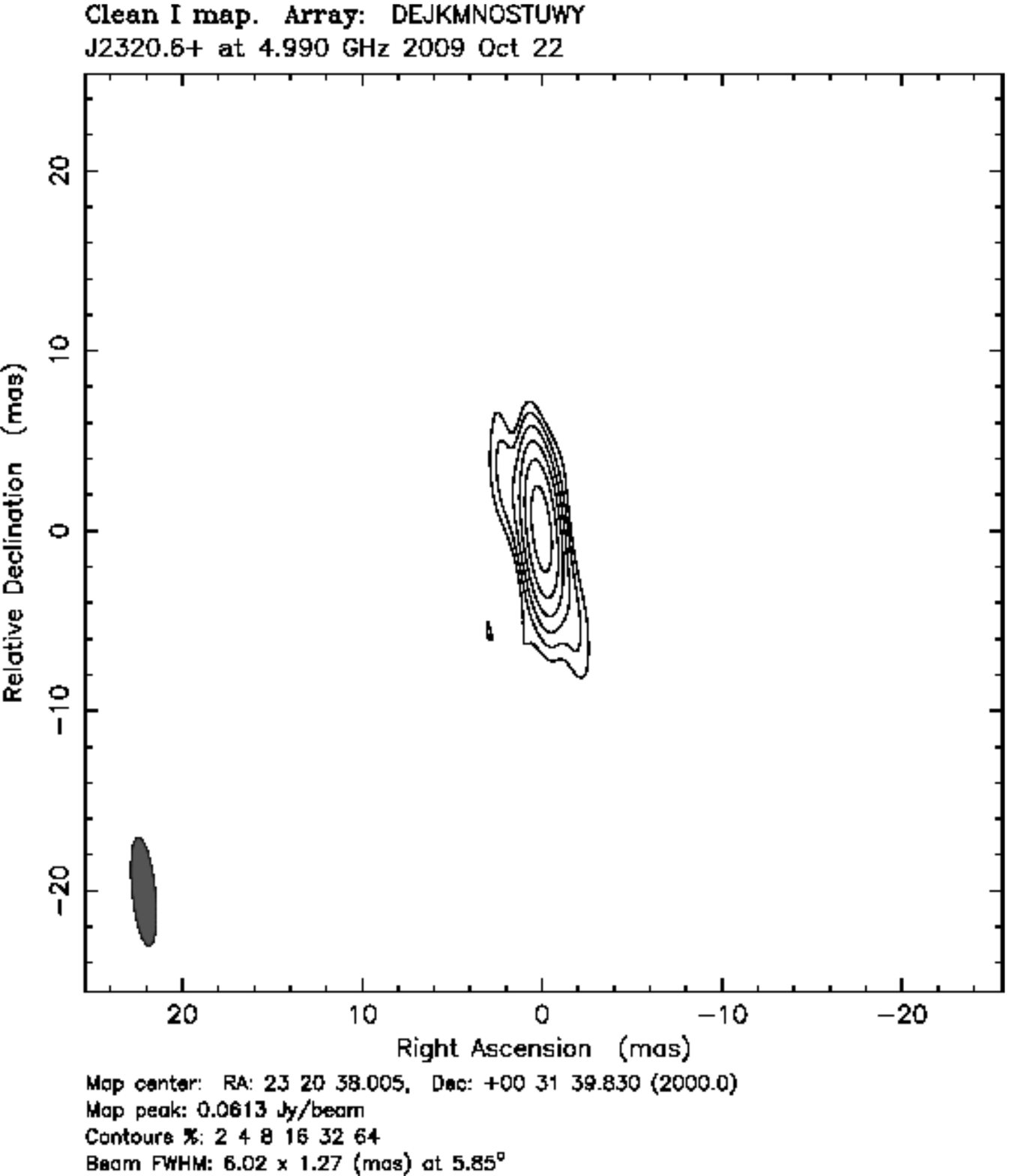} 
\includegraphics[width=7cm,clip]{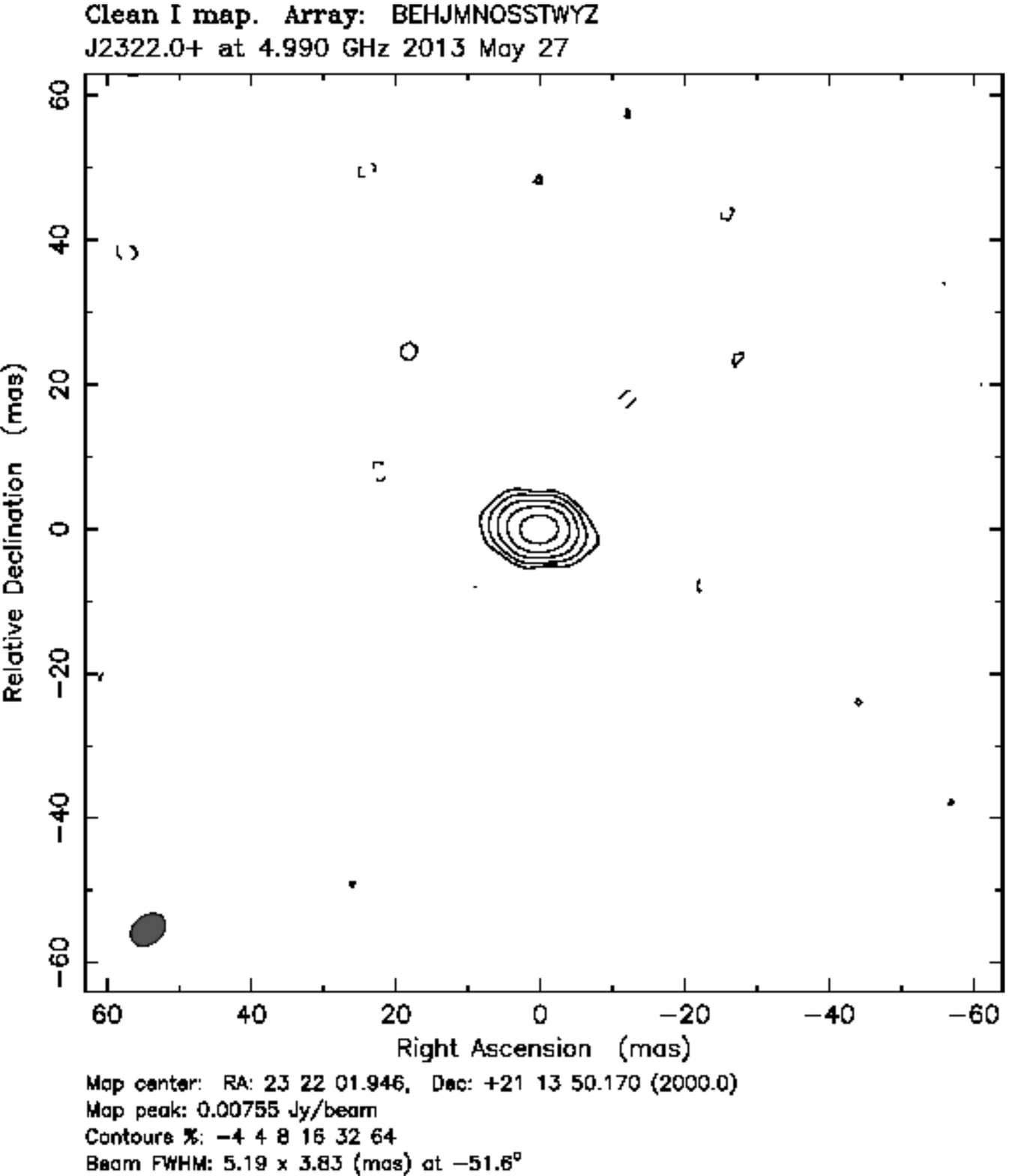}
\caption{EVN images at 5\,GHz. Continued}\label{appfig}
\end{figure*}
\begin{figure*}
\includegraphics[width=7cm,clip]{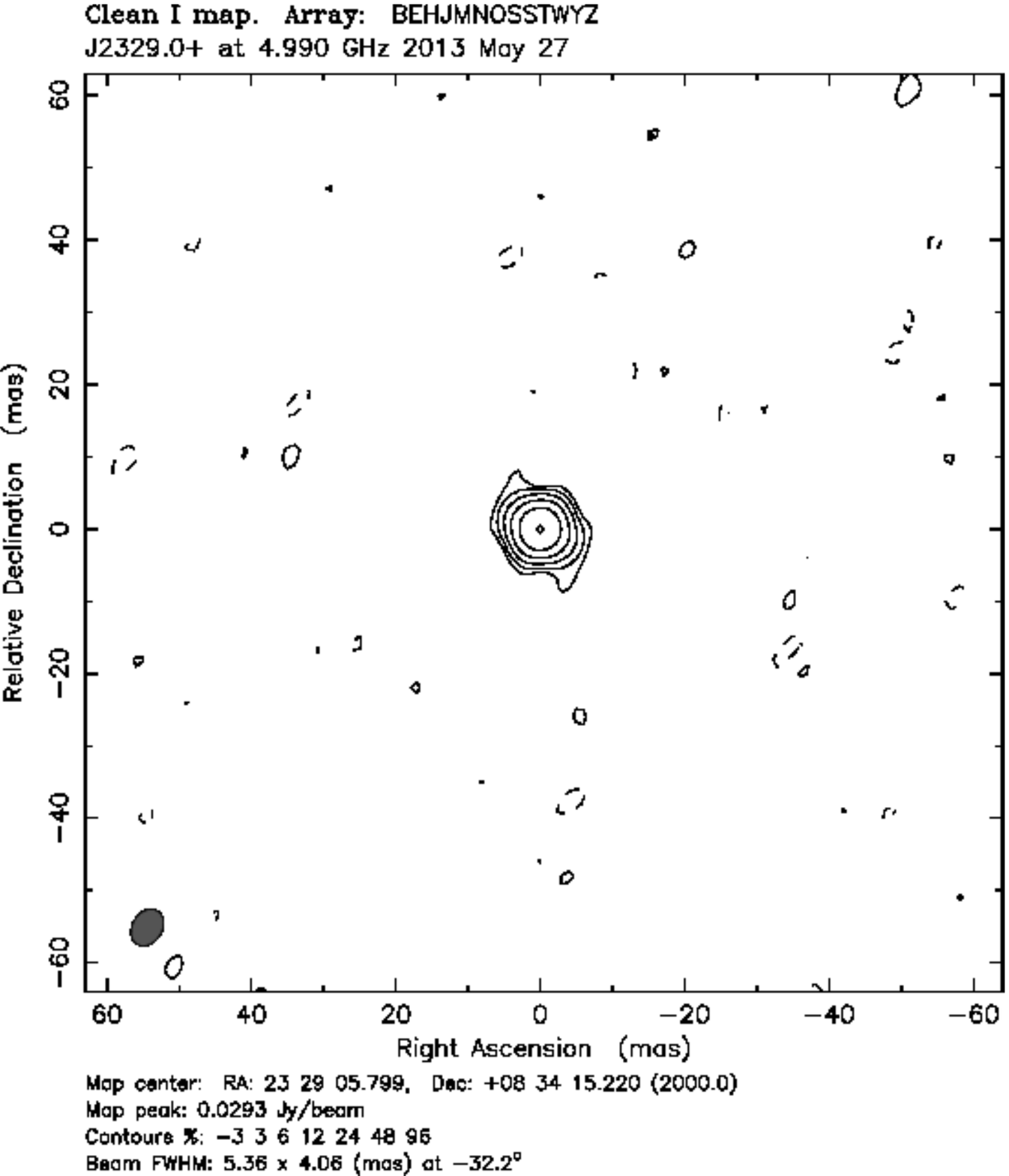}
\includegraphics[width=7cm,clip]{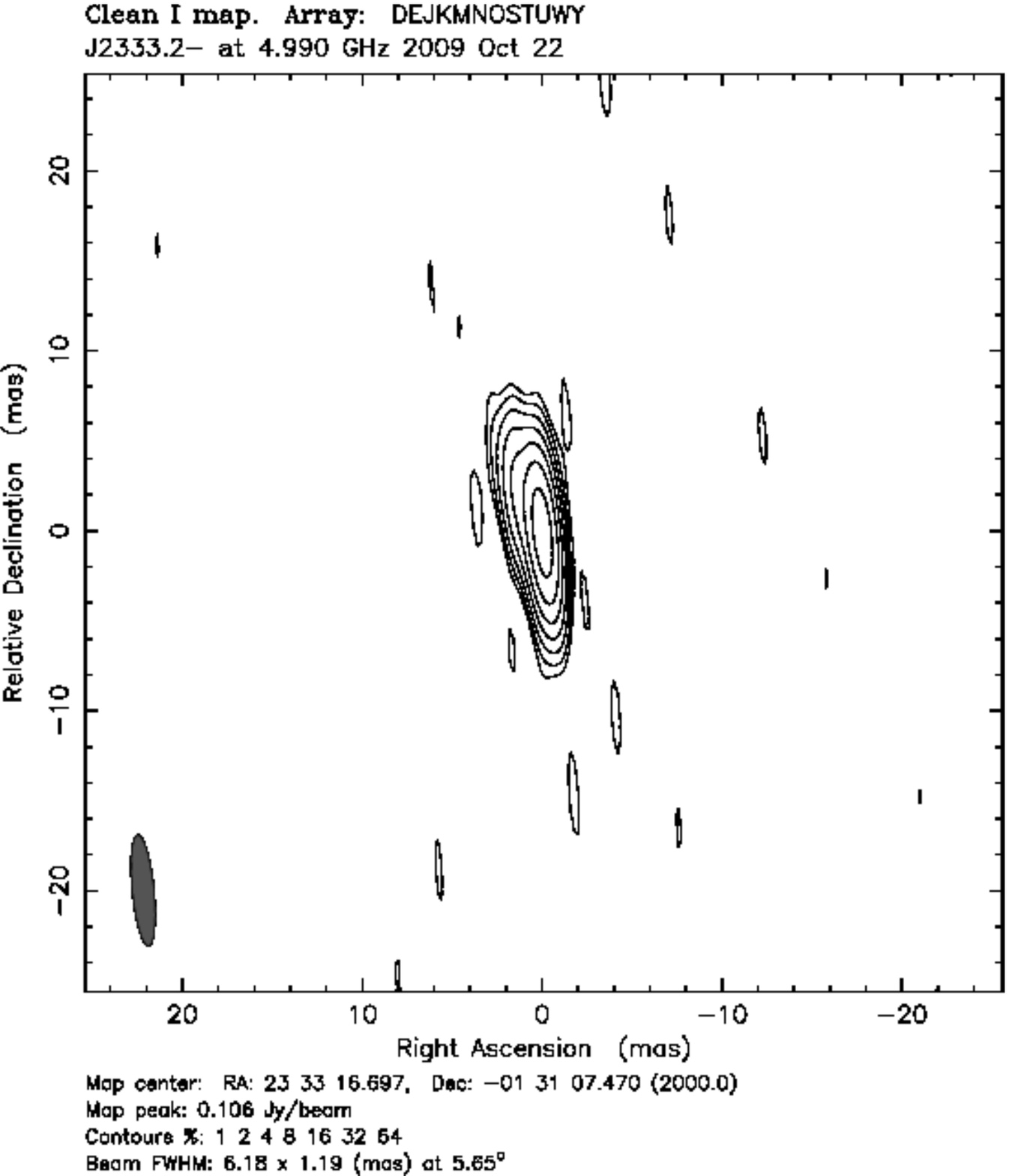} \\
\newline
\includegraphics[width=7cm,clip]{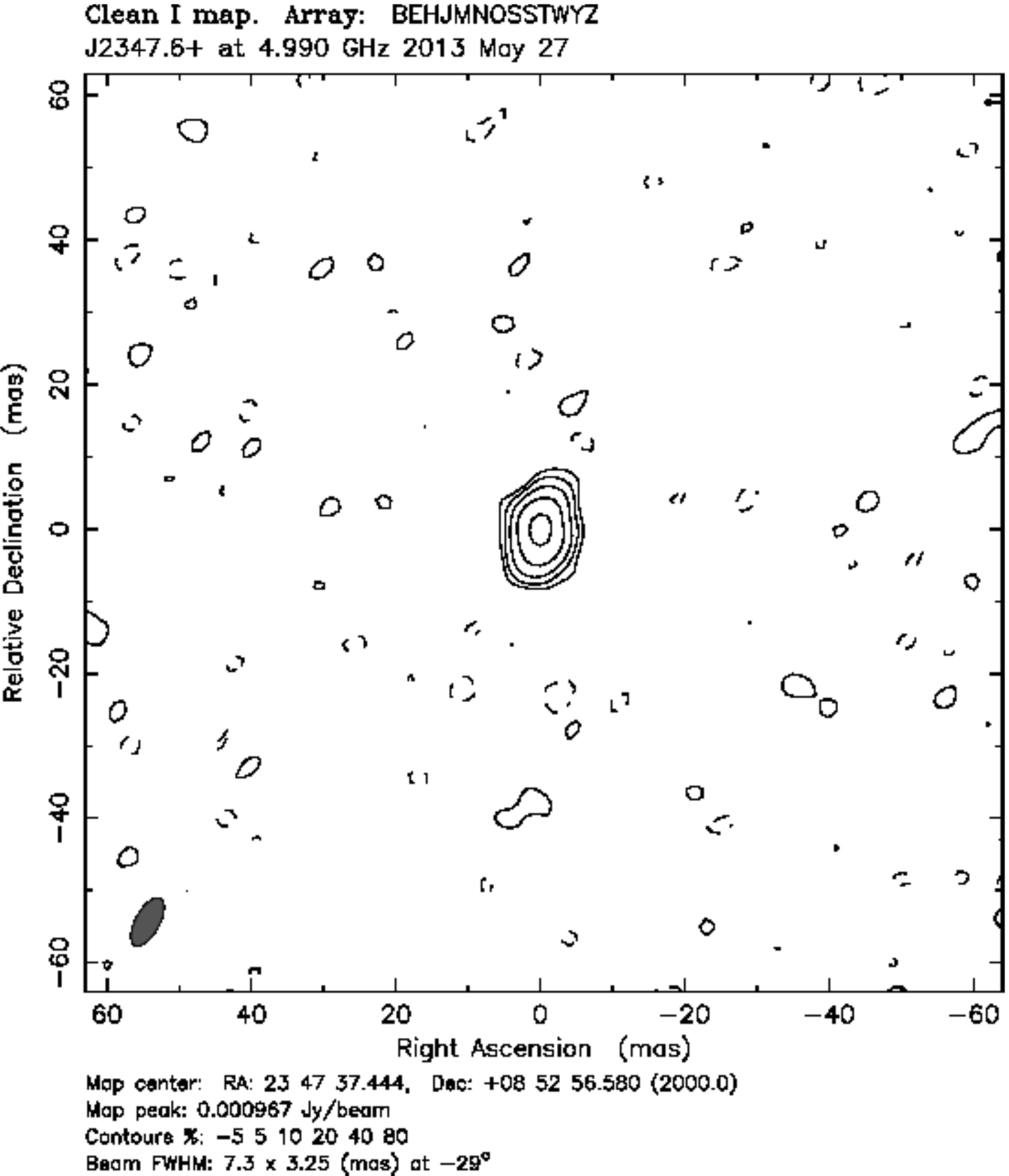}
\caption{EVN images at 5\,GHz. Continued}\label{appfig}
\end{figure*}
...
\end{appendix}


\begin{thebibliography}{}

\bibitem[Ackermann \etal (2011)]{Ackermann11}
Ackermann, M., Ajello, M., Allafort, A. et al. 2011 AJ, 741, 30

\bibitem[Becker \etal (1994)]{Becker94}
Becker R.H., White R.L. \& Helfand D.J. 1994, ASP Conf. Ser., 61, 165, 
eds. D.R. Crabtree, R.J. Hanisch \& J. Barnes

\bibitem[Condon \etal (1998)]{Condon98}
Condon, J.J., Cotton, W.D., Greisen, E.W. et al. 1998, AJ, 115, 1693

\bibitem[Cooper \etal (2007)]{Cooper07}
Cooper N.J., Lister M.L., \& Kochanczyk, M.D. 2007, ApJS, 171, 376

\bibitem[Deller \etal (2011)]{Deller11}
Deller, A.T., Brisken, W.F., Phillips, C.J. et al. 2011, PASP, 123, 275.

\bibitem[Dermer (1995)]{Dermer95}
Dermer, C.D. 1995, ApJ, 446, L63

\bibitem[Dermer \& Schlickeiser (1994)]{Dermer94}
Dermer, C.D. \& Schlickeiser, R. 1994 ApJS, 90, 945 

\bibitem[Fanti \etal (1990)]{Fanti90}
Fanti, R. Fanti, C. Schilizzi, R.T. et al. 1990 A\&A, 231,333

\bibitem[Fuhrmann \etal (2014)]{Fuhrmann14}	
Fuhrmann, L.; Larsson, S.; Chiang, J. et al. 2014 MNRAS 441, 1899

\bibitem[Gregory \etal (1996)]{Gregory96}
 Gregory, P. C., Scott, W. K., Douglas, K., Condon, J. J. 1996 ApJS, 103, 427 

\bibitem[Griffith \& Wright (1993)]{Griffith93}
Griffith, M.R. \& Wright, A.E. 1993 AJ, 105, 1666 

\bibitem[Jorstad \etal (2001)]{Jorstad01}
Jorstad, S.G., Marscher, A.P., Mattox, J.R. et al. 2001 ApJS, 134, 181 

\bibitem[Kellermann \etal (2004)]{Kellermann04}
Kellermann K.I., Lister M.L., Homan D.C. et al. 2004 ApJ, 609, 539

\bibitem[Kovalev \etal (2005)]{Kovalev05}
Kovalev, Y.Y., Kellermann, K.I., Lister, M.L. et al. 2005 AJ, 130, 2473 

\bibitem[Kovalev \etal (2009)]{Kovalev09}
Kovalev, Y.Y., Aller, H.D., Aller, M.F. et al. 2009 ApJ, 696: L17 

\bibitem[Landt \etal (2001)]{Landt01}
Landt, H., Padovani, P., Perlman, E.S. et al. 2001 MNRAS, 323, 757

\bibitem[Linford \etal (2012)]{Linford12}
Linford, J.D., Taylor, G.B., Romani, R.W. et al. 2012 ApJ, 744, 177 

\bibitem[Lister \& Marscher (1997)]{Lister97}
Lister, M.L., Marscher, A.P. 1997 ApJ, 476, 572 

\bibitem[Lister \& Homan (2005)]{Lister05}
Lister, M.L., Homan, D.C. 2005 AJ, 130, 1389 

\bibitem[Lister \etal (2009)]{Lister09}
Lister, M.L., Cohen, M.H., Homan, D.C. et al. 2009 AJ, 138, 1874

\bibitem[Lister (2010)]{Lister10}
Lister, M.L. in {\it ''Fermi meets Jansky - AGN in Radio and Gamma Rays``}, 
 T. Savolainen, E. Ros, R.W. Porcas, \& J.A. Zensus Eds. (Bonn:MPIfR), 159

\bibitem[Lister \etal (2011)]{Lister11}
Lister, M.L., Aller, M., Aller, H. et al. 2011 ApJ, 742, 27

\bibitem[Mantovani \etal (2009)]{Mantovani09}
Mantovani, F., Mack, K.-H., Montenegro-Montes, F.M. et al. 2009 A\&A, 502, 61

\bibitem[Mantovani \etal (2011)]{Mantovani11}
Mantovani, F., Bondi, M., \& Mack, K.-H. 2011 A\&A, 533, 79 

\bibitem[Nolan \etal (2012)]{Nolan12}
Nolan, P.L., Abdo, A.A., Ackermann, M. et al. 2012 ApJS, 199, 31

\bibitem[O'Dea (1998)]{O'Dea98}
O'Dea, C. 1998 PASP, 110, 493

\bibitem[Padovani \etal (2007)]{Padovani07}
Padovani, P., Giommi, P., Landt, H. \& Perlman, E.S. 2007 ApJ, 662, 182 

\bibitem[Perlman \etal (1998)]{Perlman98}
Perlman, E.S., Padovani, P., Giommi, P. et al. 1998 AJ, 115, 1253 

\bibitem[Pushkarev \etal (2010)]{Pushkarev10}
Pushkarev, A.B., Kovalev, Y.Y., \& Lister, M.L. 2010 ApJ, 722, L11

\bibitem[Shepherd \etal (1995)]{Shepherd95}
Shepherd, M.C., Pearson, T.J., \& Taylor, G.B. 1995, BAAS, 26, 305

\bibitem[Taylor \etal (2007)]{Taylor07}
Taylor, G.B., Healey, S.E., Helmboldt, J.F. et al. 2007 ApJ, 671, 1355

\bibitem[White \& Becker (1992)]{White92}
White, R.L., \& Becker, R.H. 1992 ApJS, 79, 331 

\bibitem[White, Giommi \& Angelini (1995)]{White95}
White, N.E., Giommi, P. \& Angelini, L. 1995
\newline
http://heasarc.gsfc.nasa.gov/wgcat 
%
\end{thebibliography}
\end{document}